\documentclass[aps,prd,secnumarabic,preprint,11pt,superscriptaddress,nofootinbib,floatfix]{revtex4}  

\usepackage{fancyhdr}
\usepackage{latexsym}
\usepackage{amsfonts}

\usepackage{float}
\usepackage{subfloat}
\usepackage{graphicx}
\usepackage{mathrsfs}
\usepackage{amssymb}
\usepackage{amsmath}
\usepackage{verbatim}
\usepackage{multirow}
\usepackage{color}
\usepackage{soul}
\usepackage{slashed}

\newcommand{\htb}[1]{{\color{blue} #1}}

\newcommand{\TeV}{{\ensuremath\rm TeV}}
\newcommand{\GeV}{{\ensuremath\rm GeV}}
\newcommand{\MeV}{{\ensuremath\rm MeV}}
\newcommand{\fb}{{\ensuremath\rm fb}}
\newcommand{\pb}{{\ensuremath\rm pb}}
\newcommand{\eqn}{equation}

\newcommand{\lb}{\left(}
\newcommand{\rb}{\right)}
\newcommand{\mO}{\mathcal{O}}
\newcommand{\lam}{\lambda}

\def\D0{\slash\!\!\!\!\!\!\!\!\!\:D0}
\newcommand{\HS}{\texttt{HiggsSignals}}
\newcommand{\HB}{\texttt{HiggsBounds}}
\newcommand{\HSv}[1]{\texttt{HiggsSignals-#1}}
\newcommand{\HBv}[1]{\texttt{HiggsBounds-#1}}

\newcommand{\oblique}{Altarelli:1990zd,Peskin:1990zt,Peskin:1991sw,Maksymyk:1993zm}

\newcommand{\lp}{\lambda_5}

\newcommand{\lczp}{\lambda_{345}}

\newcommand{\fr}{\frac}

\begin{document}
\bibliographystyle{hunsrt} 
\date{\today}
\title{Inert Doublet Model in light of LHC {Run I} {and astrophysical} data \vspace{2mm} \\ \vspace{0.5cm}}
\vspace*{1.0truecm}
\author{Agnieszka Ilnicka\vspace{0.2cm}}
\email{ailnicka@physik.uzh.ch}
\affiliation{Institute of Physics, University of Zurich, Winterthurstrasse 190, CH-8057 Zurich\vspace{0.2cm}}
\affiliation{Physics Department, ETH Zurich, \\ Otto-Stern-Weg 1, CH-8093 Zurich\vspace{0.2cm}}
\author{Maria Krawczyk\vspace{0.2cm}}
\email{krawczyk@fuw.edu.pl}
\affiliation{Faculty of Physics, University of Warsaw, ul. Pasteura 5, 02-093 Warsaw, Poland \vspace{0.2cm}}
\author{Tania Robens}
\email{Tania.Robens@tu-dresden.de}
\affiliation{TU Dresden, Institut f\"ur Kern- und Teilchenphysik,
Zellescher Weg 19, D-01069 Dresden, Germany\vspace{0.5cm}}

\renewcommand{\abstractname}{\vspace{0.5cm} Abstract}

\begin{abstract}
\vspace{0.5cm}

We discuss the parameter space of the Inert Doublet Model, a two Higgs doublet model with a dark matter candidate. {A}n extensive set of theoretical and experimental constraints on this model is considered, where both collider as well as astroparticle data limits, {the latter in the form of dark matter relic density as well as direct detection,} are taken into account. We discuss the effects of these constraints on the parameter space {of the model}. {In particular, we do not require the IDM to provide the full dark matter content of the universe, which opens up additional regions in the parameter space accessible at collider experiments.} The combination of all constraints leads to a relatively strong mass {degeneracy} in the dark scalar sector {for masses $\gtrsim\,200\,\GeV$}, and to a minimal scale {$\sim\,45\,\GeV$} for the dark scalar masses. {We also observe a stringent mass hierarchy $M_H^\pm\,>\,M_A$.} We propose benchmark points and benchmark planes for dark scalar pair-production for the current LHC run {being in compliance with all theoretical as well as experimental bounds}. 

\end{abstract}

\maketitle
\tableofcontents

\newpage


\section{  Introduction}
\label{Sec:Intro}
{The Inert Doublet model (IDM)  is one of the {most straightforward} extensions of the Standard Model (SM) \cite{Deshpande:1977rw,Cao:2007rm,Barbieri:2006dq}. {It is a special type of Two Higgs Doublet Model (2HDM)}. In the scalar sector it contains, in addition to the {SU(2)} doublet  $\phi_S$ responsible for the spontaneous breaking of electroweak symmetry, {a} second scalar doublet {$\phi_D$} with vanishing vacuum expectation value {($vev$)}. {This second doublet} is not involved in mass generation {via} spontaneous symmetry breaking and {does not} couple to fermions {(in this respect the model corresponds to a Type I  2HDM.).} 
{In this model  a Z$_2$ symmetry, which we label $ D$-symmetry {below}, {with the discrete transformations defined as}
}:
{
\begin{equation}
 \phi_D \to - \phi_D, \,\,
\phi_S\to \phi_S, \,\,
\text{SM} \to \text{SM},
\end{equation}
}
 is respected by the Lagrangian 
and  the vacuum.

The $\phi_S$ doublet plays the same role as the corresponding  doublet in the SM, {providing} the SM-like Higgs particle. This doublet is even under the $D$ symmetry, while the second doublet, $\phi_D$, is $D$-odd. This so called inert or dark doublet contains 4 scalars, two charged and two neutral ones, with the lightest neutral scalar being a natural DM candidate.
{This model was studied in order to provide a heavy Higgs boson \cite{Barbieri:2006dq} as well as a lighter  Higgs boson, to be produced at LHC \cite{Cao:2007rm}.  It was considered as a model with a \lq\lq{}perfect example\rq\rq{}  of {a} weakly interacting massive particle \cite{LopezHonorez:2006gr,Honorez:2010re,Dolle:2009fn,Sokolowska:2011aa}.  It leads to an interesting pattern of the Universe evolution, towards the Inert phase as given by the IDM, with one, two or three
phase transitions \cite{Ginzburg:2010wa}. Furthermore, the IDM can provide a strong first-order phase transition \cite{Hambye:2007vf,Chowdhury:2011ga,Borah:2012pu,Gil:2012ya, Blinov:2015vma},  which  is one of the Sakharov conditions needed to generate a baryon asymmetry of the Universe}.
After the discovery of a SM-like Higgs particle in 2012, many analyses have been performed for the IDM, using  Higgs {collider} data, as well as astrophysical measurements, {see e.g. {\cite{Swiezewska:2012eh,Gustafsson:2012aj,Arhrib:2013ela,Krawczyk:2013jta}}}\footnote{{Recent analyses for models which extend the IDM by an additional singlet have been performed in \cite{Banik:2014cfa,Banik:2014eda,Bonilla:2014xba,Plascencia:2015xwa}.}}. {In addition,} proposals {were made} how to search for  dark scalars at {the} LHC {in the di- or multilepton channel} \cite{Dolle:2009ft,Swaczyna,Gustafsson:2012aj}.

Recently, also the important issue of vacuum (meta)stability in the IDM has been discussed, and it was found that additional (possibly heavy)  scalars can have a strong impact on it \cite{Kadastik:2011aa,Goudelis:2013uca,Swiezewska:2015paa,Khan:2015ipa} \footnote{Similar solutions can be found in a simple singlet extension of the SM Higgs sector, cf. e.g. \cite{Pruna:2013bma, Robens:2015gla} and references therein.}.  Especially {facing} the LHC run II, the determination of the regions of parameter space which survive all current theoretical and experimental constraints is indispensible. Therefore, we here provide a survey of the IDM parameter space, after all constraints are taken into account. The main sources of these {stem from} run I LHC Higgs data as well as dark matter relic density (Planck) and direct dark matter search at LUX. {A first preliminary preview on some of our} results {was} presented in {a conference proceeding} \cite{Ilnicka:2015sra}.

\section{ The model}
\label{sec:model}
\noindent
The $D$-symmetric  potential of the IDM has the following form:
\begin{equation}\label{pot}\begin{array}{c}
V=-\fr{1}{2}\left[m_{11}^2(\phi_S^\dagger\phi_S)\!+\! m_{22}^2(\phi_D^\dagger\phi_D)\right]+
\fr{\lambda_1}{2}(\phi_S^\dagger\phi_S)^2\! 
+\!\fr{\lambda_2}{2}(\phi_D^\dagger\phi_D)^2\\[2mm]+\!\lambda_3(\phi_S^\dagger\phi_S)(\phi_D^\dagger\phi_D)\!
\!+\!\lambda_4(\phi_S^\dagger\phi_D)(\phi_D^\dagger\phi_S) +\fr{\lambda_5}{2}\left[(\phi_S^\dagger\phi_D)^2\!
+\!(\phi_D^\dagger\phi_S)^2\right],
\end{array}\end{equation}
with all  parameters real
(see e.g. \cite{Ginzburg:2010wa}). In a 2HDM with this potential different vacua can exist, e.g. a mixed one with $\langle\phi_S\rangle\neq0$, $\langle\phi_D\rangle\neq0$,  an inertlike vacuum with $\langle\phi_S\rangle=0$, $\langle\phi_D\rangle\neq0$, or {{even a charge breaking vacuum}} (see~\cite{Ginzburg:2010wa,Sokolowska:2011aa,Sokolowska:2011sb,Sokolowska:2011yi}).  
The decomposition around {the} vacuum state in the IDM is given by
\begin{equation} \label{dekomp_pol}
\phi_S = \begin{pmatrix}\phi^+\\ \frac{1}{\sqrt{2}} \lb v+h+i\xi \rb \end{pmatrix}\,,\qquad \phi_D  = 
\begin{pmatrix} H^+ \\ \frac{1}{\sqrt{2}} \lb H+iA \rb  \end{pmatrix}, 
\end{equation}
{where}
$v\,=\,246\,\GeV$ denotes the vacuum expectation value.

The first doublet, $\phi_S$, contains the  SM-like Higgs boson $h$,   with the mass  
\begin{equation}\label{Higgsmass}
M_{h}^2=\lambda_1v^2= m_{11}^2, 
\end{equation}
which, according to the LHC data, {is close to
$ 125 \textrm{ GeV}$ \cite{Aad:2015zhl}.}

The second doublet, $\phi_D$, consists of four dark (inert) scalars $H,\,A,\,H^\pm$, which do not 
couple to fermions at the tree-level. Their masses are {given as follows}:
\begin{eqnarray}\label{mass}
M_{H^\pm}^2&=&\fr{1}{2} \left(\lambda_3 v^2-m_{22}^2\right),\nonumber \\
M_{A}^2&=&M_{H^\pm}^2+\fr{1}{2}\left(\lambda_4-\lambda_5\right)v^2\ =\fr{1}{2}( \bar{\lambda}_{345}v^2-m_{22}^2), \nonumber \\
 M_{H}^2&=&
M_{H^\pm}^2+\fr{1}{2}\left(\lambda_4+\lambda_5\right)v^2\ = {\fr{1}{2}}(\lambda_{345}v^2-m_{22}^2 ).
\end{eqnarray}
\noindent
Due to an exact $D$ symmetry the lightest neutral scalar $H$ (or $A$) is stable and thereby may serve as a good dark matter candidate.\footnote{Charged DM 
 has been strongly limited by astrophysical analyses \cite{Chuzhoy:2008zy}.}
We take $H$ to be the DM candidate and so ${M_H} < M_A, M_{H^\pm}$ ({this choice implies}  $\lp<0,\ {{\lam_{45}}}=\lambda_{4}+\lambda_{5}<0$). {Note that, as the second doublet does not couple to fermions, one cannot ascribe a CP-property to the scalars of the dark sector. Therefore, in contrast to general two Higgs doublet models, in our scenario the two neutral scalars $H$ and $A$ can be treated on equal footing, and in fact the corresponding couplings correspond to each other using the replacement $\lam_5\,\leftrightarrow\,-\,\lam_5$, cf. Appendix \ref{app:fr}.}

The parameters 
\begin{equation}
\lambda_{345}=\lambda_3 + \lambda_4 + \lambda_5,\;
{\bar{\lambda}_{345}=\lambda_3 + \lambda_4 - \lambda_5}
\end{equation}
are related to {the} triple and quartic coupling  between the SM-like Higgs $h$ and the DM candidate $H$ or the 
scalar $A$, {respectively}, while $\lambda_3$ describes 
the $h$  interaction with the charged scalars $H^\pm$. The parameter $\lambda_2$ describes the quartic self-couplings of dark particles. {Note that an equivalent choice would be to take $A$ as the lightest neutral dark scalar and therefore as the dark matter candidate, which implies $\lam_5\,>\,0$. In this case, the  discussion below remains valid by replacing
\begin{\eqn*}
\lam_5\,\longleftrightarrow\,-\lam_5,\,\lam_{345}\,\longleftrightarrow\,\bar{\lam}_{345}.
\end{\eqn*}
}
{The parameter $m_{22}^2$  sets here a common {mass} scale for the dark  particles}. 
{For a list of all {relevant} Feynman rules for this model, see Appendix \ref{app:fr}.}

{After electroweak symmetry breaking,} the model has seven free parameters. Since {the} values of $v$ and {the SM-like} Higgs mass {are} fixed by experimental data, there are {in total} 5 {free} parameters  {defining} {the scalar sector of the IDM.} {Typical choices are either} physical parameters $(M_H,M_A,M_{H^{\pm}}, \lam_2, \lam_{345} )$ or  potential parameters $(m_{22}^2,\lam_2, \lam_3, \lam_4, \lam_5)$. We here choose to work in the physical basis.

\section{ Constraints}\label{sec:constraints}
As has been widely discussed in the literature, the IDM is subject to numerous constraints, 
both from theoretical {{conditions}} as well as from experimental results. We here briefly remind the reader of these constraints and refer to the literature \cite{Cao:2007rm, Agrawal:2008xz, Gustafsson:2007pc, Dolle:2009fn, Dolle:2009ft, LopezHonorez:2006gr, Arina:2009um, Tytgat:2007cv, Honorez:2010re, Krawczyk:2009fb, Kanemura:1993hm, Akeroyd:2000wc, Swiezewska:2012ej, Lundstrom:2008ai, Gustafsson:2010zz,Gustafsson:2012aj,Modak:2015uda} for further details.

\subsection{ Theoretical constraints}\label{sec:thconst}

\begin{itemize}
\item{}First, we require the potential to be bounded from below, such that there is no field configuration for which $V\,\rightarrow\,-\infty$. This directly leads to the following conditions on the couplings \cite{Nie:1998yn}
\begin{\eqn}
\lam_1\,>\,0,\,\lam_2\,>\,0,\,\lam_3+\sqrt{\lam_1 \lam_2}>0,\,\lam_{345}+\sqrt{\lam_1 \lam_2}\,>0,
\end{\eqn}
 {with $\lam_{345}$ as defined above.} The above relations hold on tree level and neglect higher order contributions which {could} (de)stabilize the electroweak vacuum. For a more detailed discussion of such effects, cf. e.g. \cite{Goudelis:2013uca, Swiezewska:2015paa}. 
\item We also require the scalar {$2\,\rightarrow\,2$} scattering matrix\footnote{{We consider electroweak gauge boson scattering in the high energy limit, where {the corresponding Goldstone modes are used in accordance} {with} the equivalence theorem \cite{Chanowitz:1985hj}.}} to be unitary. {We do this by the means of perturbative unitarity \cite{Lee:1977eg}, where we} {consider} all scattering matrices {{for scalars}}  with specific hypercharge and isospin and {require} the eigenvalues to be $|\L_i|\,\leq\,8\,\pi$. This follows the generic prescription as e.g. given in \cite{Ginzburg:2005dt}, where we make use of the routine implemented in the Two Higgs Doublet Model Calculator (2HDMC) tool \cite{Eriksson:2009ws}.
\item{}Furthermore, we require all quartic Higgs coupling to be perturbative, i.e. to take absolute values $\leq\,4\,\pi$. {We apply this limit both to the coupling parameters in the potential, $\lam_i$, as well as the couplings stemming from vertex Feynman rules. \footnote{{The latter is implemented as a standard constraint in \cite{Eriksson:2009ws}.}}}
\item In the IDM, two {minima} can coexist. In order to guarantee the {{inert}} vacuum to be global, we therefore 
require \cite{Ginzburg:2010wa, Gustafsson:2010zz,Swiezewska:2012ej}
\begin{\eqn}\label{eq:invac}
\frac{m_{11}^2}{\sqrt{\lam_1}}\,\geq\,\frac{m_{22}^2}{\sqrt{\lam_2}}.
\end{\eqn}
\end{itemize}
\subsection{Experimental constraints}\label{sec:expconst}
Apart from the theoretical constraints discussed above, several experimental constraints exist which impose further bounds on the models parameter space.
\begin{itemize}
\item{}We set the mass of the SM-like Higgs boson $h$ to $$M_h=125.1\,\GeV \label{Eq:mhexp},$$ in agreement with the results from the LHC experiments \cite{Aad:2015zhl}
 $M_h\,=\,125.09\,\pm\,0.24\,\GeV$.
\item{}We furthermore require the {total} width of the 125\,\GeV~ Higgs to obey an {upper} limit \cite{Khachatryan:2014iha,Aad:2015xua}
\begin{\eqn*}
\Gamma_\text{tot}\,\leq\,22\,\MeV.
\end{\eqn*}
{The validity of the above constraint relies on the assumption that no new particles enter in the loop-induced ggh vertex and the corresponding continuum contributions\footnote{{See the original proposal in \cite{Caola:2013yja} as well as the discussion in \cite{Englert:2014aca}}.}; both of these assumptions are fulfilled in the IDM. Obviously, another contamination of the proposal in \cite{Caola:2013yja} would be additional new physics diagrams leading to the same final states; in our case, this could come from e.g. $AA$ production and successive decays with negligible missing transverse energy. As discussed below, contributions from these final states are usually relatively small, and we therefore consider these to be negligible. A more detailed discussion is in the line of future work.} 

\item{}
Furthermore, we take into account  strong bounds {provided by} the total width of the electroweak gauge bosons {(cf. e.g. \cite{Agashe:2014kda})}, {in a following simple form:}

\begin{\eqn}\label{eq:gwgz}
M_{A,H}+M_{H^\pm}\,\geq\,m_W,\,M_A+M_H\,\geq\,m_Z,\,2\,M_{H^\pm}\,\geq\,m_Z.
\end{\eqn} 

\item{} We {also} require a $2\,\sigma$ {(i.e. $95 \%$ C.L.)} agreement with electroweak precision observables, parametrized through the electroweak oblique parameters $S,T,U$ \cite{\oblique}.

{The constraints from the electroweak oblique parameters $S$, $T$ and $U$ are included} by evaluating
\begin{align}
\chi^2_\mathrm{STU} = \mathbf{x}^T \mathbf{C}^{-1} \mathbf{x},
\end{align}
with $\mathbf{x}^T = (S - \hat{S}, T - \hat{T}, U - \hat{U})$.  {The observed parameters are given by~\cite{Baak:2014ora}
\begin{align}
\hat{S} = 0.05,\quad \hat{T} = 0.09,\quad \hat{U} = 0.01,
\end{align}
while} the \emph{unhatted} quantities denote the model predictions, which we obtained from interfacing with the publicly available code 2HDMC \cite{Eriksson:2009ws}. The covariance matrix reads~\cite{Baak:2014ora}\footnote{{We here use the best linear unbiased estimator, cf. \cite{Baak:2012kk}. See also \cite{Lopez-Val:2014jva}.}}
\begin{align}
(\mathbf{C})_{ij} = \left(\begin{array}{ccc}
0.0121 & 0.0129 & -0.0071 \\
0.0129 & 0.0169 & -0.0119 \\
-0.0071 & -0.0119 & 0.0121 \\
\end{array}\right).
\end{align}
We then require $\chi^2_\mathrm{STU} \le 8.025$, corresponding to a {maximal} $2\sigma$ deviation given the three degrees of freedom.
 \item{} {In order to evade bounds from long-lived charged particle searches,} we conservatively set an upper limit on the charged scalar lifetime of $\tau\,\leq\,10^{-7}\,s$, to guarantee decay within the detector. This translates to {an upper bound on the total decay width of the charged scalar $H^\pm$ of} $\Gamma_\text{tot}\,\geq\,6.58\,\times\,10^{-18}\,\GeV$. 
{\item{}
 A bound on the lower mass of $M_{H^\pm}$ has been derived in \cite{Pierce:2007ut}. Although this bound does lack a dedicated analysis within the current models' framework, 
{we take} $M_{H^\pm}\,\geq\,70\,\GeV$ {as a conservative lower limit}.
\item{}We also {require agreement with} the  null-searches from the LEP, Tevatron, and LHC experiments. We do this via the publicly available tool {\HBv4.2.1}~\cite{Bechtle:2008jh, Bechtle:2011sb, Bechtle:2013wla}, which includes results from all relevant experimental searches until summer 2014.
\item{} 
 We require agreement within $2\,\sigma$ for the 125 GeV~ Higgs signal strength measurements. For this, we make use of the publicly available tool {\HSv1.4.0}~\cite{Bechtle:2013xfa}, {and require $\Delta \chi^2\,\leq\,11.3139$}, corresponding to a $95 \%$ confidence level\footnote{{We used the signal strength parametrization of the \texttt{latestresults-1.3.0-LHCinclusive} data sample within \HS~.}}}.

\item{}Finally, {some} collider searches for dark matter, e.g. in supersymmetric models, could be reinterpreted within our model\footnote{Several tools {exist}  which  allow for recasting experimental results in such frameworks, cf. e.g. \cite{Drees:2013wra,Conte:2014zja}.}. While doing this is  beyond the scope of the current work, we use one {important reinterpretation of} a LEP analysis \cite{EspiritoSanto:2003by} within the IDM framework \cite{Lundstrom:2008ai}. This particularly rules out all regions where
\begin{\eqn}\label{eq:leprec}
M_A\,\leq\,100\,\GeV,\,M_H\,\leq\,80\,\GeV,\,\, \Delta M {(A,H)}\,\geq\,8\,\GeV,
\end{\eqn}
simultaneously.
\item{}
After taking into account all the above limits we are outside of the region excluded due to the {recent} reinterpretation of the supersymmetry analysis from LHC run I \cite{Belanger:2015kga}.

\item{} We apply dark matter relic density limits obtained by the Planck experiment \cite{Planck:2015xua}:
\begin{\eqn}\label{eq:planck}
\Omega_c\,h^2\,=\,0.1197\,\pm\,0.0022.
\end{\eqn}
In this work, we do not require the dark matter candidate of the IDM to provide the full relic density, but use it as an upper limit\footnote{{In such a scenario, additional dark matter candidates would be needed in order to account for the missing relic density; {cf. e.g. \cite{Cheung:2012qy} for a dedicated discussion of such scenarios within a supersymmetric setup.}}} {(cf. e.g. also the discussion in \cite{Bonilla:2014xba}).}
Being conservative, we require
\begin{\eqn}\label{eq:planck_up}
\Omega_c\,h^2\,\leq\, 0.1241,
\end{\eqn}
which corresponds to not overclosing the universe at $\sim\,95\,\%$ confidence level. {Additionally,  we {identify regions which are within a} 2 $\sigma$ {band} from the central value {and} thus {reproduce the} observed DM density {at $95\%$ confidence level}.}

\item{} To respect the most stringent direct detection limits from dark matter-nucleon scattering  provided by the LUX experiment, (see Fig. 1 \cite{Akerib:2013tjd}),
\begin{figure}[!tb]
\begin{minipage}{0.4\textwidth}
\includegraphics[width=\textwidth]{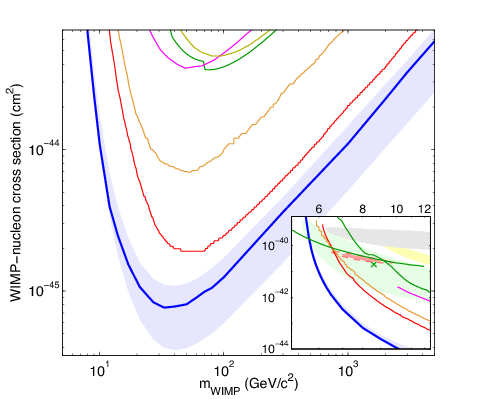}
\end{minipage}
\caption{\label{fig:luxpap}  Upper limit on dark matter-nucleon scattering cross-sections from the LUX experiment as a function of the dark matter mass. Taken from \cite{Akerib:2013tjd}.}
\end{figure}%
we use an approximation in form of an analytic formula 
\begin{eqnarray}\label{eq:luxlim}
\frac{\sigma_\text{max}\lb M_{H}\rb}{[\text{cm}^2]}&=&1.86934\,\times\,10^{-46} + 1.1236\times\,10^{-47}\,{M_H {/ \GeV}} + (-1.32312\times 10^{-52})\,\lb M_H {/\GeV}\rb^2 \nonumber \\
&& + \frac{1.85167\times10^{-28}}{(-9.35439\times 10^{15})\,\lb M_H {/\GeV}\rb + 1.12151\times 10^{15}\,\lb M_H {/\GeV} \rb^2},
\end{eqnarray}
which,
{for ${M_H}\,\geq\,10\,\GeV$, reproduces the actual exclusion limits on the $1-8\,\%$ level.} 
\end{itemize}
{The above exclusion constraints, together with the respective constraint applied for each criterium, are again summarized in Table \ref{tab:constraints}. The statistical treatment of search constraints from limits provided by \HB~and \HS~have been described at great length in \cite{Bechtle:2008jh, Bechtle:2011sb, Bechtle:2013wla,Bechtle:2013xfa}, which we refer the reader to for more details. All above constraints are applied assuming no correlation between the different limits.}
\begin{table}
\begin{tabular}{l|l|l}
Constraint& Criterium& Used Tools\\ \hline
Higgs width of $M_h$&$\Gamma_h\,\leq\,22\,\MeV$&2HDMC\\
no electroweak gauge boson decay into new states&Eqn.(\ref{eq:gwgz})&2HDMC\\
$S,T,U$&$\Delta\chi^2\,\leq\,8.025$, assuming 3 d.o.f&2HDMC\\
decaying $H^\pm$&$\Gamma_{H^\pm}\,\geq\,6.58\,\times\,10^{-18}\,\GeV$&2HDMC/ Madgraph\\
lower bound on $M_{H^\pm}$&$M_{H^\pm}\,\geq\,70\,\GeV$&--\\
agreement with collider searches&see \cite{Bechtle:2008jh, Bechtle:2011sb, Bechtle:2013wla}&\HB\\
agreement with signal strength measurement&$\Delta\,\chi^2\,\leq\,11.3139$, assuming 5 d.o.f.&\HS\\
recast LEP searches&Eqn.~(\ref{eq:leprec})&--\\
recast LHC searches&not applicable after other constraints&--\\
agreement with dark matter relic density (upper limit)&$\Omega\,h^2\,\leq\,0.1241$ (2 $\sigma$ deviation)&MicrOmegas\\
agreement with LUX upper bound on direct detection&Eqn.(\ref{eq:luxlim})&MicrOmegas
\end{tabular}
\caption{\label{tab:constraints} Experimental constraints, exclusion criterium applied, and tools used for the calculation of the respective criterium for the experimental constraints discussed in Section \ref{sec:expconst}. No correlation was assumed between the different rows listed above.}
\end{table}

{Note that, for a multi-component dark matter scenario, the above upper limit depends on the actual DM relic density for the specific point in parameter space; therefore, we have to introduce a rescaling factor, leading to the (relic density dependent) limit
\begin{\eqn}\label{eq:rescale}
\sigma\,(M_H)\,\leq\,\sigma^\text{LUX}(M_H) \times\,\frac{\Omega^\text{Planck}}{\Omega (M_H)},
\end{\eqn}
where we use $M_H$ as a short-hand notation to describe the dependence on a specific parameter point \footnote{{See also \cite{Cao:2007fy,Cheung:2012qy,Belanger:2014bga,Belanger:2014vza,Badziak:2015qca}.}}. The above rescaling, however, only applies in a multi-component dark matter scenario. We will therefore use the hard LUX limit in the remainder of the discussion, and comment on the effect of assuming a multi-component DM scenario in Section \ref{sec:darkmc}.
}

\subsection{Comparison with previous {analyses} }

Before the discussion of the scan setup and respective results, it is worthwhile to compare the list of constraints as given above with previous scan results from the literature. We restrict ourselves to work after the discovery of the Higgs boson, which set the SM-like Higgs mass $M_h\,\sim\,125\,\GeV$ accordingly \cite{Swiezewska:2012eh,Arhrib:2013ela, Krawczyk:2013jta, Swiezewska:2012ej,Goudelis:2013uca}  (see appendix \ref{app:bms} for a more detailed {comparison}).

{We summarize the major differences as follows}
\begin{itemize}
\item{}In this work, we apply the most updated bounds from collider and astroparticle physics. We consistently use limits from the Higgs signal strength by the means of a combined fit, as implemented in \HS, {and combine these with the most updated constraints from astrophysical data.}
\item{}We specifically identify regions where single constraints function as {\sl decisive}, and point to direct correlations between theoretical or experimental constraints and limits in two dimensional parameter planes, whenever possible. {This} will help to directly determine the effects of improved measurements on the models parameter space, {and contrast these to} regions {that} are excluded per se due to theoretical constraints, and therefore will not be affected by possible future improvements in experimental precision.
\item{}{We use dark matter relic density as an {\sl upper limit}, which opens up regions with intermediate dark scalar masses $\lesssim\,600\,\GeV$, which promise interesting from a collider perspective.}
\item{}Finally, none of the studies provide detailed benchmark points for the current LHC run, which are highly requested by the experimental community in order to tune and optimize search strategies\footnote{We thank the conveners of the Higgs Cross Section working group for encouraging us to select and present benchmark points for our model.}. {Most} {p}revious studies of LHC phenomenology of this model, although presenting a generic idea, mainly discuss parameter points which are no longer in compliance with current experimental bounds {(cf. discussion in Appendix \ref{app:bms}).} In addition to providing an explicit scan, the specification of benchmark points and planes for the present LHC respecting all present bounds run is a major outcome of the present study.

\end{itemize}

\section{Scan setup}

As an input {for our scan} {we chose} the physical parameters: the {dark}  scalars' masses and couplings $\lam_2$ and $\lczp$,  corresponding to {self-}interaction in the dark sector and interaction between the {dark matter and SM-like Higgs boson}, respectively:
{
\begin{\eqn}
M_H,\,M_A,\,M_{H^{\pm}},\, \lam_2,\, \lam_{345}.
\end{\eqn}
} 
The mass {of the SM-like Higgs particle was always fixed to the value $M_h\,=\,125.1\,\GeV$.} 
{We} let all other mass parameters float in the regime $[0; 1\,\TeV]$,
where we however enforce $M_H$ to be the smallest of all dark scalars masses. We also impose a direct hard lower limit of $M_{H^\pm}\,\geq\,70\,\GeV${, as well as a minimal mass difference $M_{H^\pm}-M_H\,\geq\,100\,\MeV$ (cf. discussion in Section \ref{sec:collim}) }. {Furthermore, for $M_H\,\leq\,80\, \GeV$, the combination of limits from electroweak gauge boson decays widths and direct LEP searches leads to a strongly constrained region in the ($M_H,\,M_A$) plane, which we also used an input in our scans (cf. discussion in Section \ref{sec:collim})}. 
 If not stated otherwise, $\lam_2\,\in\,[0;4.5]$ and $\lam_{345}\,\in\,[-1.5;2]$. The range for $\lam_{345}$ chosen here is motivated by a set of pre-scans, where we did not find any allowed parameter points outside this range. {If not stated otherwise, points were generated using flat distributions in parameter space.}

Our {exclusion} scan was performed in three steps. All points, whether allowed or excluded, are kept; all exclusion criteria for a specific parameter point are memorized.
\begin{itemize}
\item{}In the first step, we
 test all theoretical constraints as discussed in Section \ref{sec:thconst}, as well as the total width of the $125\,\GeV$ Higgs $h$, the charged scalar, {and} EW gauge bosons. {We furthermore include constraints from electroweak precision observables.} {All quantities are calculated using 2HDMC.} Points which have passed all the{se} constraints will be labeled "OK step 1".
\item{}Only  points which have passed the above constraints {are then}  checked against limits from the Higgs collider searches and signal strength via \HB~/ \HS. Points which are also accepted by {these} cuts are in set "OK step 2".
\item{} Points which passed {all} {above}  bounds {(unless specifically stated differently)}, are then furthermore tested against dark matter, i.e. Planck and LUX, data. {This is done by processing through MicrOmegas {(version 4.2.3)}\cite{MO2013} {and confronting with Eqns.~(\ref{eq:planck_up}) and (\ref{eq:luxlim}) respectively.} }Points which pass these constraints are labeled "OK step 3". 
\end{itemize}

{Points allowed by the above constrains ("OK step 3")} were then used to calculate their cross-sections for dark scalar pair production {at the} LHC. {We refer to Section \ref{sec:crossx} for further details.}  

{In the following, whenever we quote hard limits on quantities, we want to emphasize that in each case we have performed extensive {\sl additional} scans where the respective relation was explicitly violated. For constraints resulting from step 2 or 3, we have for each of these generated at least $10^4$ such points after step 1. All statements concerning hard limits should be read in this spirit. The above numbers refer to general scan results reported in Section \ref{sec:results}.}

{Finally, we want to comment on the general strategy of our scan. In our results, we have used $95 \% C.L.$ {\sl exclusion bounds} whenever applicable, i.e. especially for bounds from direct collider searches as well as direct hard limits\footnote{{For similar approaches, cf. \cite{Pruna:2013bma, Robens:2015gla} for analyses with a singlet-extended Higgs sector, or \cite{Heisig:2013rya} for a study within a supersymmetric model.}}. We neglect correlations between these different bounds, and especially do not try to determine a best fit point within the parameter space\footnote{{Studies in this direction have e.g. been performed in \cite{Arhrib:2013ela}.}}. Therefore, the density of points presented in our results section should be read having in mind we applied a flat scan procedure with the parameter bounds discussed above{, and care must be taken to not interpret them in any probabilistic way}. The main goal of our work is the identification of regions in the five-dimensional parameter space which are still viable after all current constraints are taken into account, but we do not aim at a probabilistic interpretation within these regions.}

\section{Exclusions}
As stated in Section \ref{sec:model}, the IDM has 5 free parameters and not all of the bounds can be translated into clear {limits} within the 2-dimensional exclusion plots. However, some of the constraints lead to direct limits on the {parameter space}, and in the following we briefly list those and the relevant physical reason for it. 
{In the following, plots containing information about exclusion during step 1 correspond to $10^3$ {or more} parameter points; all others contain $10^3$ points belonging to the "OK Step 1" set.} {Furthermore, color coding usually refers to the constraints {\sl excluding} certain points in parameter space; points which {\sl survive} all constraints shown in a certain plot are presented in red. For more details, see the respective figure captions.}

\subsection{{Limits from theoretical constraints}}

{From the bounds tested in the first step of the scan, the requirements of perturbativity as well as positivity set relatively strong limits in the $\lb \lam_2,\,\lam_{345}\rb$ plane, as is  visible from Fig. \ref{fig:l345}.  Here, the most obvious constraint stems from the {perturbativity} limit on the quartic coupling of the dark matter candidate $H$: $\lam_{HHHH}\,=\,3\,\lam_2$, and therefore $\lam_2\,\leq\,\frac{4}{3}\,\pi\,\approx\,4.19$.
{Furthermore,} positivity clearly forbids $\lam_{345}\,\leq\,-1$ in the whole region considered here. There is an additional {lower bound on} $\lam_2$, stemming from positivity of the potential, which is dependent on the value of $\lam_{345}$.

None of the other theoretical constraints tested in step 1 of our scan, although  putting strong limits, lead to clear distinctions in two-dimensional parameter planes. {However, }the step 2 and step 3 constraints, {which} will be discussed below, put such clear bounds in the $\lb\lam_{345},M_{H^\pm} \rb$ and $\lb M_H,\,\lam_{345}\rb$ planes, respectively.%
\begin{figure}
\centering
\includegraphics[width=0.45\textwidth]{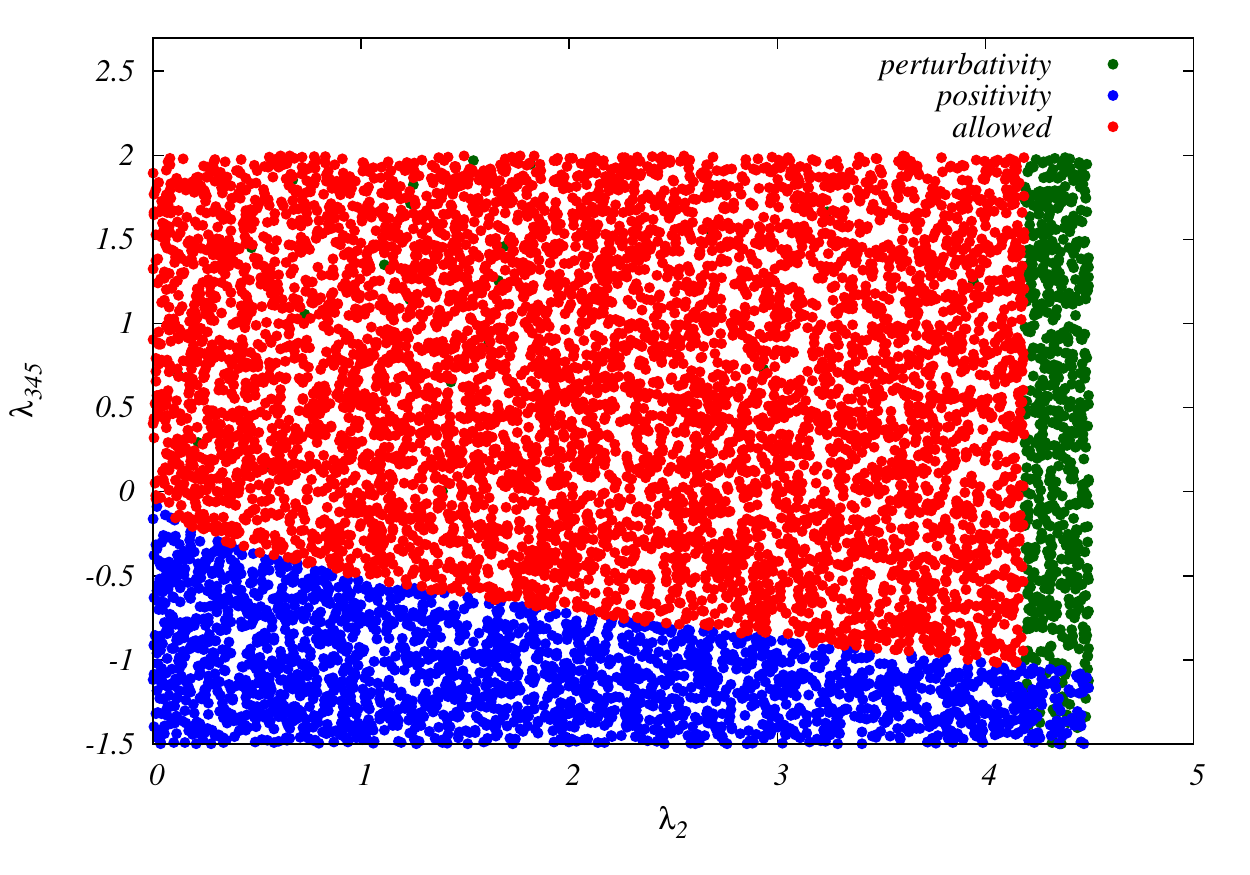}
\caption{ \label{fig:l345} Limits from positivity and perturbativity constraints in the $(\lam_2,\,\lam_{345})$ plane. We here set $\lam_{345}\,\leq\,{2}$.  The allowed points belong to the set "OK step 1". 
}
\end{figure}%
\begin{figure}
\centering
\begin{minipage}{0.45\textwidth}
\includegraphics[width=\textwidth]{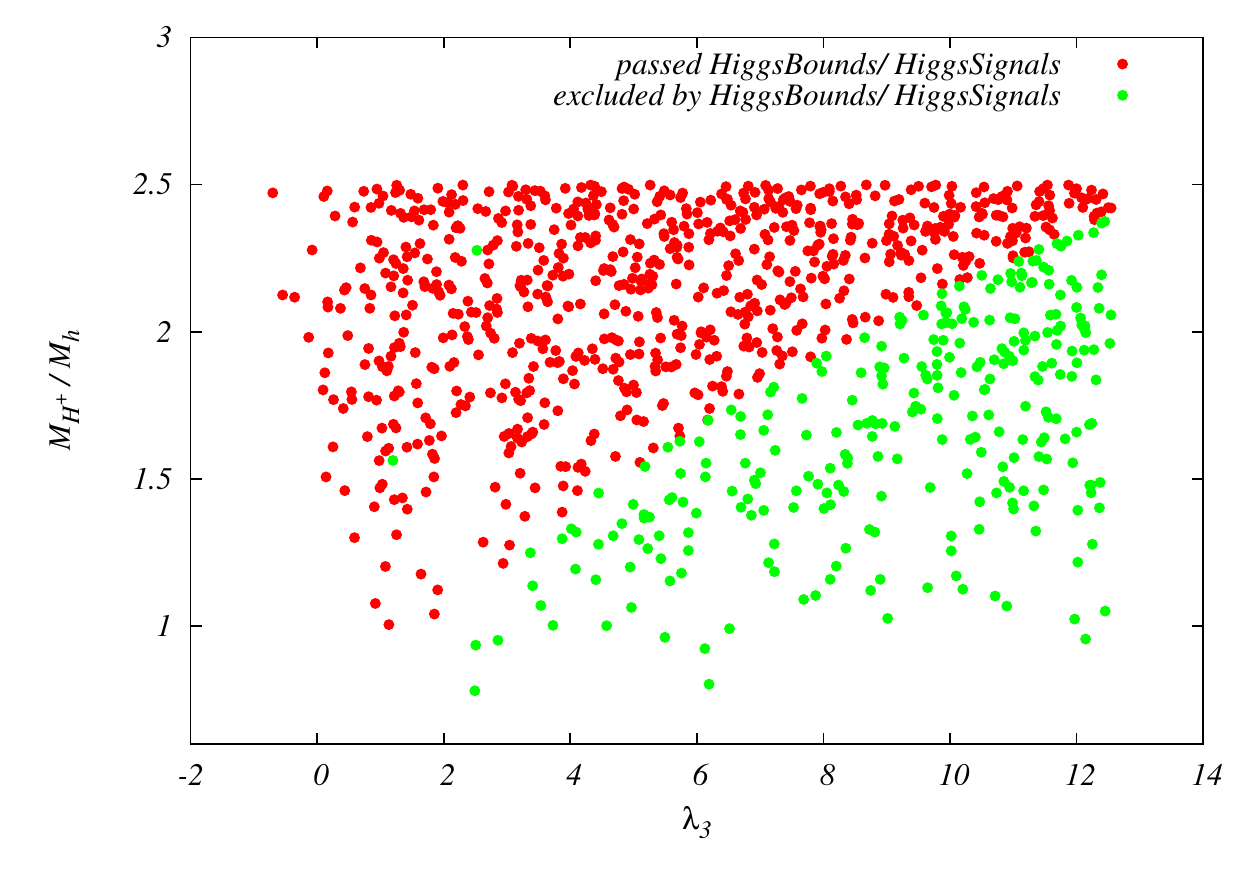}
\end{minipage}
\begin{minipage}{0.45\textwidth}
\includegraphics[width=\textwidth]{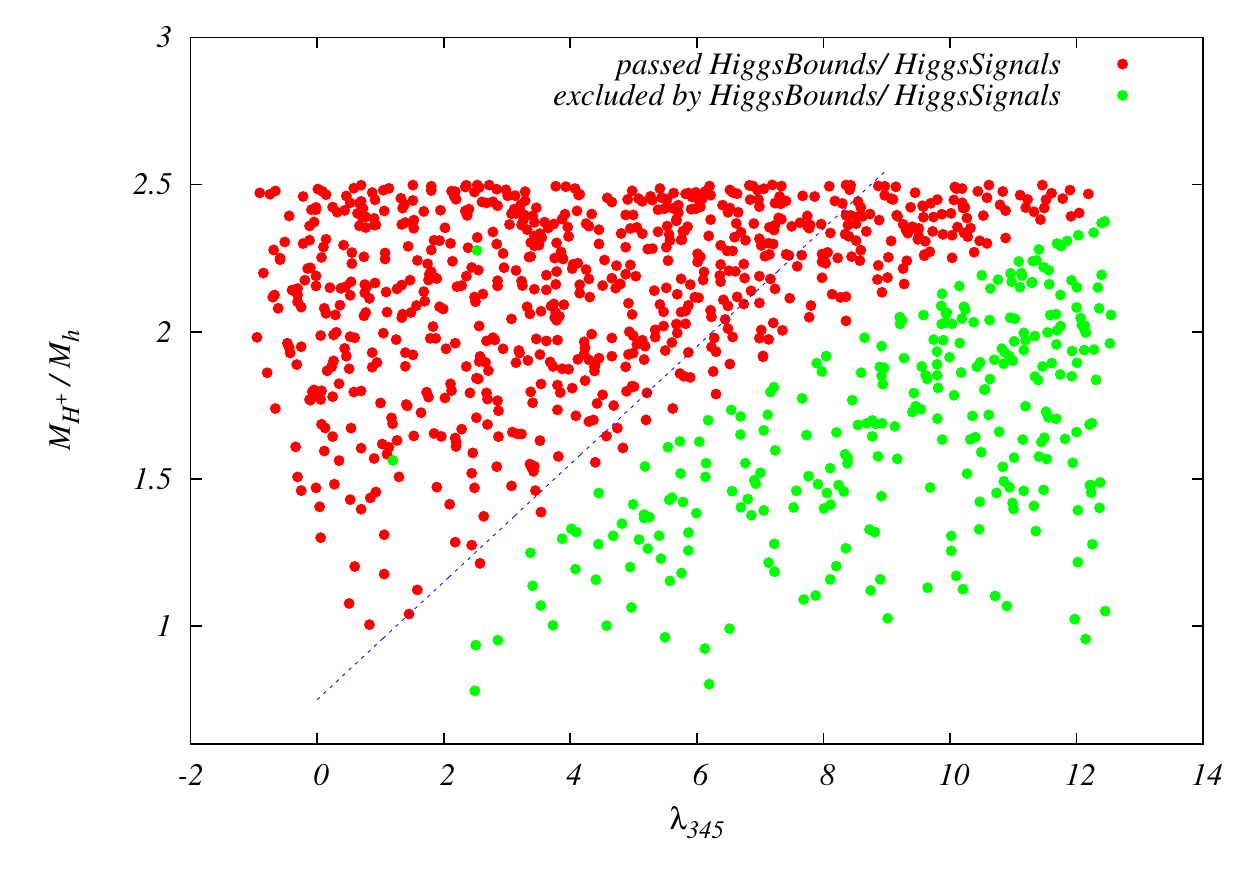}
\end{minipage}
\caption{\label{fig:hbhs}{Step 1} Points which are ruled out by LHC direct searches {and Higgs Signal strength},
 {in the ($\lam_3,\frac{M_{H^\pm}}{M_h}$) {\sl (left)} and ($\lam_{345},\frac{M_{H^\pm}}{M_h} $) {\sl (right)} plane}. {The line shown here} corresponds to Eqn.~(\ref{eqn:H+form}). {Green points for small $\lam_3$ or $\lam_{345}$ values and large mass mass ratios are forbidden by the branching ratio $h\,\rightarrow\,H\,H$, with $M_H\,<\,M_h/2$. Forbidden points in the green (lower) triangular shape are excluded from the $h\,\rightarrow\,\gamma\,\gamma$ branching ratio. } } 
\end{figure}
\subsection{Limits from colliders}\label{sec:collim}

At the second step of the scan, we consider bounds from direct {scalar} searches as well as the 125 \GeV~ Higgs coupling strengths, where we employ \HB \, and \HS, respectively. These lead to direct bounds} on $\lam_{345}$ and $\lam_3$  together with $M_{H^\pm}$ or $M_H$.  
\begin{itemize}
\item {The limits implied by \HB~ and \HS, being sensitive to the $h$ decaying to two photons,} are most obvious when considering the $(\lam_{345}\, , \frac{M_{H^\pm}}{M_h})$ plane. {In order to investigate this dependence in more detail, we allow $\lam_{345}$ to take values $\in[-2;4\,\pi]$, {exceeding the allowed range for negative values of $\lam_{345}$ discussed above and} 
keeping in mind that this parameter space will be subjected to much stronger constraints once dark matter direct searches are taken into account.}

{The observed limits} can be explained by the fact that the virtual contribution to the $h\,\rightarrow\,\gamma\gamma$ decay stemming from the charged Higgs loop contains terms proportional to $\frac{\lam_3}{M_{H^\pm}^2}\,f\,\lb M_{H^\pm}^2 \rb$ \cite{Arhrib:2012ia, Swiezewska:2012eh}.
{A direct limit {in} {F}igure \ref{fig:hbhs}, given by
\begin{\eqn}
\label{eqn:H+form}
\frac{M_{H^\pm}}{M_h}\,\geq\,0.2\,\,\lam_{345}+0.75,
\end{\eqn}
determines a region of parameters always allowed by} the $h\,\rightarrow\,\gamma\gamma$ branching ratio\footnote{{See also Appendix \ref{app:rgg} for limit setting by $h\,\rightarrow\,\gamma\gamma$ only.}}.
The forbidden points {in this figure}, which {apparently} violate {the above bound},  correspond to parameter points where $M_H\,\leq\,M_h/2$ and are ruled out by limits on the invisible decays of the 125 \GeV~ Higgs.
\item {{Also the  total width of the $125\,\GeV$ Higgs {{boson $h$}} clearly sets}} strong limits as soon as $M_H\,\leq\,{M_h}/{2}$, 
what is visible from {F}igure \ref{fig:widtheffs}. A relatively narrow stripe with $|\lam_{345}|\,\lesssim\,0.05-0.1$ survives where the additional, {ie. invisible},  decay mode is suppressed. 
\begin{figure}
\centering
\begin{minipage}{0.45\textwidth}
\includegraphics[width=\textwidth]{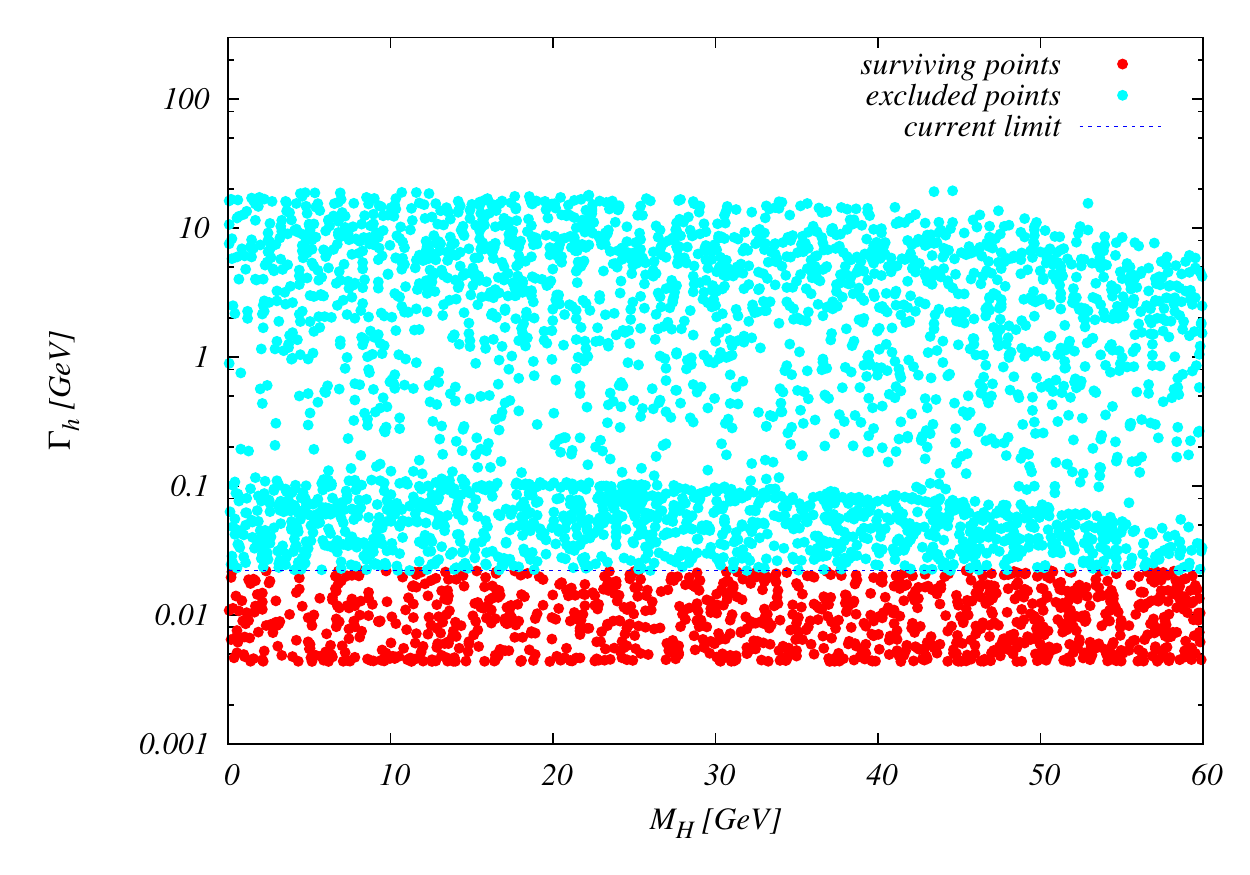}
\end{minipage}
\begin{minipage}{0.45\textwidth}
\includegraphics[width=\textwidth]{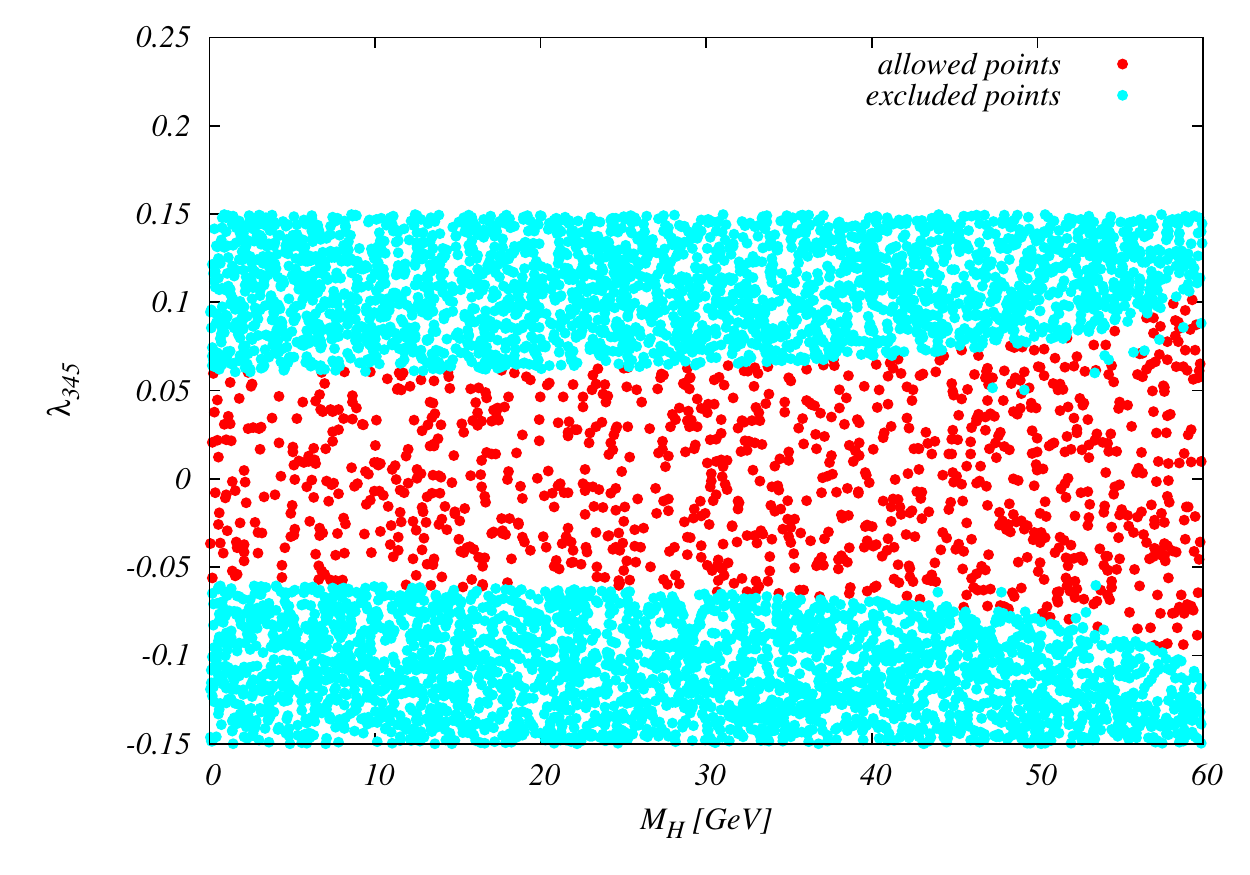}
\end{minipage}
\caption{\label{fig:widtheffs} {\sl Left:} Total width of the 125 \GeV~ Higgs $h$ as a function of the dark matter mass $M_H$, together with the current limit {$\Gamma_h\,\leq\,22\, \MeV$}. 
 {\sl Right:} $\lam_{345}$ as a function of the dark matter mass $M_H$, with points which do (red) or do not ({cyan}) survive {the condition on {$\Gamma_h$}; 
points which survive} have $|\lam_{345}|\,\leq\,0.1$. {Points with $|\lam_{345}|\,\leq\,0.15$ lead to maximal widths $\mO\lb 0.1\,\GeV \rb$.}}
\end{figure}%
Note however that the{se} limits for $\lam_{345}$ are {completely} superseded by {bounds from} direct DM search{es}, {discussed below.}

\item
{Another important constraint stems from the total decay width of the electroweak gauge bosons, which {are} implemented according to Eqn.~(\ref{eq:gwgz}).}
Here 
we need to distinguish three {cases} where a combination from LEP searches \cite{Lundstrom:2008ai} and the decay widths of $W$ and $Z$ lead to quite narrow regions in parameter space:
\begin{itemize}
\item{}$M_H\,\in\,[0;41\,\GeV]$: $M_A\,\geq\,100\,\GeV$, 
\item{}$M_H\,\in\,[41;45\, \GeV]$: $M_A\,\in\, [m_Z-M_H;M_H+8\,\GeV]$ or $M_A\,\geq\,100\,\GeV$,
\item{}$M_H\,\in[45; 80\, \GeV]$: $M_A\,\in\,[M_H;M_H+8\,\GeV]$ or $M_A\,\geq\,100\,\GeV$.
\end{itemize}
\item{}
We take as a benchmark lifetime $\tau\,\leq\,10^{-7} {s}$ to ensure {$H^+$} decay inside the detector. In {the} {IDM}, this leads to constraints of the parameters entering the possible decay modes
\begin{\eqn*}
H^\pm\,\rightarrow\,W^\pm\,\lb A/H\rb.
\end{\eqn*}
{As the respective couplings are determined by the electroweak SM sector (cf. appendix \ref{app:fr}), this {basically} puts constraints on the mass {hierarchy.}}
We found that this requirement rules out regions for which
\begin{\eqn}\label{eq:chwlim}
M_A\,\geq\,M_{H^\pm} \; {\rm{and}} \;M_{H^\pm}-M_H\,\leq\,100\,\MeV,
\end{\eqn}
simultaneously hold. We have therefore enforced 
\begin{\eqn*}
M_{H^\pm}-M_H\,\geq\,100\,\MeV
\end{\eqn*}
throughout our scan.

\end{itemize}

\subsection{{Limits from astrophysical measurements}}

 {In fact} the strongest limits on the {IDM} parameter space stem from requiring agreement with the dark matter {{measurements}}, included {at step 3} in the scan.  
In general the direct dark-matter-nucleon scattering\footnote{{See \cite{Belanger:2008sj} for a detailed discussion of the calculation of the scattering cross-section within MicrOmegas.}}  leaves a quite narrow allowed strip in the ($\lam_{345},\,{M_{H}}$) plane { especially in {the} low dark matter mass region}, cf. {F}igure \ref{fig:l345_om}. For this, rough estimates of upper LUX limits translated to the values of $\lam_{345}$, {valid for high mass region $M_H \gtrsim M_h$ \footnote{For lower {$M_H$}, the limit given by {the} fit is slightly too strong, however in this range strong constraints on low $\lam_{345}$ from too high relic density start to play an important role {and this {slight overconstraint} has no impact on {the} results}. }
\begin{eqnarray}\label{eq:l345min}
\lam_{345}^\text{low}&\sim&0.03985\,-\,6.786\,\times\,10^{-4}\,M_H-4.828\,\times\,10^{-7}\, M_H^2,
\end{eqnarray}
\begin{equation}
\lam_{345}^\text{up} = - \lam_{345}^\text{low} \ \  \text{as} \ \ \sigma_{\text{DirDet}} \sim |\lam_{345}|^2.
\end{equation}
\begin{figure}[!tb]
\centering
\includegraphics[width=0.45\textwidth]{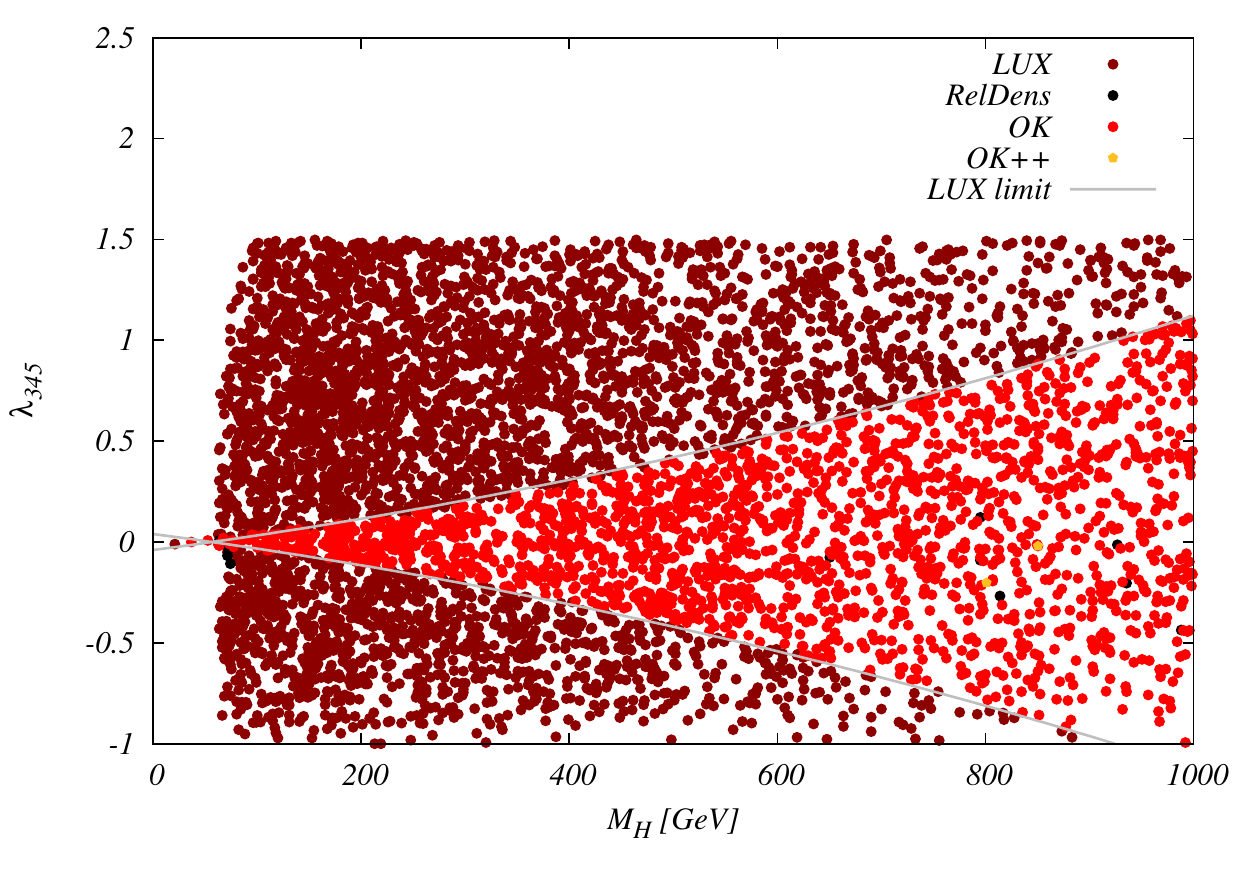}
\caption{ \label{fig:l345_om} 
{Results of step 3 scan on set "OK step 2" points}: allowed (red) and forbidden {(dark red)} regions in the $(M_H,\,\lam_{345})$ {(mass in  GeV)}  plane from the direct DM detection from LUX (dark red) and relic density from Planck measurements (black).  {Additionally points which reproduce {the} whole observed relic density were marked {in gold} (OK++). } 
}
\end{figure}%
\begin{figure}[!tb]
\centering
\begin{minipage}{0.45\textwidth}
\includegraphics[width=\textwidth]{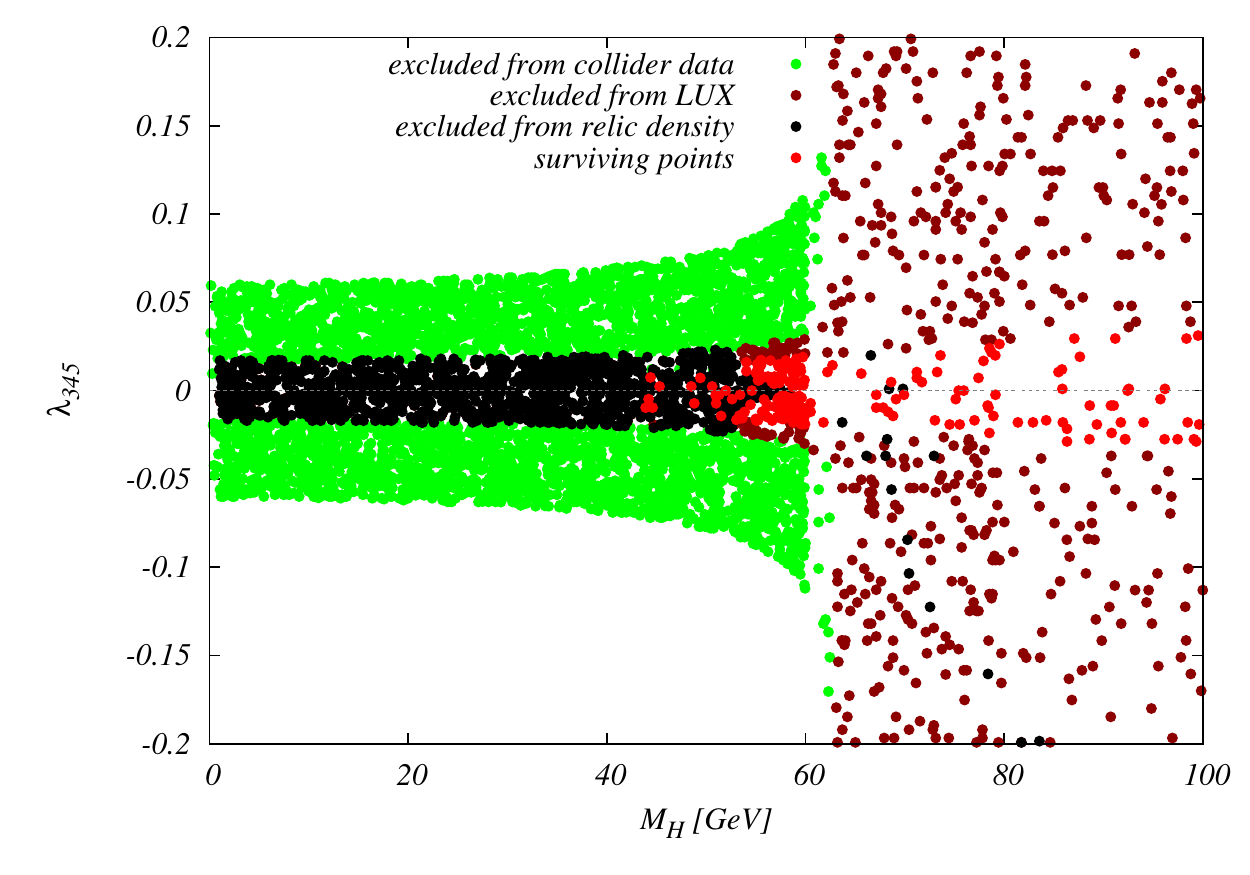}
\end{minipage}
\caption{\label{fig:l345bel100} {Step 1} points below 100 \GeV after the application of collider, {i.e. \HB~/\HS,} and dark matter limits. {Points below 60 \GeV~ were sampled more frequently.} }
\end{figure}%
\begin{figure}[!tb]
\centering
\begin{minipage}{0.45\textwidth}
\includegraphics[width=\textwidth]{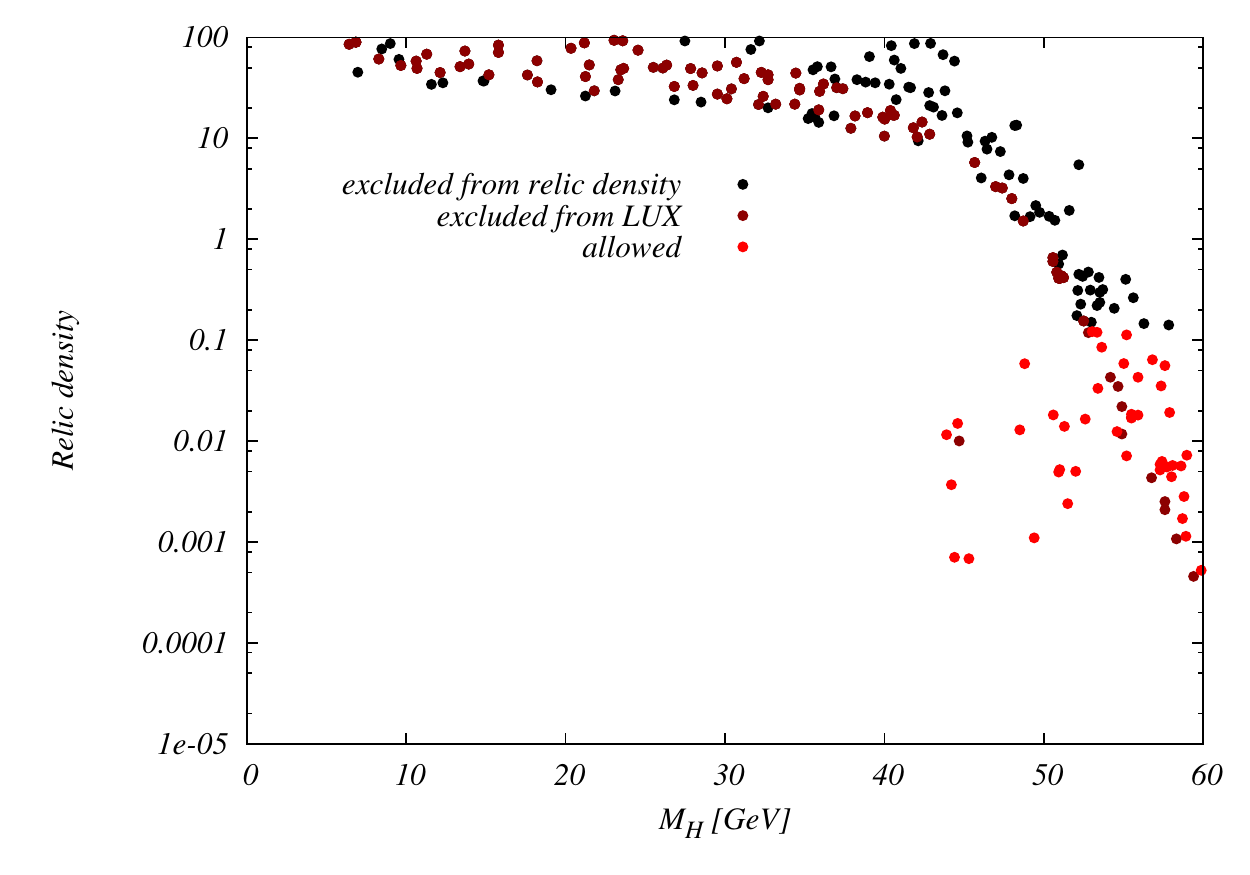}
\end{minipage}
\begin{minipage}{0.45\textwidth}
\includegraphics[width=\textwidth]{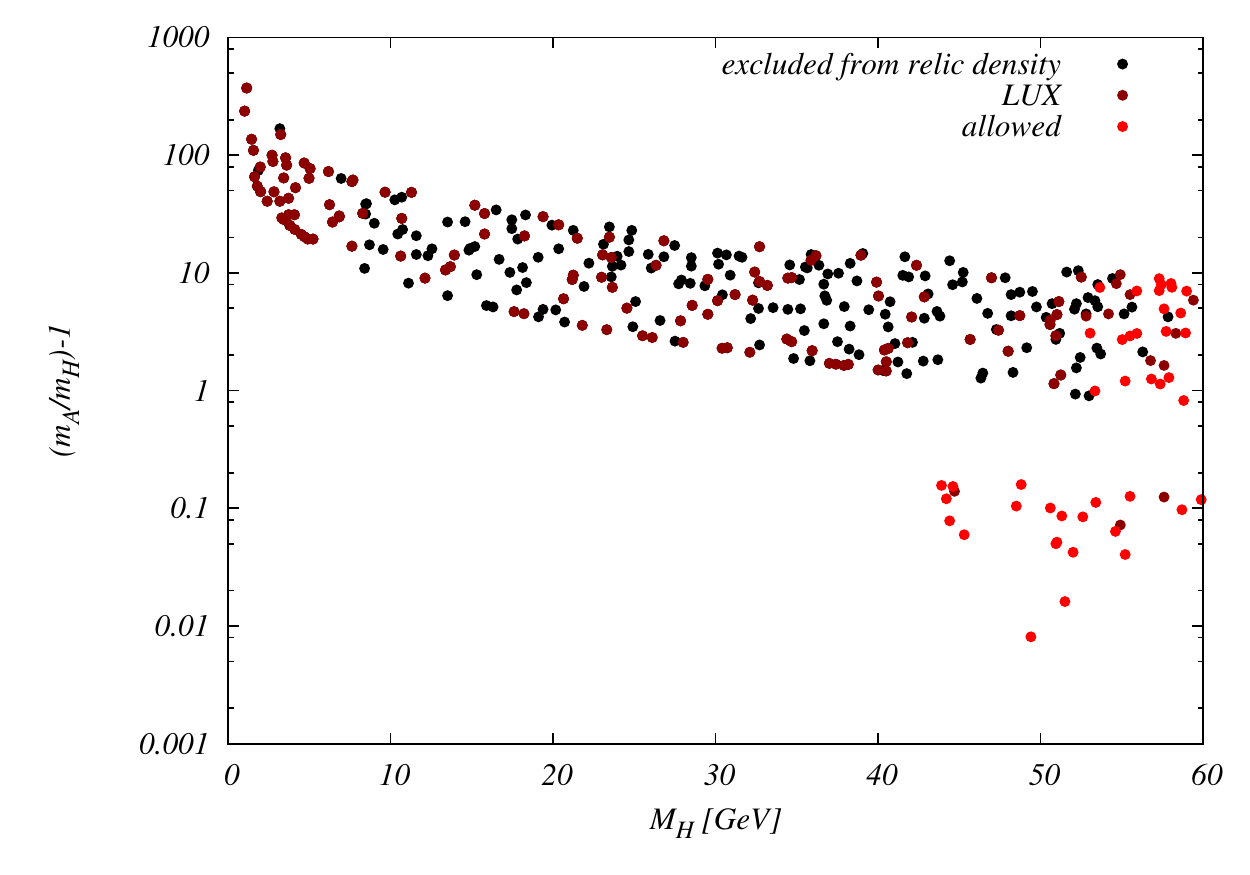}
\end{minipage}
\caption{\label{fig:dm_lowmass} DM density as a function of $M_H$ {\sl (left)} as well as relative difference of $M_A$ and $M_H$ as a function of $M_H${\sl (right)}. The exceptional points all exhibit a strong degeneracy between $M_A$ and $M_H$, leading to dominance of co-annihilation and thereby a smaller dark matter relic density. }
\end{figure}%
{Constraints from dark matter relic density (cf. Eqn.~(\ref{eq:planck_up})), only start playing a role in either very high or very {low DM} mass regions. {As before, the region where $M_H\,\leq\,M_h/2$, which allows for the invisible decay of the 125 \GeV~Higgs, is of special interest.} {Here} LUX data put strong limits on the  ($M_H,\,\lam_{345}$) {plane}, (cf. {F}igure \ref{fig:luxpap}). At step 3 of our scan we found quite strong constraints on the allowed parameter space for $M_H\,\leq\,54\,\GeV$, cf. Fig.~\ref{fig:l345bel100}, basically ruling out all points for which $M_H\,\lesssim\,{45}\,\, \GeV$.} 
{This can be traced back to the fact that} the dark matter density rises {strongly} with smaller masses,  {rising} above 100 at DM mass{es} around 10 GeV, cf. Fig.~\ref{fig:dm_lowmass} {\sl (left)}. However, a couple of exceptional points prevail. For these, the masses $M_A$ and $M_H$ are degenerate, with differences $\mO\lb 10\% \rb$, {cf. Fig. \ref{fig:dm_lowmass} {\sl (right)}}. In this case, the dominant channels are co-annihilation {into $q\,\bar{q}$ final states}, leading to a much smaller total dark matter density.
The tension between the collider limits on the BR($h\rightarrow HH$),  allowing only {for} very low values of $\lam_{345}$, {cf. Fig. \ref{fig:l345bel100}} and relic density measurements is visible in this region. On the other hand, for larger dark matter masses \, {50} $\GeV \leq  M_H \leq 600\,\GeV$, {relic density data} poses no constraint if we do not additionally require agreement with the {$\sl$ lower} bound on the relic density. For higher masses, however, this measurement again becomes quite constraining, as {shown in  F}igure \ref{fig:omdep}. In summary, we can therefore say the dark matter {data} poses a lower bound on the mass of the dark matter candidate of $M_H\,\gtrsim\,{45}\,\GeV${, which obviously induces the same lower limit for $M_A$ and $M_{H^\pm}$}. {This is one of {the} main results of our paper.}\\
\begin{figure}
\centering
\begin{minipage}{0.45\textwidth}
\includegraphics[width=\textwidth]{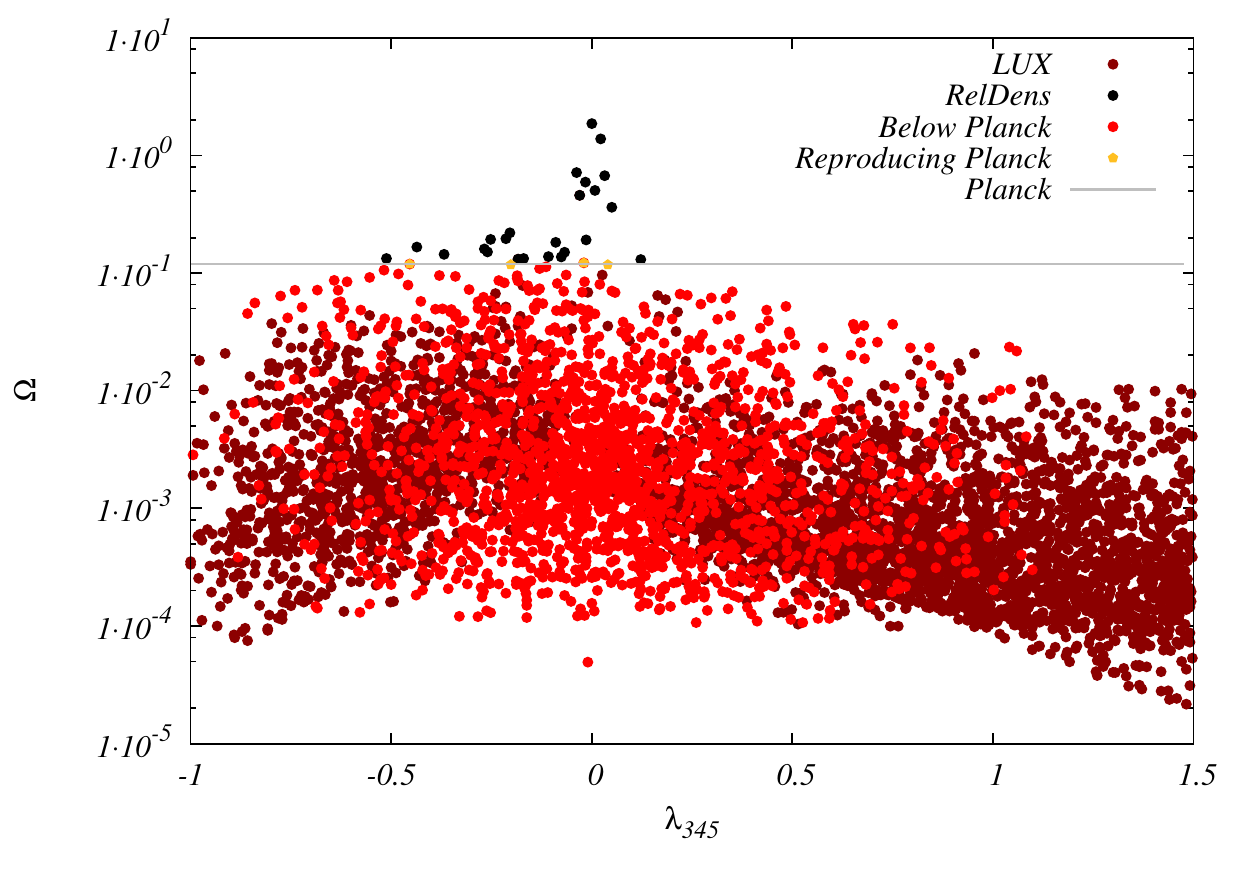}
\end{minipage}
\begin{minipage}{0.45\textwidth}
\includegraphics[width=\textwidth]{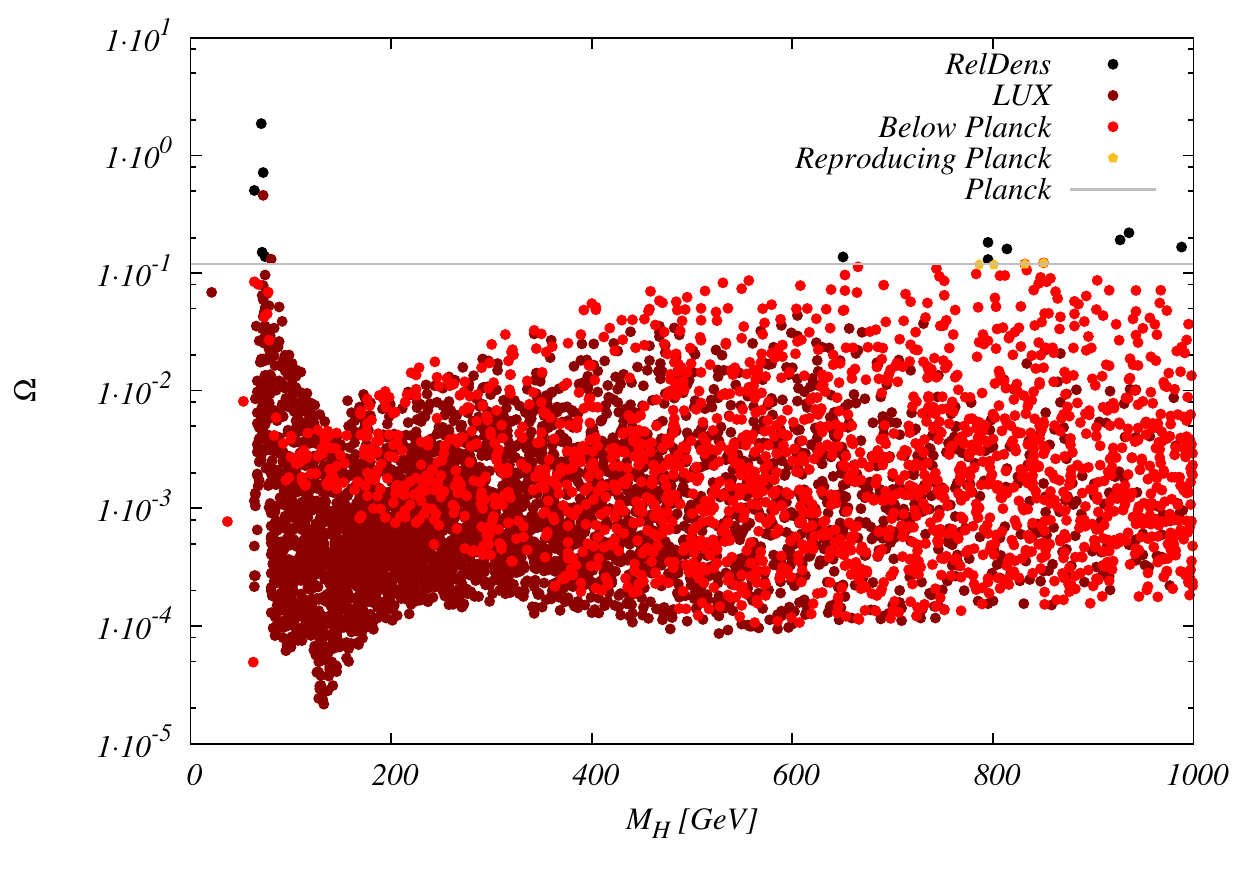}
\end{minipage}
\caption{\label{fig:omdep} Values for relic density after all other constraints have been taken into account {(i.e. after step 2)}, as a function of $\lam_{345}$ {\sl (left)} as well as $M_H$ {\sl (right)}. The additional line signifies agreement with {\sl lower} limit on the relic density at $95\,\%$ C.L., {specified in Eqn. (\ref{eq:planck}).} {In the dark scalar sector, a mass degeneracy was enforced (see main body of text for details).} }
\end{figure}%
\begin{figure}
\centering
\begin{minipage}{0.45\textwidth}
\includegraphics[width=\textwidth]{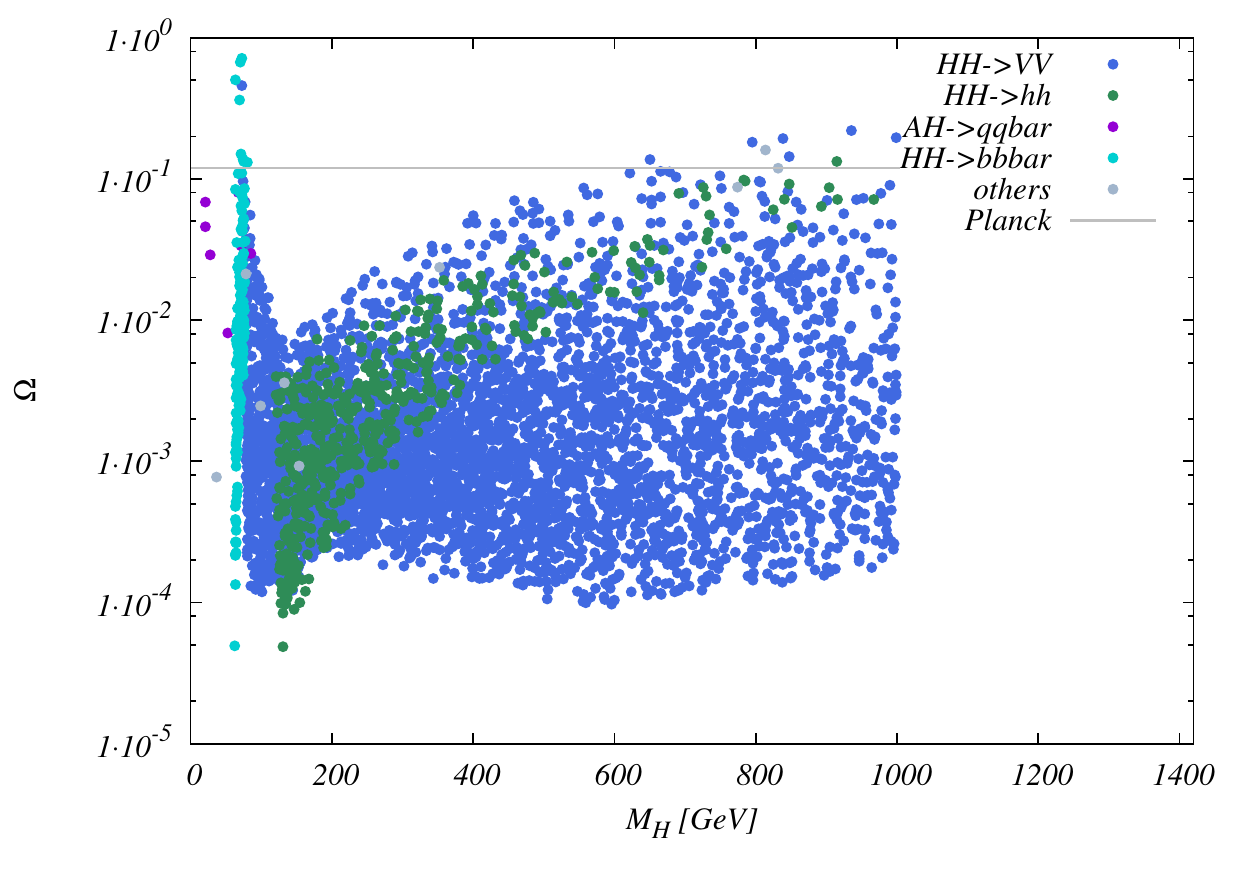}
\end{minipage}
\begin{minipage}{0.45\textwidth}
\includegraphics[width=\textwidth]{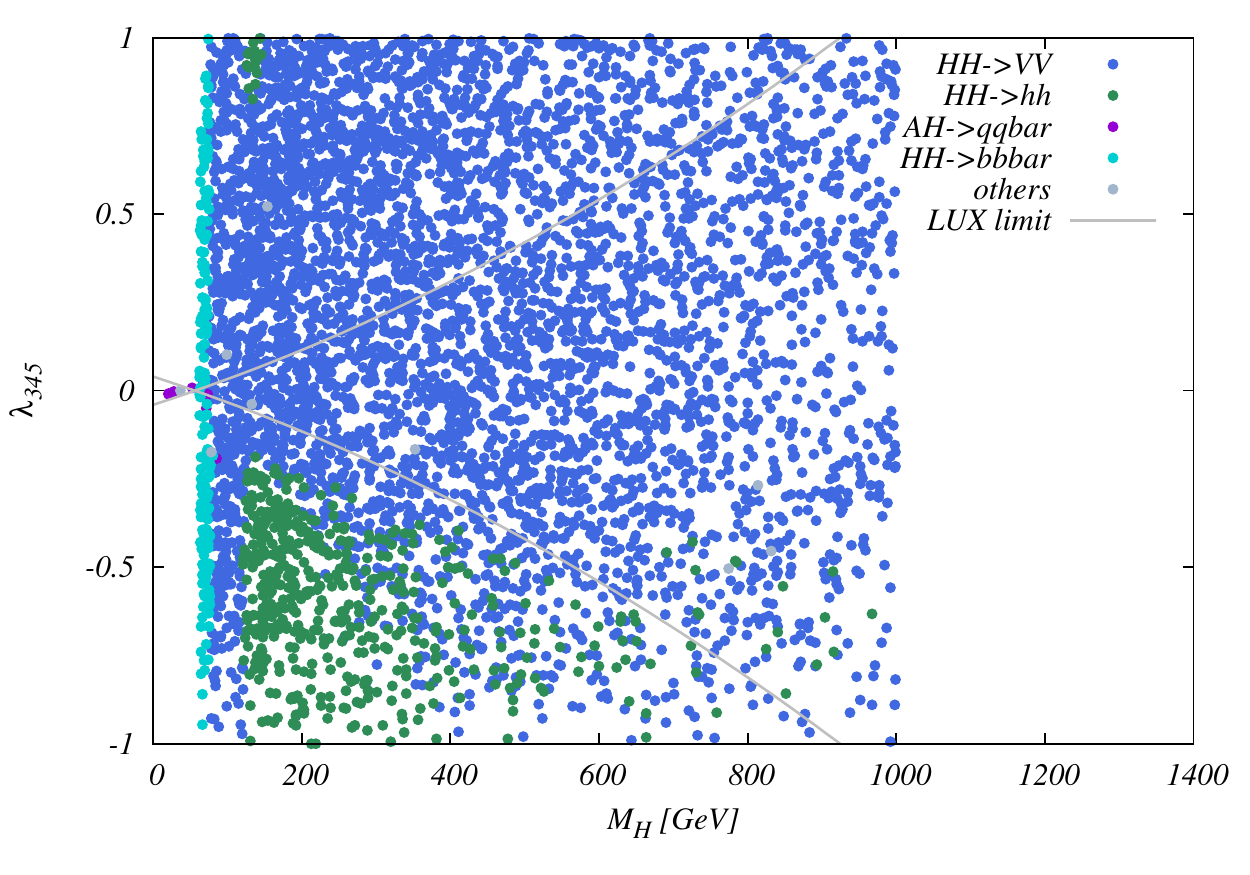}
\end{minipage}
\caption{\label{fig:dm_chan}
 {{\sl Left:} The relic density as a function of dark particle mass with the} 
{dominant production channel (see text for explanation)}. 
{ {\sl Right:} Leading channel contribution in $\lb M_H,  \lam_{345}\rb$ plane. All points are from set "OK step {2}". {Scan points as in Fig.~\ref{fig:omdep}.}
} }
\end{figure}%
\begin{figure}
\centering
\begin{minipage}{0.45\textwidth}
\includegraphics[width=\textwidth]{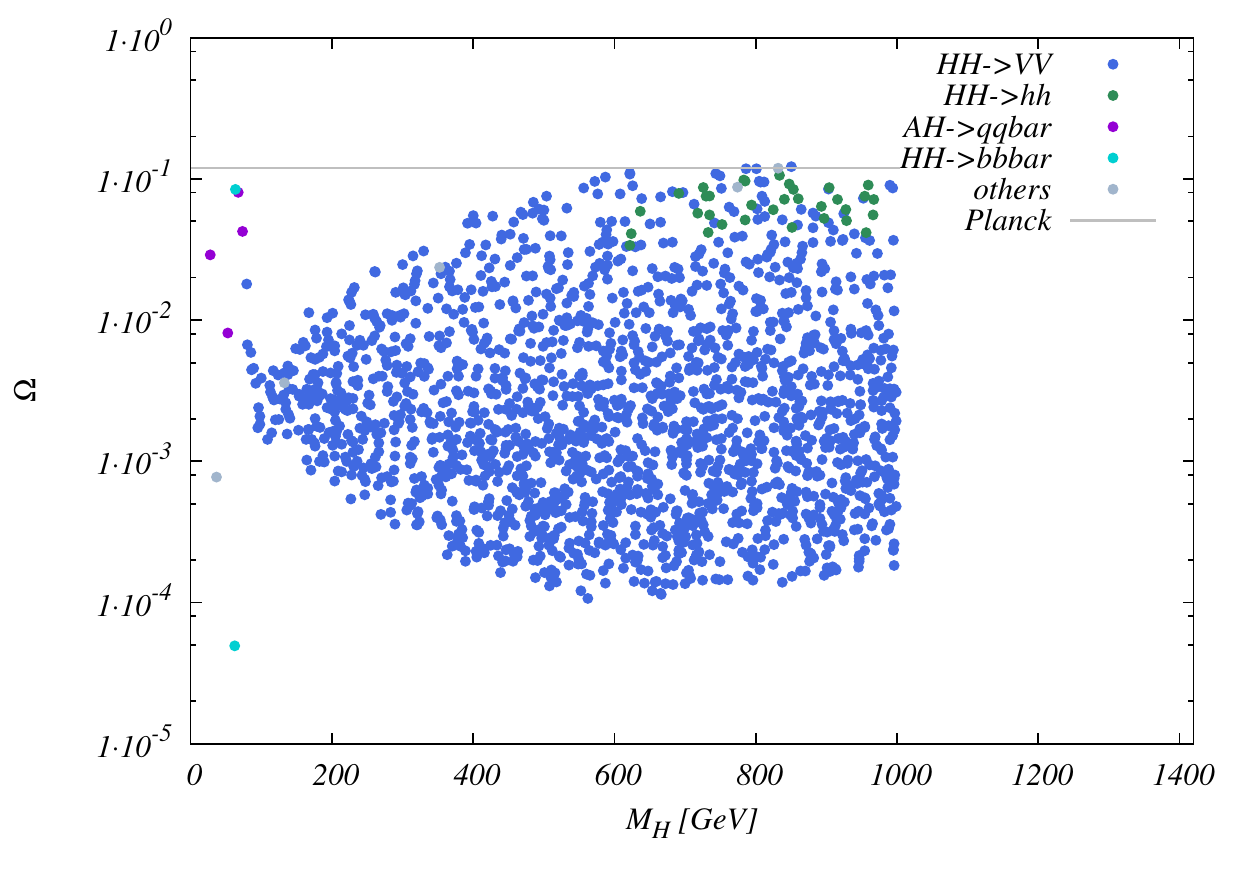}
\end{minipage}
\caption{\label{fig:dm_chan_all} As Fig.~\ref{fig:dm_chan} (left), but now additionally applying limits on $\lam_{345}$ from the LUX direct search. In the parameter space allowed after all constraints, pair-annihilation into vector bosons dominates. 
 }
\end{figure}%

}

To further exemplify the constraints stemming from dark matter relic density, in Fig.~\ref{fig:omdep} we plot the dependence of it as a function of $\lam_{345}$ (left) as well as the dark scalar mass $M_H$ {\sl (right)}, for points which have passed the constraints in the second step of our scan. {For Figs.~\ref{fig:omdep}, \ref{fig:dm_chan}, and \ref{fig:dm_chan_all}, we have enforced a maximal hierarchy $\frac{M_{H^\pm}}{M_H}\,\lesssim\,2,\,\frac{M_{A}}{M_H}\,\lesssim\,2$, to enforce a larger number of allowed points in the low mass region.}
To better understand the major contributions to the dark matter density, in Fig.~\ref{fig:dm_chan}, we plot the channel identified as dominant\footnote{{"Dominant" here signifies the largest contributions; next leading channels, even if contributing in a similar range, are not shown.}} contribution for the dark matter density,  as a function of the dark matter mass, for points which have survived after step 2 of our scan. The respective cross-sections mainly depend {on} electroweak gauge couplings as well as $\lam_{345}$. One can observe (cf. e.g. \cite{Sokolowska:2011sb}) that pair-annihilation to electroweak gauge bosons as well as SM-like Higgses dominate in large regions of parameter space, while in the low mass range (co) annihilation into $d\,\bar{d}$ and $b\,\bar{b}$ final states dominate. Once the limit from LUX is applied on $\lam_{345}$, the dominant channel is annihilation into vector boson final states, with a few parameter points {corresponding to} {annihilation} into $hh$ final states, {resulting from parameter points with $\lam_{345}\,\lesssim\,-0.5$ and $M_H\,\gtrsim\,600\,\GeV$, cf. Fig. \ref{fig:dm_chan}}.

Generally, dark matter relic density is well below the value measured by the Planck collaboration (cf. Eqn.~\ref{eq:planck}).
In fact, we found that the measured {95\% CL} value of the relic density can only be reproduced in either the low mass ($M_H\,\leq\,60\,\GeV$) or high mass ($M_H\,\gtrsim\, 600\,\GeV$) region\footnote{{Similar results have been obtained in \cite{Abe:2014gua}.}}. However, to exactly match the Planck measurement within a $95\%$ confidence level allowance requires an extreme fine tuning, and generally points with masses $\gtrsim\,500\,\GeV$ are relatively challenging for collider searches. We will comment on a low scale scenario which can in principle reproduce $\Omega_c$ in Section \ref{sec:crossx}. \\

\subsection{{Results }}\label{sec:results}
In this section, we summarize the results we have obtained from the scan as described in the previous sections, and provide scatter plots for allowed and excluded regions in several planes depending on two or three of the independent input parameters
$M_H,\,M_A,\,M_{H^\pm},\,\lam_{345}$. {The order of plotting always corresponds to the legends read from top to bottom in the respective scatter plot.} {Furthermore, as we here use results for a generic flat scan with masses up to 1 \TeV~ for $M_H$, some specific features of the low mass region are not prominent. For example, a general flat scan usually features dark masses $M_H\,\gtrsim\,200\,\GeV$.}
We omit $\lam_2$, as this obeys quite strong limits from positivity and perturbativity (cf. Fig. \ref{fig:l345}) but otherwise does not have an impact on quantities related to current collider and astrophysical measurements.\\

{
As we have seen in the previous sections, in the general scan\footnote{The region where $M_H\,\leq\,M_h/2$ is special and is treated separately.} the most obvious constraints stem from LUX limits on the coupling $\lam_{345}$. In Figs.~\ref{fig:l345plots1} and \ref{deltas}, {left}, we show the respective constraints and allowed points in the dark scalar mass {versus $\lam_{345}$} planes. We have seen that the parameter space opens up for larger masses $M_H$. This naturally leads to a higher density of allowed points in the higher mass region\footnote{Using a non-flat distibution in the scan parameters could in principle reduce this, but is beyond the scope of the present work.}. The triangular shape of allowed regions {appears} in Fig.~\ref{fig:l345plots1}, which gets slightly distorted in the other dark scalar mass plots, cf. Figs.~\ref{fig:l345plots1} {\sl (right)} and \ref{deltas} {\sl (left)}.\\
\begin{figure}
\centering
\begin{minipage}{0.45\textwidth}
\includegraphics[width=\textwidth]{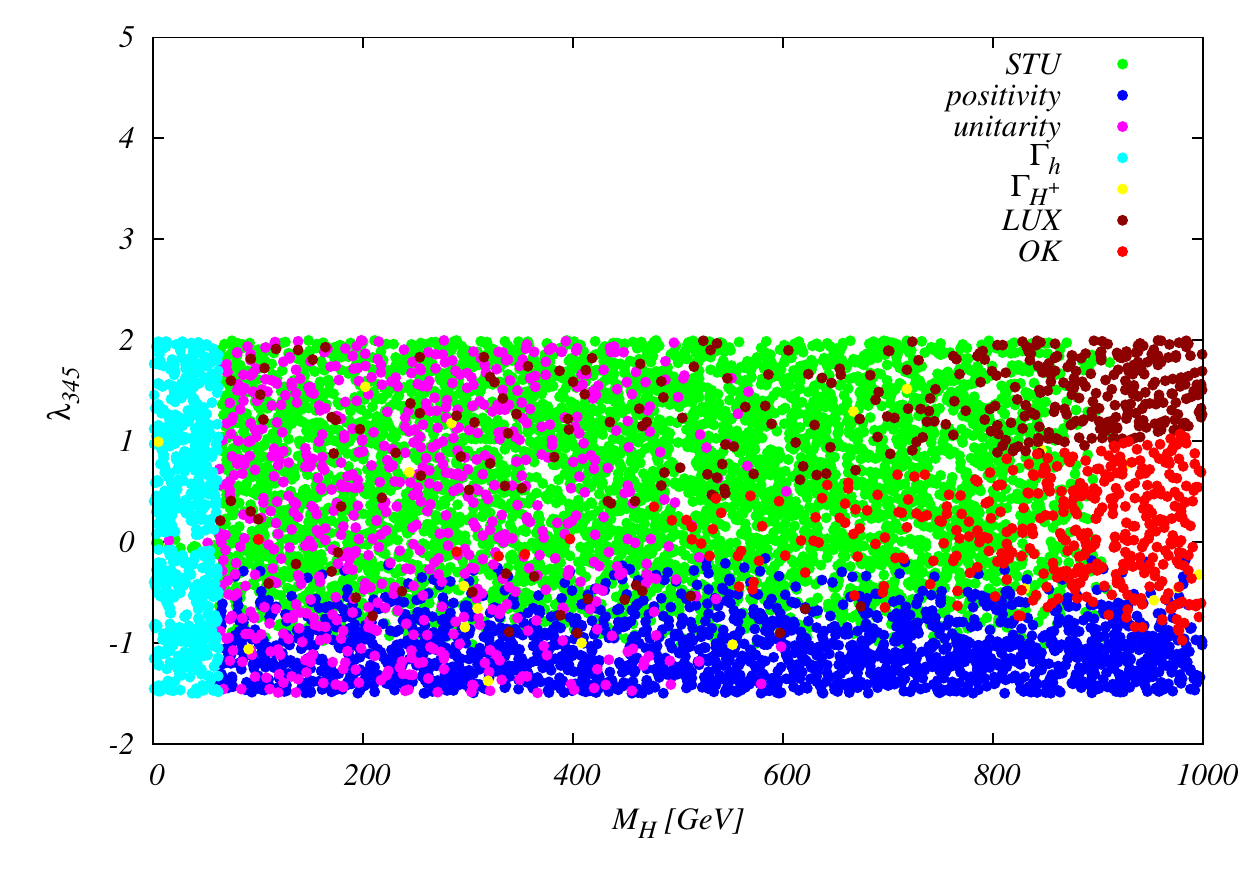}
\end{minipage}
\begin{minipage}{0.45\textwidth}
\includegraphics[width=\textwidth]{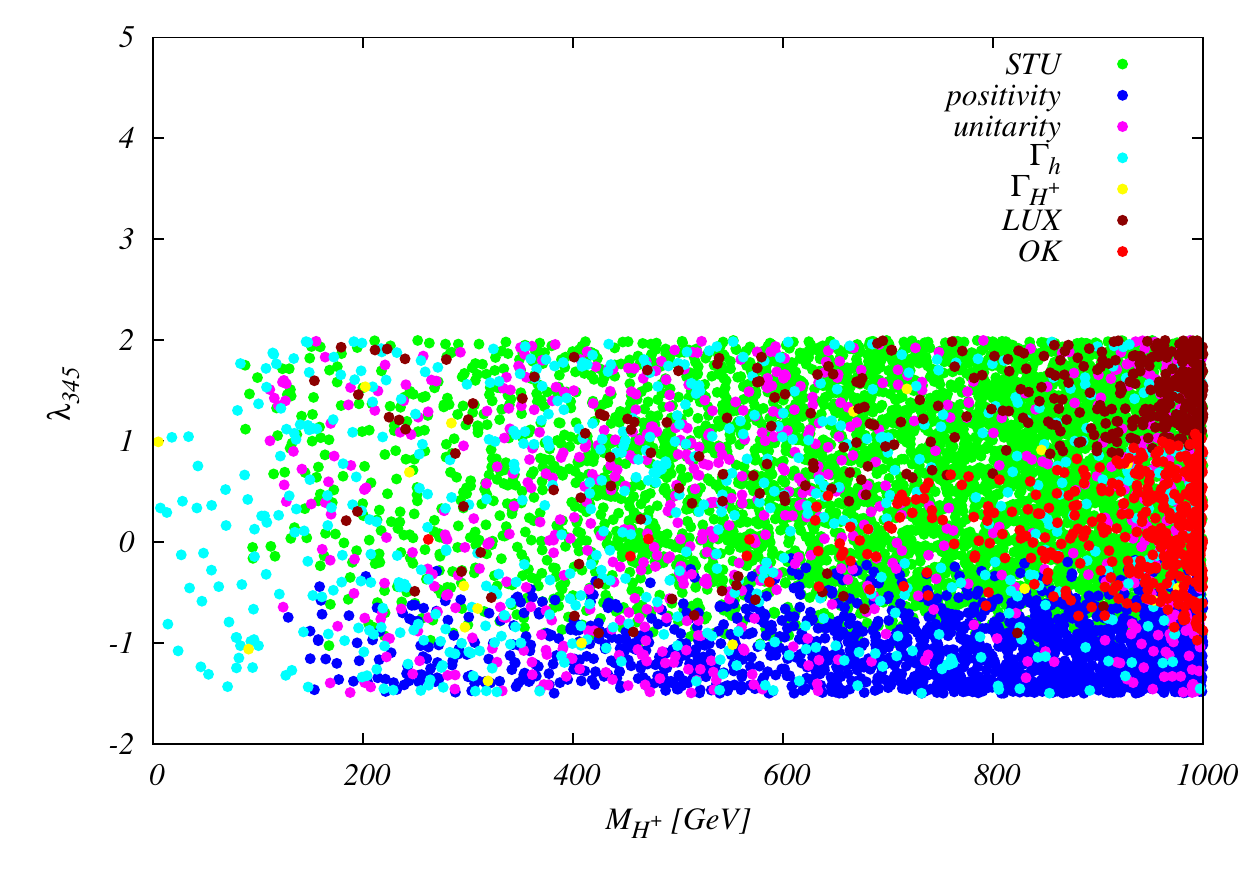}
\end{minipage}
\caption{\label{fig:l345plots1} $M_H$ vs $\lam_{345}$ {{\sl (left)}and $M_{H^\pm}$ vs. $\lam_{345}$ {\sl{( right)}}}. All exclusion limits are shown, with the scan steps as described in the main body of the text.}
\end{figure}%
\begin{figure}
\centering
\begin{minipage}{0.45\textwidth}
\includegraphics[width=\textwidth]{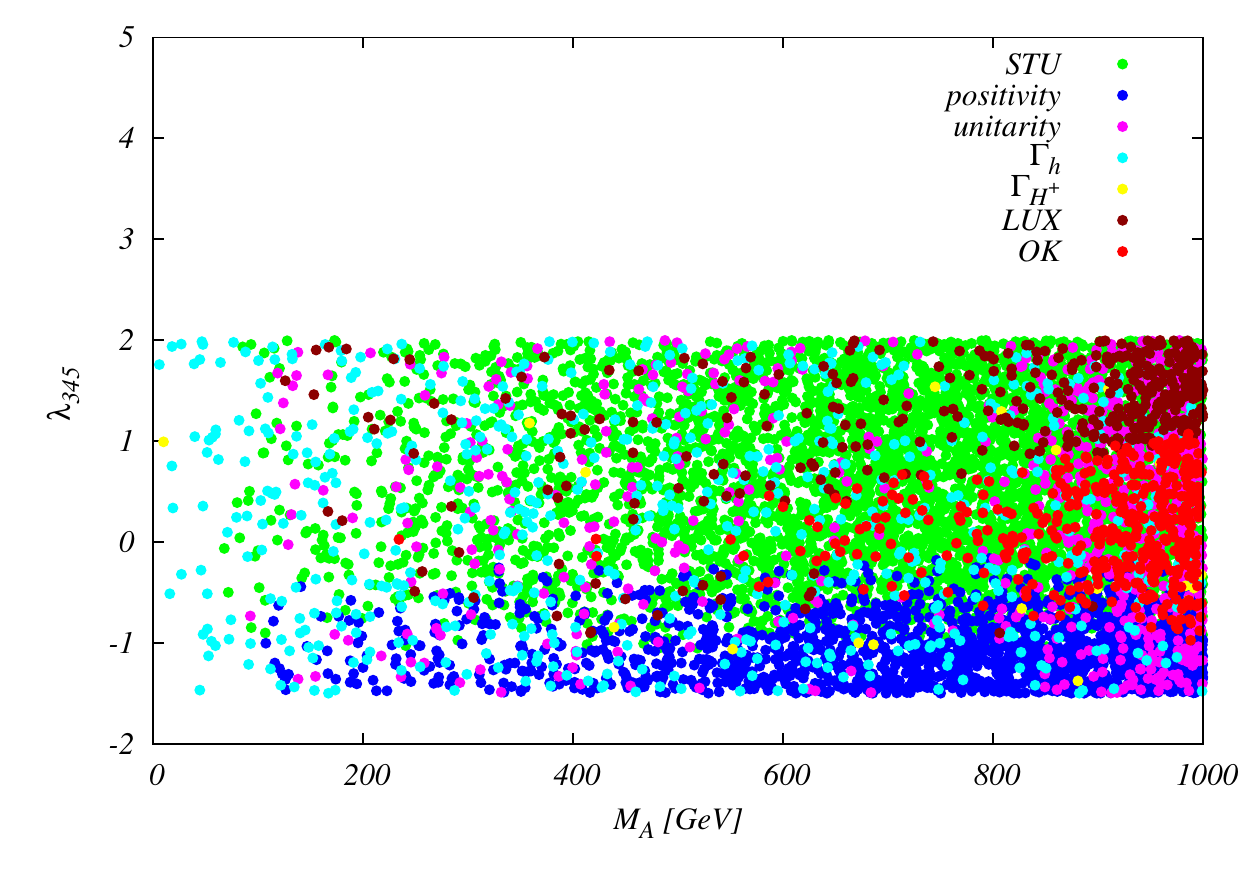}
\end{minipage}
\begin{minipage}{0.45\textwidth}
\includegraphics[width=\textwidth]{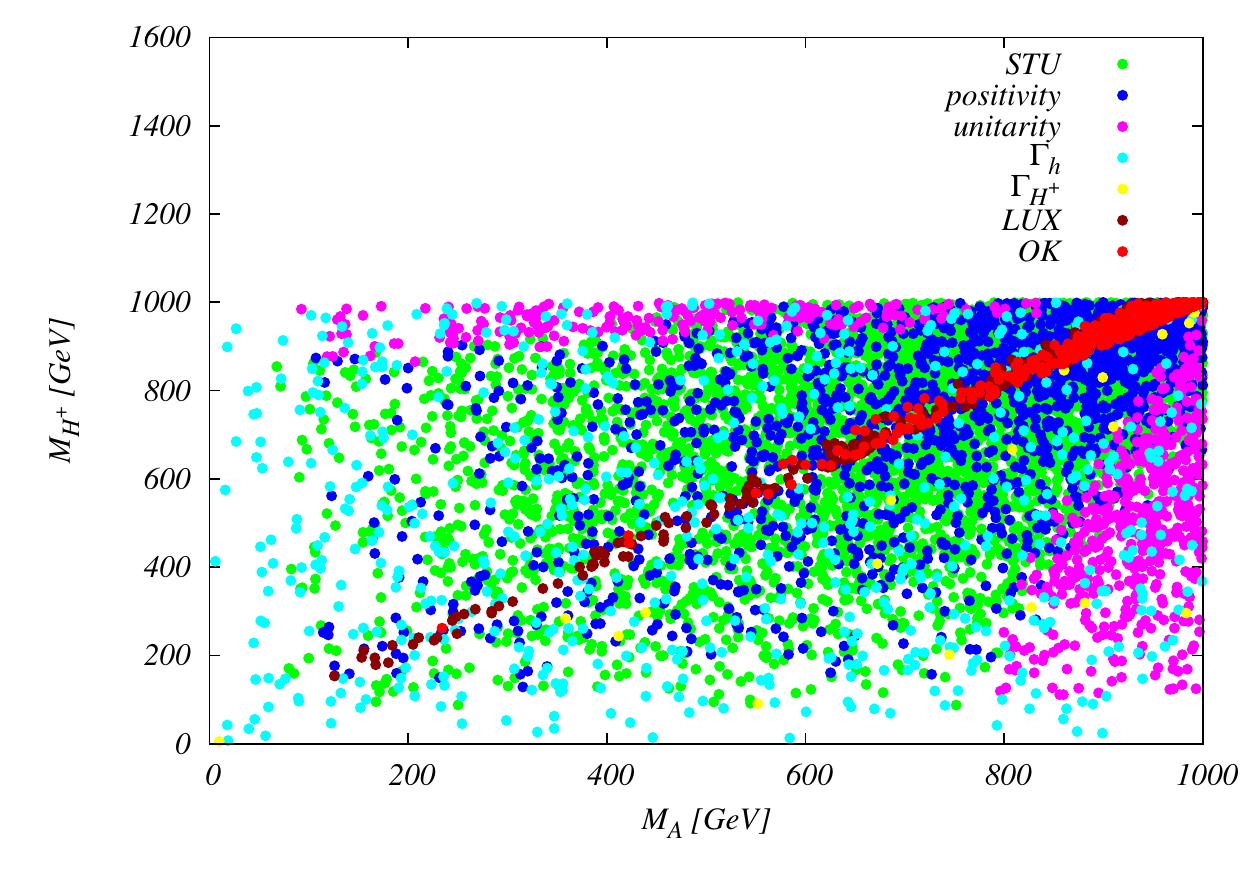}
\end{minipage}
\caption{\label{deltas} {$M_A$ vs $\lam_{345}$ {\sl (left)} and} $M_A$ vs $M_{H^\pm}$ {\sl (right)}. {All exclusion limits are shown, with the scan steps as described in the main body of the text.}}
\end{figure}
\begin{figure}
\centering
\begin{minipage}{0.45\textwidth}
\includegraphics[width=\textwidth]{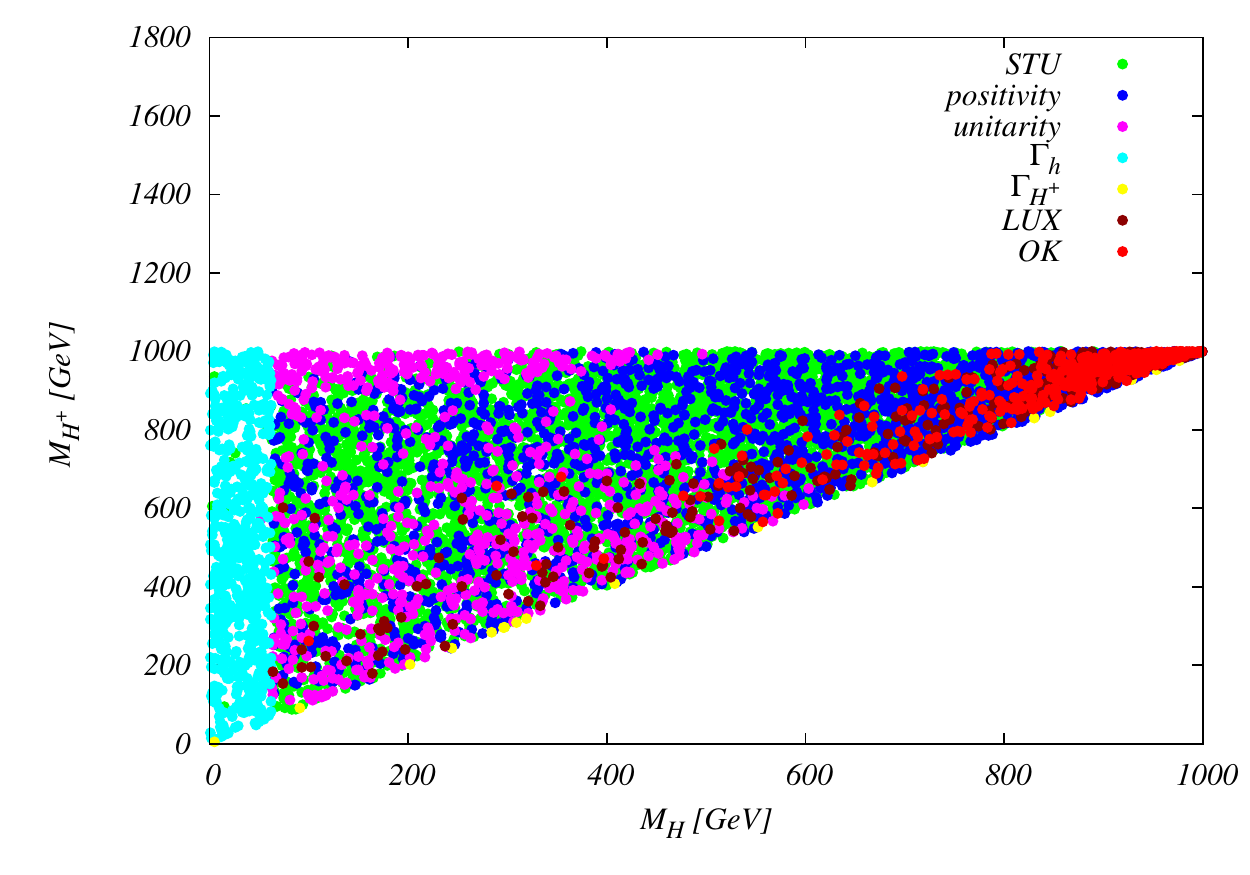}
\end{minipage}
\begin{minipage}{0.45\textwidth}
\includegraphics[width=\textwidth]{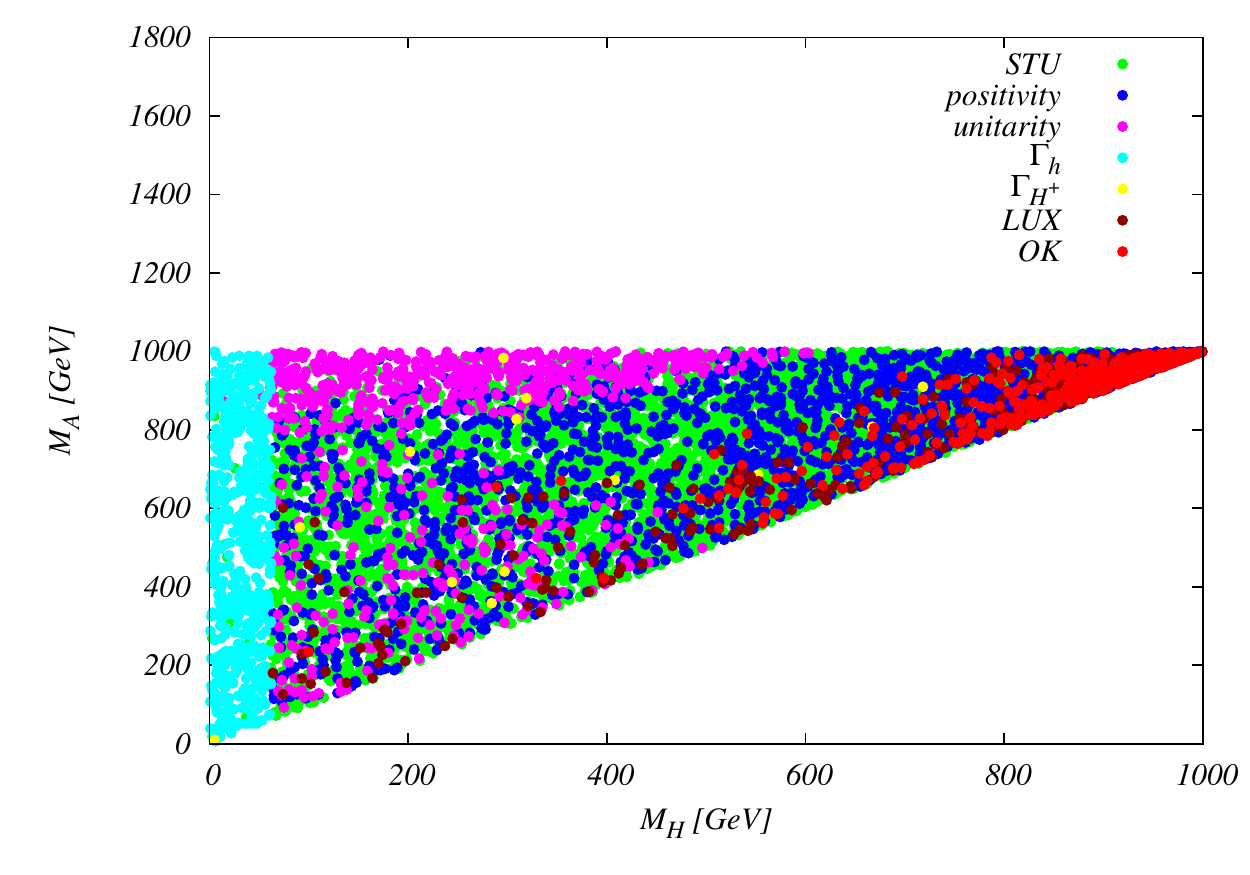}
\end{minipage}
\caption{\label{fig:MH} $M_H$ vs $M_{H^\pm}$ {\sl (left)} and $M_H$ vs $M_A$ {\sl (right)} {All exclusion limits are shown, with the scan steps as described in the main body of the text.}}
\end{figure}

A second major {new} result {of our study} is that we found that {all} dark scalar masses are quite degenerate {for $M_H\,\gtrsim\,200\,\GeV$}. This becomes apparent in Figs.~\ref{deltas} and \ref{fig:MH}, where we plot our results in the dark scalars' mass planes\footnote{{For $M_A$ and $M_{H^\pm}$, a similar result has already been presented in \cite{LopezHonorez:2010tb}.}}. In contrast to the constraints on $\lam_{345}$, we here could not pin down a single constraint which leads to this hierarchy, but rather found this to be the result of the interplay of several constraints. Fig.~\ref{fig:rats} shows the ratio of these masses, which {we found} to be {$\lesssim\,2$} for dark masses {$\gtrsim\,200\,\GeV$}\footnote{{In the low-mass region, with $M_H\,\lesssim\,100\,\GeV$, ratios of $M_A/M_H\,\sim\,10$ and $M_{H^\pm}/M_H\,\sim\,10$ are perfectly viable, while $M_{H^\pm}/M_A\,\lesssim\,2.4$.}}. {We also found that $M_{H^\pm}\,\geq\,M_A$ for {\sl all} points in our scan after step 3, cf. e.g. Fig.~\ref{fig:rats}}.
\begin{figure}
\centering
\begin{minipage}{0.45\textwidth}
\includegraphics[width=\textwidth]{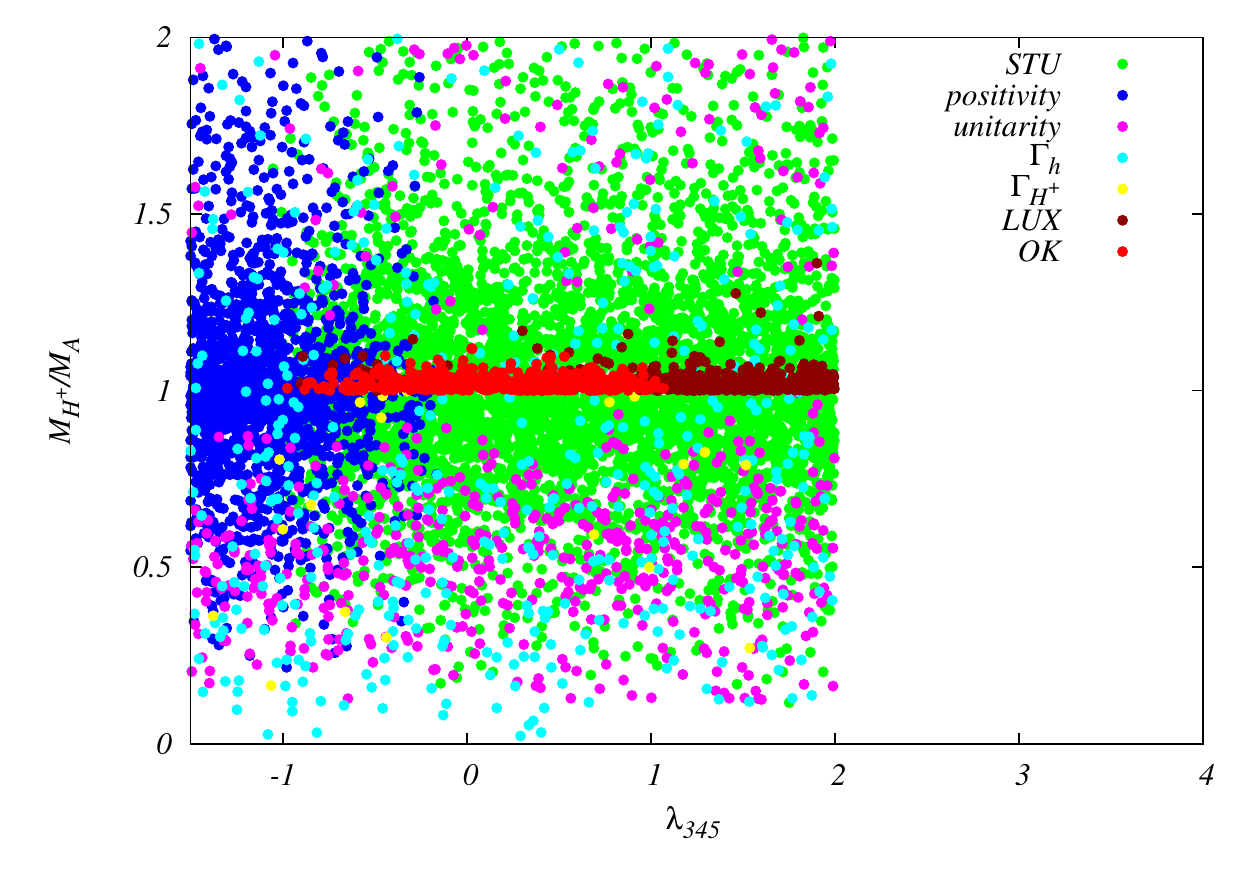}
\end{minipage}
\begin{minipage}{0.45\textwidth}
\includegraphics[width=\textwidth]{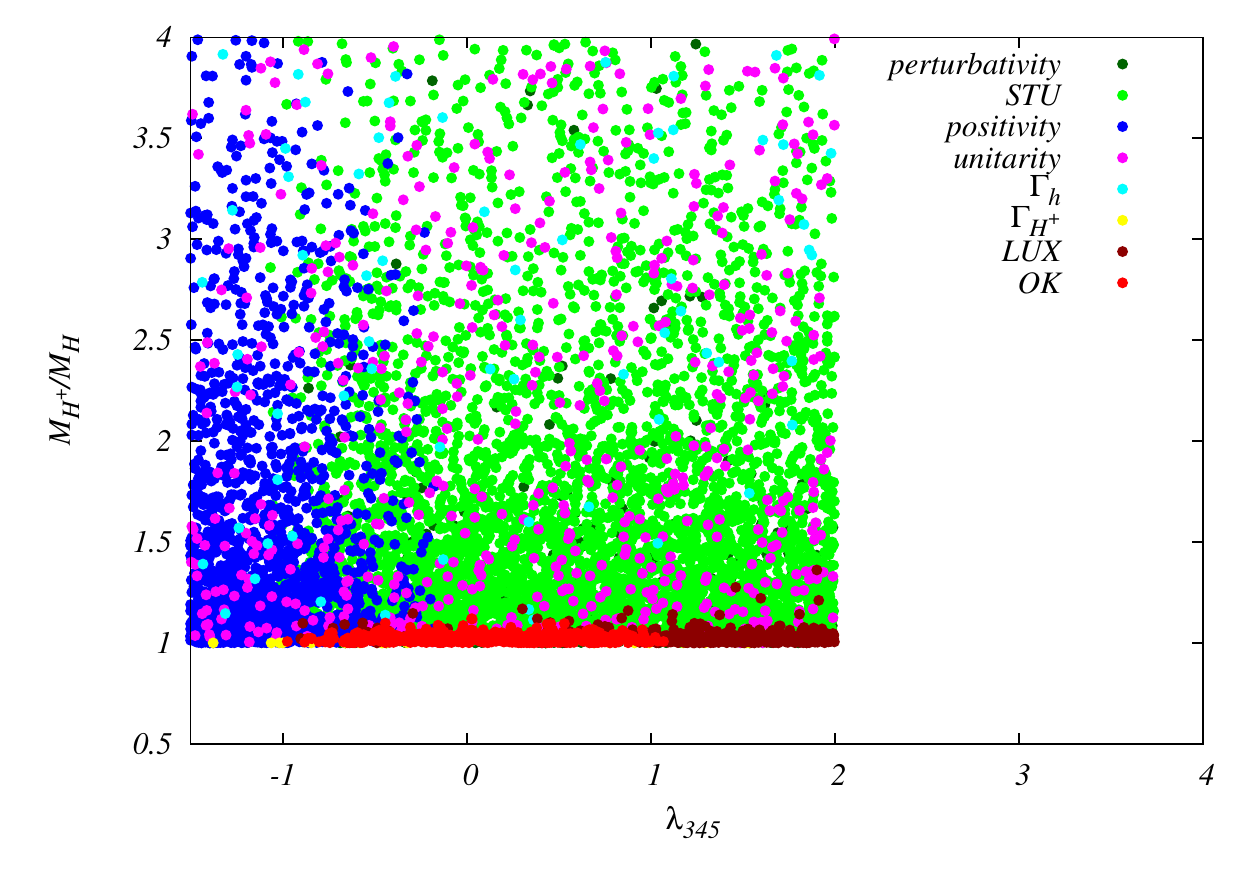}
\end{minipage}
\caption{\label{fig:rats} $\frac{M_{H^\pm}}{M_A}$ {\sl (left)} and $\frac{M_{H^\pm}}{M_H}$ {\sl (right)} vs $\lam_{345}$ {All exclusion limits are shown, with the scan steps as described in the main body of the text. }}
\end{figure}

Generically, the available parameter space closes rapidly for lower masses of $M_H$, mainly due to the strong constraints from the LUX experiment, cf. Fig.~\ref{fig:l345_om}. Therefore, a general scan does not find many allowed points in {lower} mass ranges, {which}} can of course be cured by enforcing $10^3$ valid points after step 1 {or introducing more specific scan boundaries}. However, even in this case a very low fraction survives, cf. Fig.~\ref{fig:l345bel100}. On the other hand, for relatively low masses the relic density {value}, {as given in Eqn.~(\ref{eq:planck})}, can be achieved. {For the low-mass dark matter region, where $M_H \lesssim M_h/2$, the interplay of the 125 \GeV~ Higgs width and constraints from relic density (cf. Fig. \ref{fig:dm_lowmass}) lead to an extreme finetuning. The lowest allowed mass for the DM candidate was found to be $M_H \sim 45\, \GeV$.} A more dedicated discussion of this region is in the line of future work.

\section{\label{sec:crossx} {Benchmark points and planes} for dark scalar pair-production at a 13 \TeV~ LHC}
After having provided constraints on the models parameter space, an obvious question is whether this model can be tested at the current run of the LHC. Therefore, we here briefly comment on the collider phenomenology, concentrating on a 13 \TeV~ proton-proton collider. {In the following, we provide {\sl total production cross-sections} for pair production of the dark scalars, which can already give a hint of regions which could eventually be accessible at a 13 \TeV~collider; however, the results presented here do not contain a dedicated phenomenological study, including e.g. simulation of SM background or kinematic distributions and cuts on the decay products of the dark scalars. Although this is obviously an important topic to investigate, we consider this an independent extension of our current study, which is in the line of future work, with the current results as input.}\\

The $D(Z_2) $ symmetry forces dark particles to be-pair produced; therefore, production cross-sections for 
\begin{\eqn}
p\,p\,\rightarrow\,(A\,A);(H^\pm\,H^\mp);(H\,A);{(H^\pm\,H);(H^\pm\,A)}
\end{\eqn}
are of interest here. We always obtain a pair of dark matter scalars $H$ in the final states, so the collider signature is in the form of
\begin{\eqn}
p\,p\,\rightarrow\,X\,+\slashed{E}_T,
\end{\eqn}
where $X$ denotes {either a single or several} SM particle(s) and $\slashed{E}_T$ stands for missing transverse energy (MET). In principle, the process
\begin{\eqn}
p\,p\,\rightarrow\,H\,H
\end{\eqn}
could also be investigated; however, at leading order this would lead to a collider signature of missing transverse energy which requires special treatment, and we therefore disregard this process here.

\subsection*{Benchmark planes}
Taking all the results presented above into account, we have calculated the leading order pair-production  cross-sections for dark sector scalars using MadGraph5 \cite{Alwall:2011uj} with the IDM UFO model file \cite{Goudelis:2013uca}, where we have additionally implemented the $ggh$ effective vertex{, which describes the one-loop induced $ggh$ coupling in the $m_\text{top}\,\rightarrow\,\infty$ limit in a standard way}\footnote{This is basically an adoption from the implementation in \cite{Maina:2015ela}; TR wants to thank E. Maina for useful comments.}.  We {generally} find the following ranges for {the} cross-sections {of the} pair-produced particles 
\begin{eqnarray}\label{eq:allprocs}
p\,p\,\rightarrow\,H\,A&:&\,\leq\,0.03\,\pb,\nonumber\\
p\,p\,\rightarrow\,H^\pm\,H&:&\,\leq\,0.03\,\pb,\nonumber\\
p\,p\,\rightarrow\,H^\pm\,A&:&\,\leq\,0.015\,\pb,\\
p\,p\,\rightarrow\,H^+\,H^-&:&\,\leq\,0.01\,\pb,\nonumber \\
p\,p\,\rightarrow\,A\,A&:&\,\leq\,{0.0015}\,\pb.\nonumber
\end{eqnarray}

These processes are all s-channel mediated according to
\begin{eqnarray}\label{eq:smind}
q\,\bar{q}\,\rightarrow&Z\,\rightarrow&H\,A,\nonumber\\
q\,\bar{q}\,\rightarrow&\gamma,\,Z&\rightarrow\,H^+\,H^-,\nonumber\\
q\,\bar{q}'\,\rightarrow&W^\pm&\rightarrow\,(H^\pm H),\,(H^\pm A)
\end{eqnarray}
{where $\bar{q}'$ denotes a different quark flavour. The production of $H^+\,H^-$ and $A\,A$ are (additionally) mediated by }
\begin{\eqn}\label{eq:l345ind}
(g\,g),\,(q\,\bar{q})\rightarrow\,h\,\rightarrow\,(H^+\,H^-),\,(A\,A).
\end{\eqn}

{All processes} in (\ref{eq:smind}) {are} purely SM gauge coupling induced; the BSM dependence of the production cross-sections therefore purely stems from the masses of the produced particles via phase space. The process{es} in (\ref{eq:l345ind}) depend on the coupling{s} $\lam_{3}$ {and $\bar{\lam}_{345}$, respectively}\footnote{{See Appendix \ref{app:fr} for the relevant Feynman rules.}}; in the parameter range allowed after the scan, we found that this contribution is relatively small compared to the "SM{-gauge bosons} induced" production processes in (\ref{eq:smind}). {For the channel $AA$, on the other hand, the dominant contribution is from gluon-induced Higgs production. However,} we found that the production cross-sections for direct pair-production for $AA$ final states is relatively small, cf. (\ref{eq:allprocs}), and therefore neglect this process in the following discussion. 
In {F}igure{s} \ref{fig:xsec} {and \ref{fig:xsec2}} we show results for pair-production total cross-sections at a 13 \TeV~LHC for the dominant modes in the respective mass planes. {They} constitute the {\sl benchmark planes} which should be investigated in the current LHC run.
\begin{figure}
\begin{minipage}{0.45\textwidth}
\includegraphics[width=\textwidth]{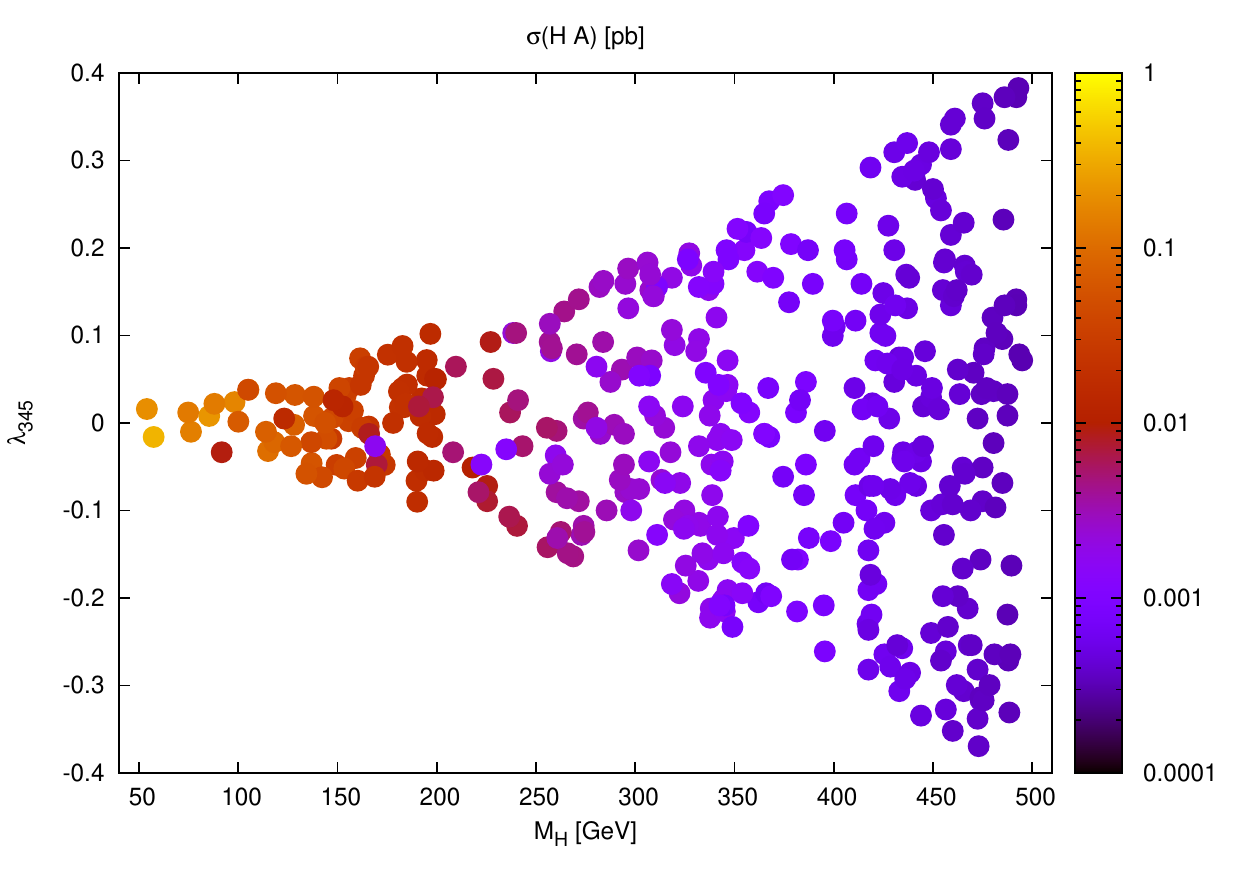}
\end{minipage}
\begin{minipage}{0.45\textwidth}
\includegraphics[width=\textwidth]{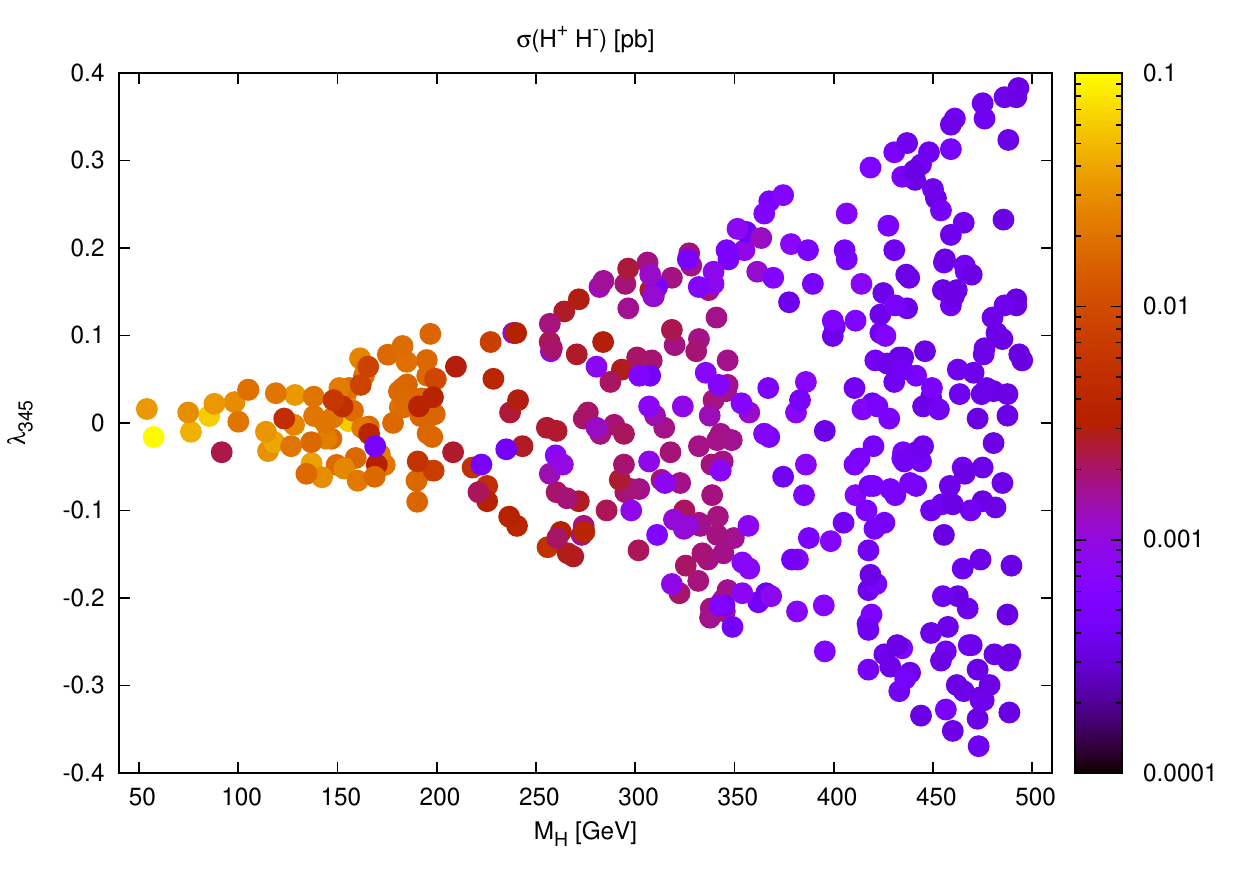}
\end{minipage}
\begin{minipage}{0.45\textwidth}
\includegraphics[width=\textwidth]{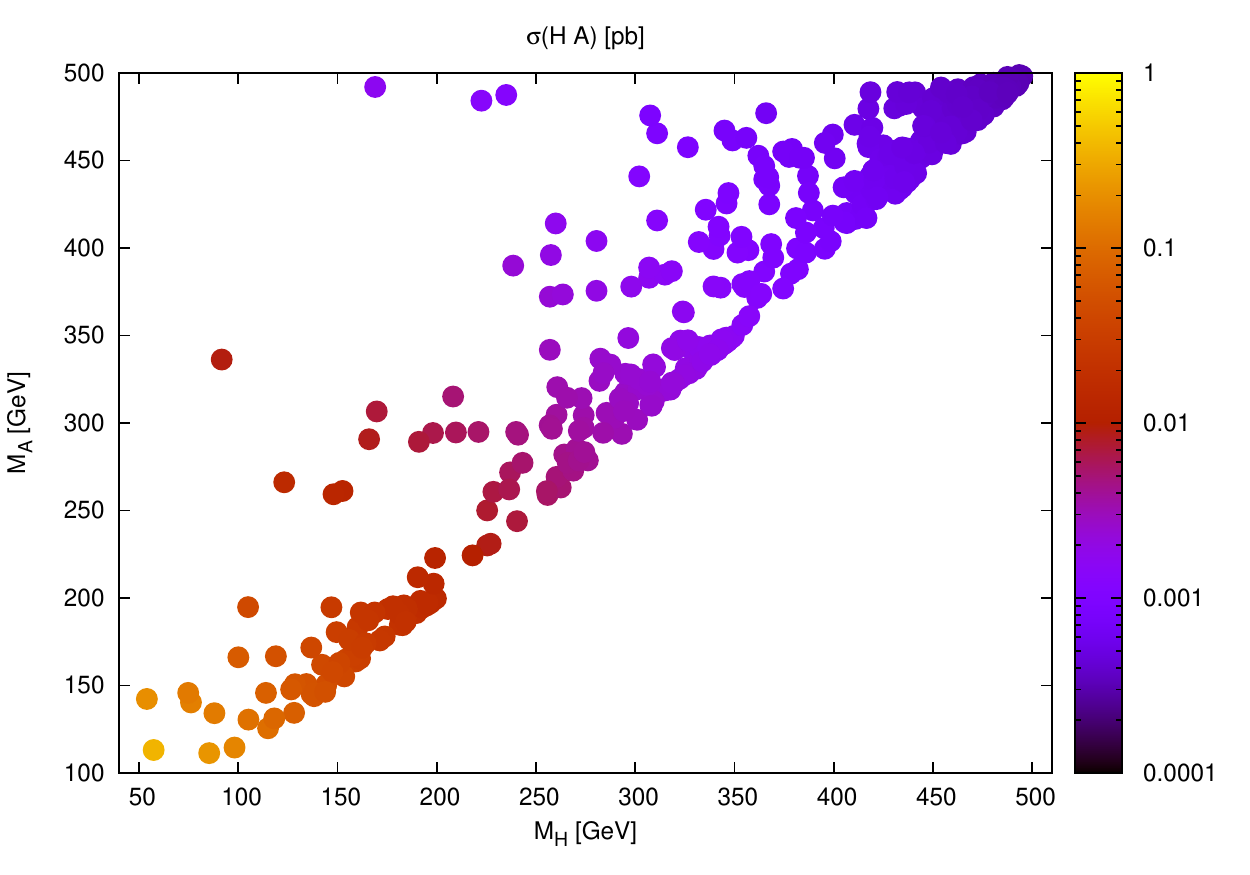}
\end{minipage}
\begin{minipage}{0.45\textwidth}
\includegraphics[width=\textwidth]{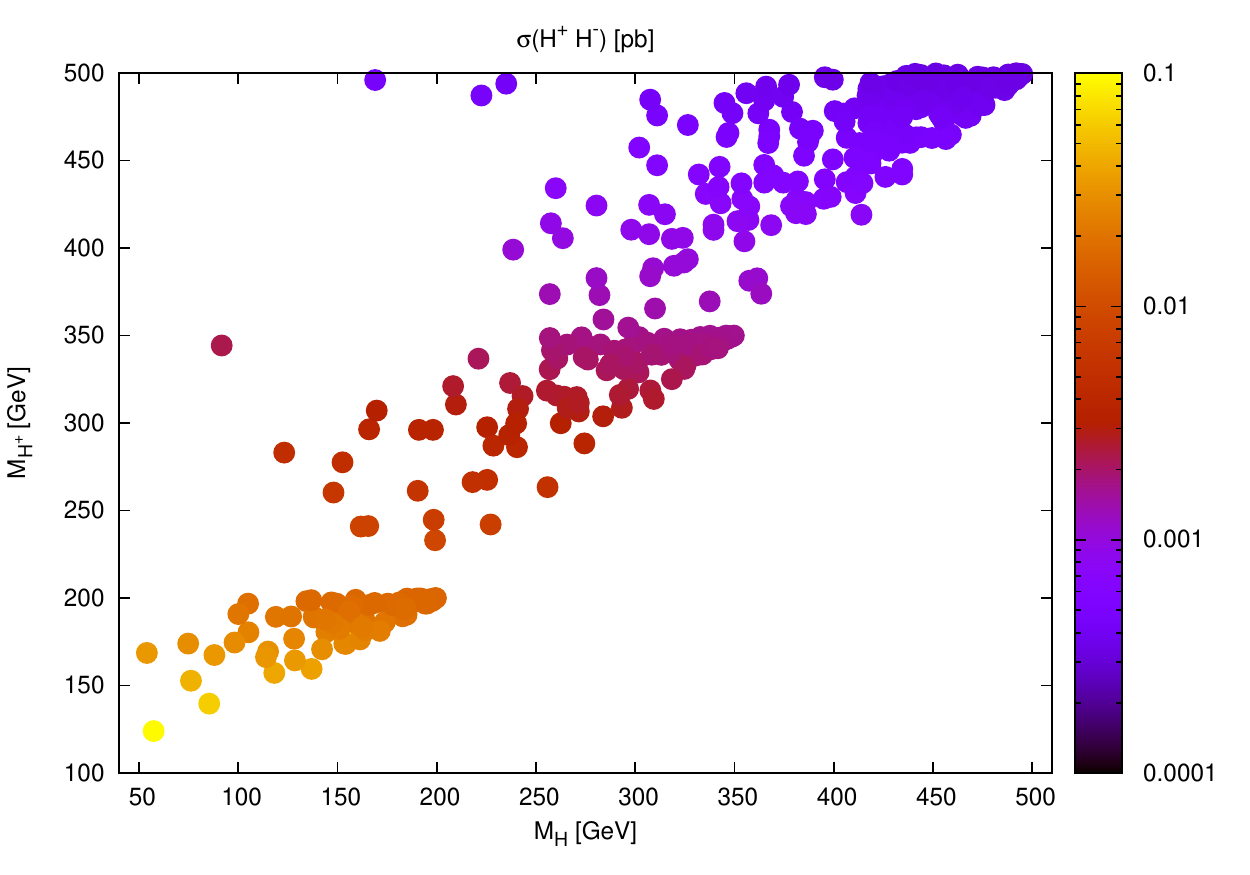}
\end{minipage}
\caption{\label{fig:xsec} {\bf Benchmark planes}: Production cross-sections in \pb~ at a 13 \TeV~ LHC for $H A$ {\sl (left)} and $H^+\,H^-$ {\sl (right)}, in the $M_H,\,\lam_{345}$ plane {\sl (upper low)} and in the $M_H,\,M_A$ or $M_H,\,M_H^{\pm}$ plane {\sl (lower row)}. }
\end{figure}
\begin{figure}
\begin{minipage}{0.45\textwidth}
\includegraphics[width=\textwidth]{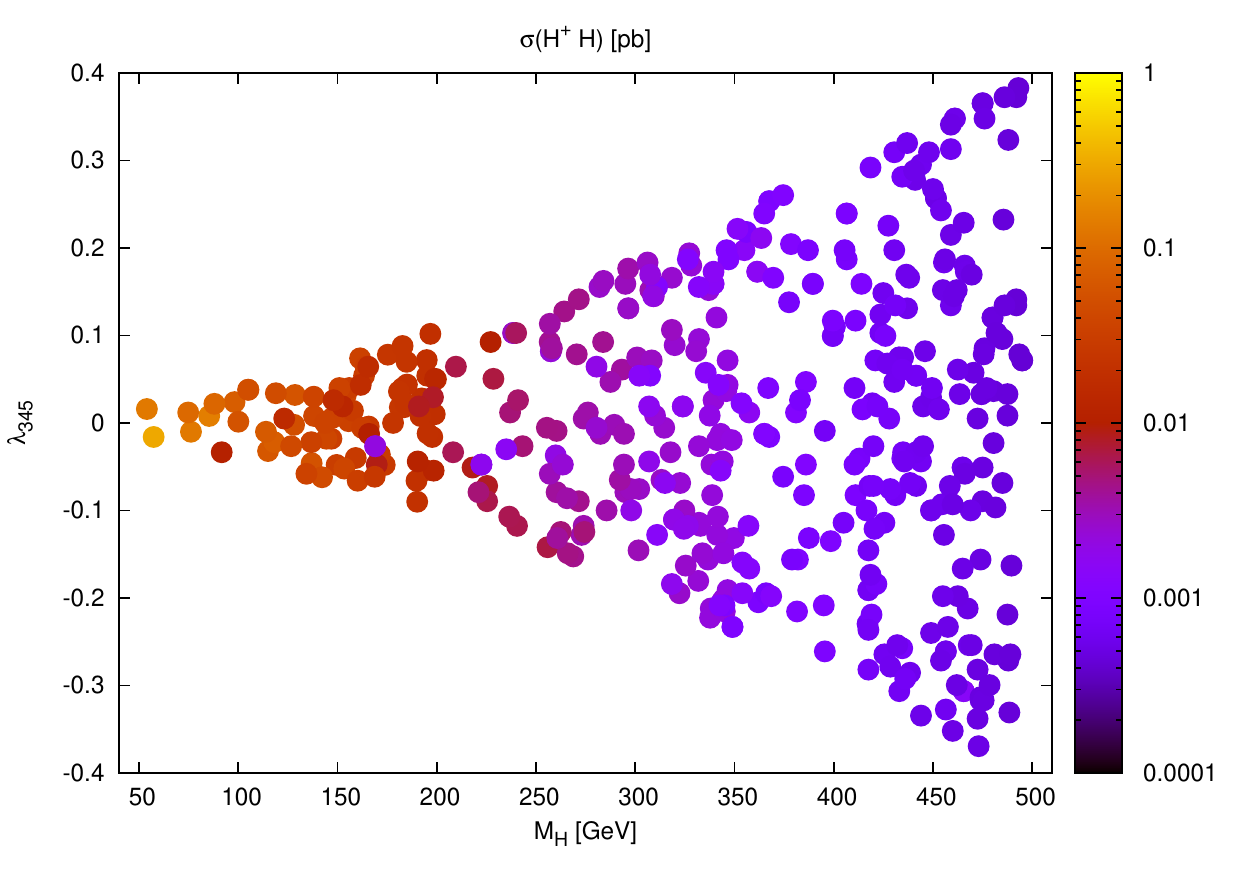}
\end{minipage}
\begin{minipage}{0.45\textwidth}
\includegraphics[width=\textwidth]{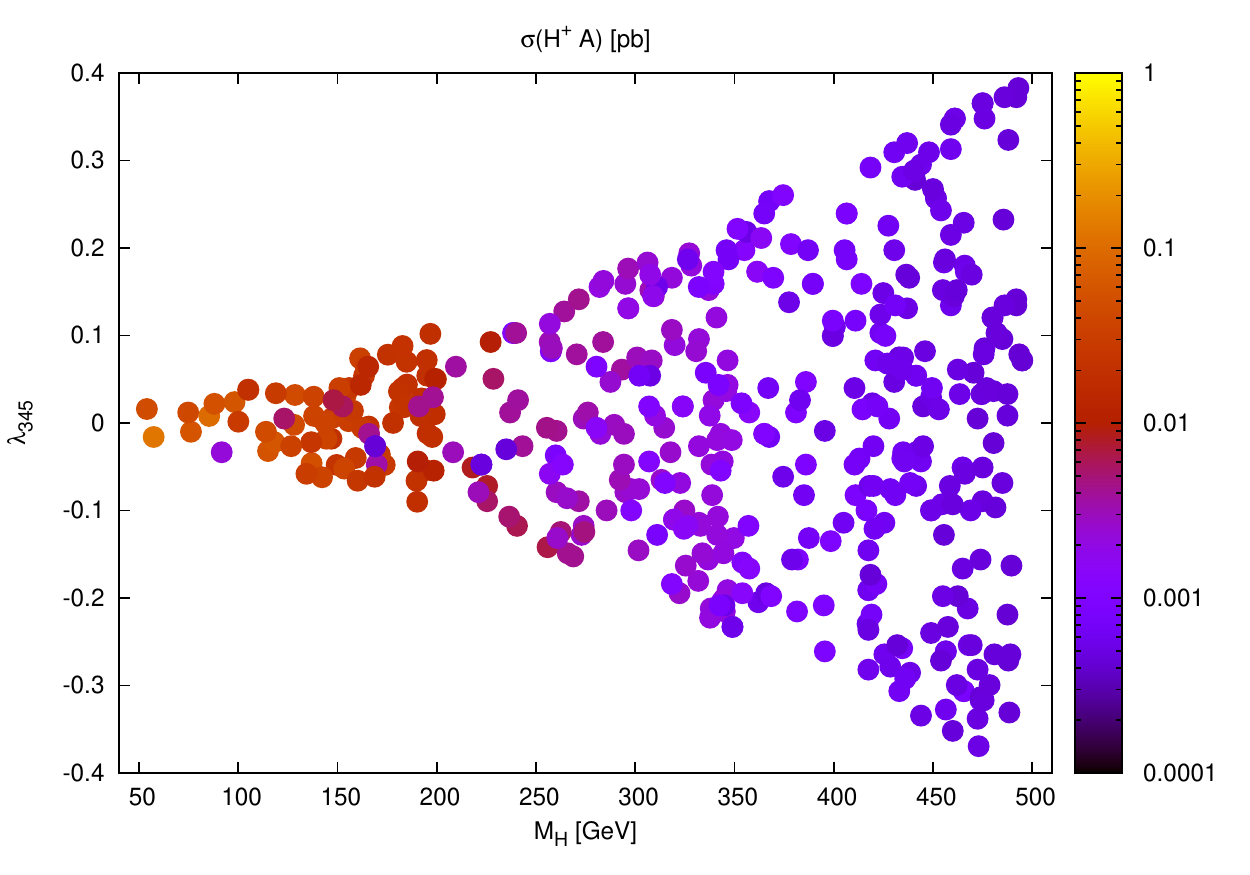}
\end{minipage}
\begin{minipage}{0.45\textwidth}
\includegraphics[width=\textwidth]{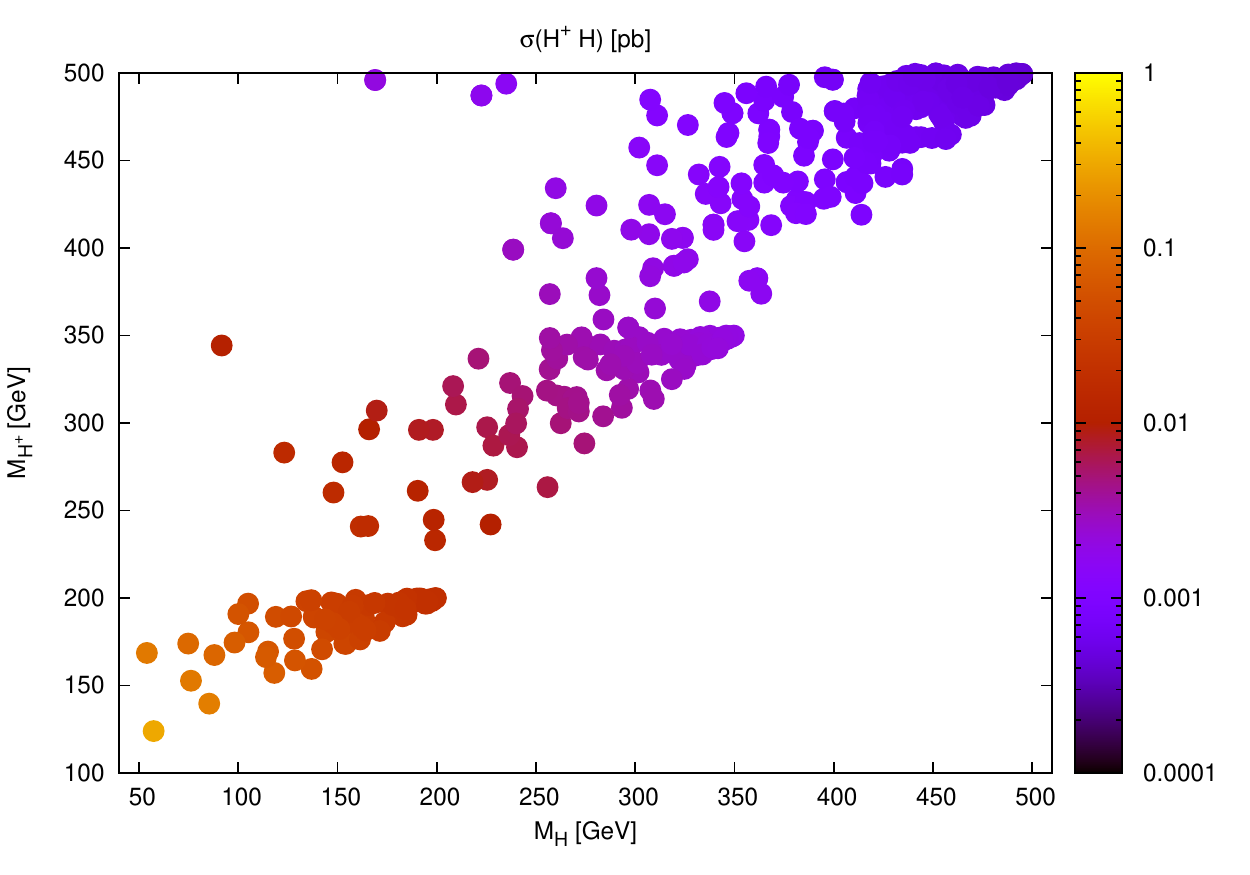}
\end{minipage}
\begin{minipage}{0.45\textwidth}
\includegraphics[width=\textwidth]{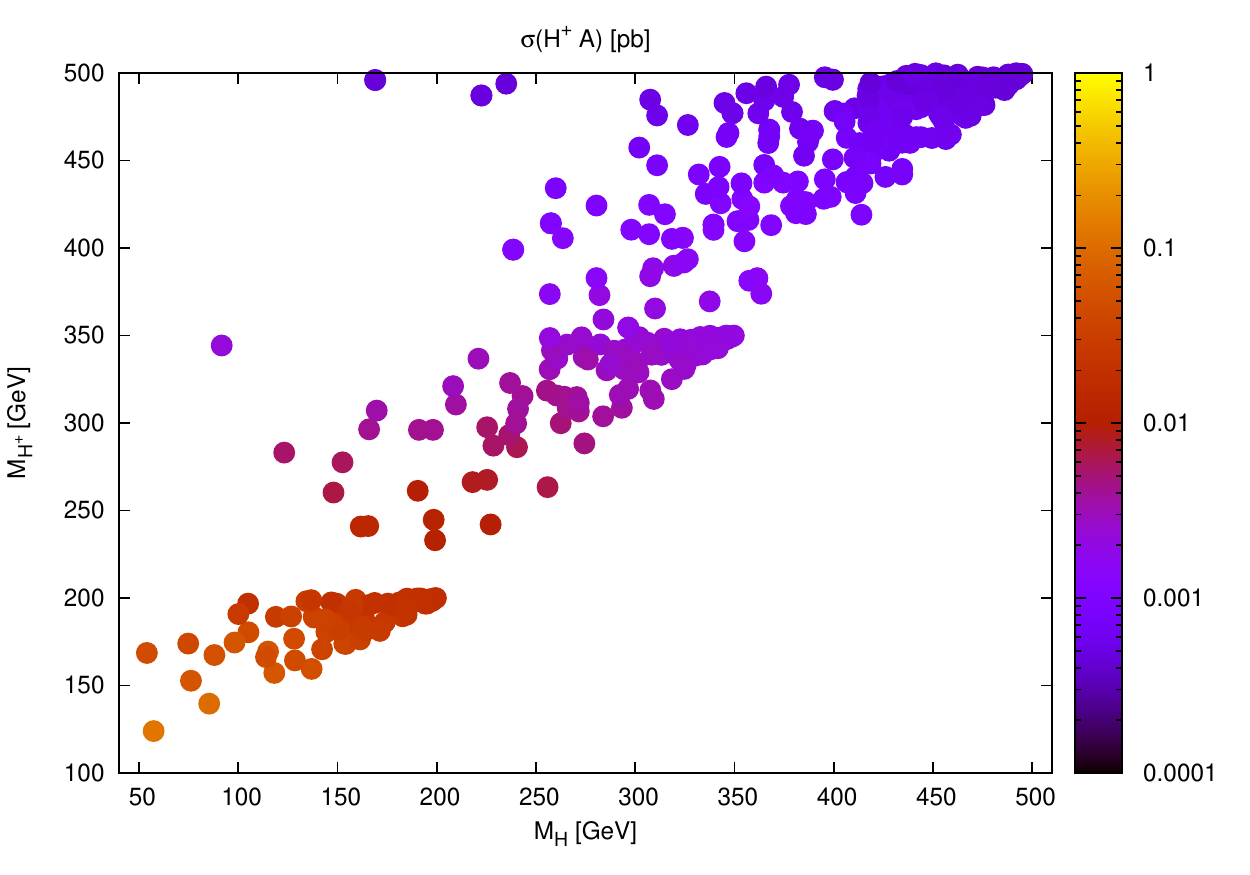}
\end{minipage}
\caption{\label{fig:xsec2} {\bf Benchmark planes}: Production cross-sections in \pb~ at a 13 \TeV~ LHC for $H^\pm H$ {\sl (left)} and $H^\pm\,A$ {\sl (right)}, in the $M_H,\,\lam_{345}$ plane {\sl (upper low)} and in the $M_{H^\pm},\,M_H$ plane {\sl (lower row)}. }
\end{figure}
We can see that the main factor driving the cross-sections values are masses of dark {particles} and their {hierarchy}.

For collider phenomenology, the decays of the unstable particles $H^\pm,\,A$ need to be taken into account. In our scan
\begin{\eqn*}
\text{BR}\,\lb A\,\rightarrow\,Z\,H \rb\,=\,1,
\end{\eqn*}
while for most parameter points the decay
\begin{\eqn}\label{eq:hpdec}
H^\pm\,\rightarrow\,W^\pm\,H
\end{\eqn}
dominates; the second decay mode is given by $H^\pm\,\rightarrow\,W^\pm\,{A}$.  For the process (\ref{eq:hpdec}), we plot the respective branching ratio in the $(M_H,\,M_{H^\pm})$ plane in Fig. \ref{fig:brhpm}. All branching ratios have been obtained using 2HDMC. The electroweak gauge bosons decay as in the SM.
\begin{figure}
\centering
\begin{minipage}{0.45\textwidth}
\begin{center}
\includegraphics[angle=-90, width=\textwidth]{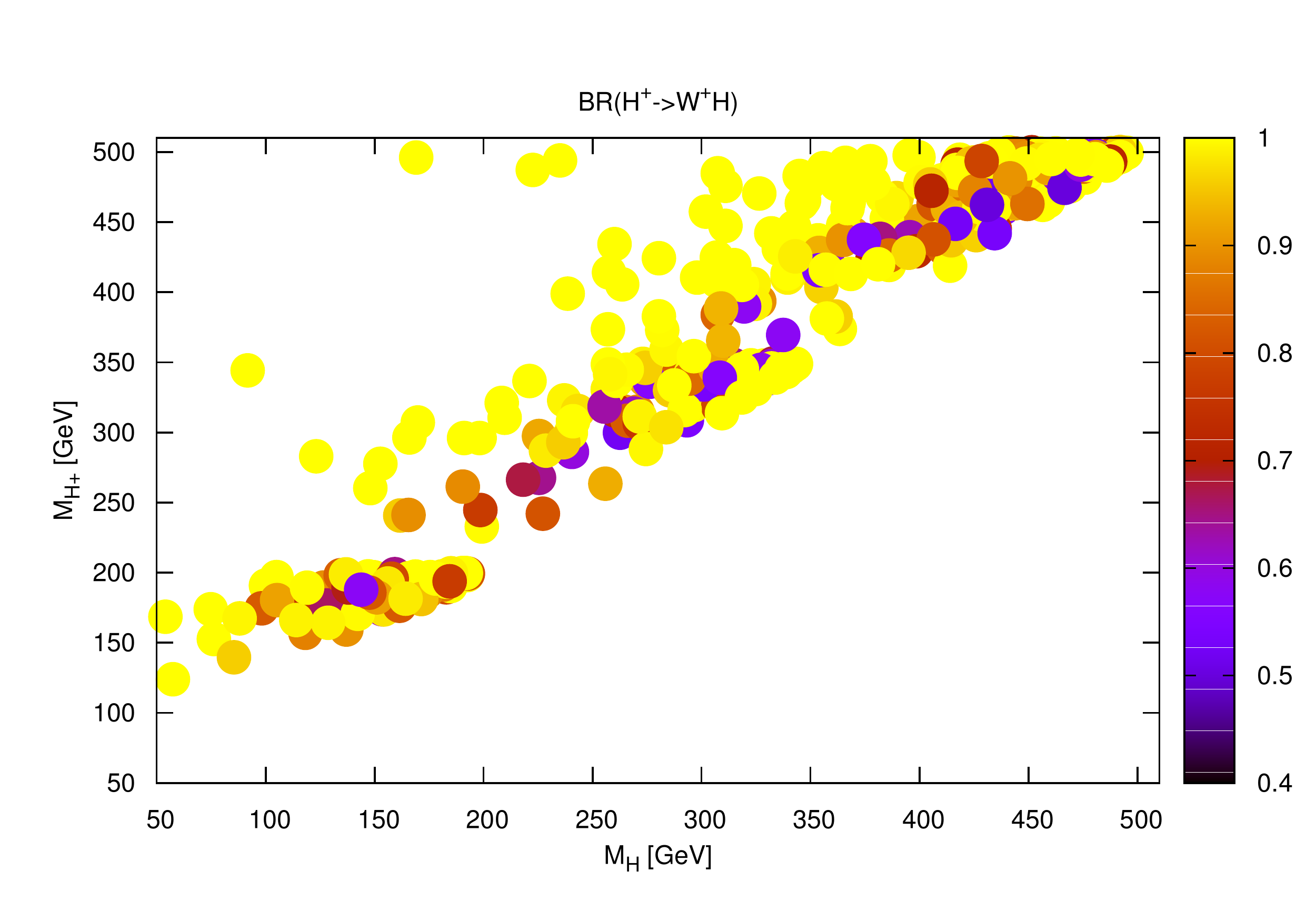}
\end{center}
\end{minipage}
\caption{\label{fig:brhpm} { Branching ratio of charged scalar decay into W boson and dark matter (H). The only other channel is $H^{\pm} \rightarrow A W^\pm$. } }
\end{figure}
As before, the couplings involved all stem from the electroweak gauge sector.

{Before proceeding to the discussion of benchmark points, we want to briefly comment on the results reported here. As discussed above, we do not perform a dedicated phenomenological study, which is clearly beyond the scope of the present work; especially for SM backgrounds, dedicated searches at 13 \TeV, including the latest higher order predictions for such processes, would be needed, which requires an extensive computational setup clearly beyond the scope of the present study. However, even from the numbers quoted above one can draw conclusions as follows
\begin{itemize}
\item{}According to the current design strategy at LHC Run II (cf. e.g. \cite{lhc_lumi}), at $13/ 14~ \TeV$ the LHC aims at a total integrated luminosity of $\sim\,100\,\fb^{-1}$ before the second long shutdown (LS2, in 2019) and a total integrated luminosity of about $300\,\fb^{-1}$ before the third LS (in 2023/4).
\item{}These numbers can be easily translated to number of events which could be produced for these two time benchmarks: for about 100 total events produced at LS2 (LS3), a minimal total production cross-section of 1 (0.3) \fb~ would be needed.
\item{}Considering the numbers presented above, this means that a priori all channels discussed here can and should be investigated by the experiments at LHC run II. However, in practise, for LS2 only scenarios for dark masses $\lesssim\,400\,\GeV$ might provide large enough cross-sections to render feasible predictions.
\end{itemize}
Although a more dedicated study might provide more information, we consider the above already an important input for the explicit design of experimental search strategies at LHC Run II.
}

\subsection*{Benchmark points}
In addition to the benchmark planes presented above, we also provide some {\sl benchmark points} for our model which should be considered for investigation at the present LHC run\footnote{{See e.g. \cite{Dorsch:2014qja,Hespel:2014sla,Haber:2015pua} for benchmark points for general 2HDMs, {as well as Appendix \ref{app:bms} for previous benchmark suggestions in the IDM which are excluded by now.}}}; {other viable benchmark points for masses $\leq\,1\,\TeV$ have e.g. been presented in \cite{Gustafsson:2012aj,Garcia-Cely:2013zga,Goudelis:2013uca}}. For this, we fix the dark scalars masses. {The couplings $\lam_2,\,\lam_{345}$, when kept within the allowed ranges discussed above, do not have a significant influence on the production and decay processes discussed here\footnote{{In this respect, the benchmarks above are also safe from constraints when vacuum stability and perturbativity are promoted to higher scales, as e.g. in \cite{Goudelis:2013uca}.}}. For these, we therefore give the allowed ranges which should be kept for consistency reasons.} Cross-sections have been produced as described above. In the list of benchmarks given below, we provide respective production cross-sections for dark scalar pair-production at a 13 \TeV~ LHC.

\begin{itemize}
\item {\bf Benchmark I: low scalar mass}
\begin{\eqn*}
M_H=57.5\,\GeV,\,M_A=113.0\,\GeV,M_{H^\pm}= 123,\,\GeV,\,\lam_2\,\in[0;4.2],\,|\lam_{345}|\,\in[-0.002;0.015] 
\end{\eqn*}
$H A: 0.371 (4) \pb,\,H^+\,H^-: 0.097 (1) \pb\,${$,H^\pm H: 0.3071 (4) \pb,\,H^\pm\,A: 0.1267 (1) \pb$} \\
A decays 100 $\%$ to $Z\,H$, and $H^+\,\rightarrow\,W^+\,H$ with a BR $\geq\,0.99$.
\item {\bf Benchmark II: low scalar mass}
\begin{\eqn*}
M_H=85.5\,\GeV,\,M_A=111.0\,\GeV,M_{H^\pm}= 140,\,\GeV,\,\lam_2\,\in[0;4.2],\,|\lam_{345}|\,\in[0.;0.015] 
\end{\eqn*}
$H A: 0.226 (2) \pb, H^+ H^-: 0.0605 (9) \pb$ {$,\,H^\pm H: 0.1439 (2) \pb,\,H^\pm\,A: 0.1008 (1) \pb$}\\
A decays 100 $\%$ to $Z\,H$, $H^+\rightarrow\,W^+\,H (A)$ with a BR $\sim 0.96 (0.04)$.
\item {\bf Benchmark III: intermediate scalar mass}
\begin{\eqn*}
M_H=128.0\,\GeV,\,M_A=134.0\,\GeV,M_{H^\pm}= 176.0,\,\GeV,\,\lam_2\,\in[0;4.2],\,|\lam_{345}|\,\in[0.;0.05] 
\end{\eqn*}
 $H\,A: 0.0765 (7)\pb$, $H^+\,H^-: 0.0259 (3) \pb${$,\,H^\pm H: 0.04985 (5) \pb,\,H^\pm\,A: 0.04653 (5) \pb$};\\
A decays 100 $\%$ to $Z\,H$,  $H^+\rightarrow\,W^+\,H (A)$ with a BR $\sim 0.66 (0.34)$
\item {\bf Benchmark IV: high scalar mass, mass degeneracy}
\begin{\eqn*}
M_H=363.0\,\GeV,M_A= 374.0\,\GeV,M_{H^\pm}= 374.0\,\GeV,\,\lam_2\,\in[0;4.2],\,|\lam_{345}|\,\in[0.;0.25] 
\end{\eqn*}
$H,A: 0.00122(1) \pb$, $H^+ H^-:  0.00124 (1) \pb${$,\,H^\pm H: 0.001617 (2) \pb,\,H^\pm\,A: 0.001518 (2) \pb$};\\
A decays 100 $\%$ to $Z\,H$, and $H^\pm$ 100 $\%$ to $W^\pm H$ 
\item {\bf Benchmark V: high scalar mass, no mass degeneracy}
\begin{\eqn*}
M_H=311.0\,\GeV,M_A= 415.0\,\GeV, M_{H^\pm}\,=\,447.0\,\GeV,\lam_2\,\in[0;4.2],\,|\lam_{345}|\,\in[0.;0.19] 
\end{\eqn*}
$H,A: 0.00129 (1) \pb$, $H^+ H^-: 0.000553 (7) \pb${$,\,H^\pm H: 0.001402 (2) \pb,$ $\,H^\pm\,A: 0.0008185 (8) \pb$};\\
A decays 100 $\%$ to $Z\,H$,  $H^+\rightarrow\,W^+\,H$ with a BR $\gtrsim 0.99$ 
\end{itemize}
While benchmarks I and II are exceptional points in a sense that the allowed parameter space is extremely constrained in the low mass region, benchmarks III to V are more typical, as these parts of the parameter space are more highly populated. Furthermore, for scenario IV the production cross-sections for $HA$ and $H^+ H^-$ have similar order of magnitude. {Of all benchmarks above, only the first one is able to produce the correct {95\% CL} relic density (Eqn.~\ref{eq:planck}).}
 
IDM searches {at the} LHC {call for a detailed investigation} of {the} best experimental strategy, as final states with IDM particles can have different topologies. {For example}, in {the} two lepton + MET channel, {the emitted leptons can have significantly different kinematic properties{, namely}:}
\begin{itemize}
\item{}$HA\rightarrow HHZ \rightarrow HH l^+l^-$ with two boosted leptons, {with an invariant mass around $m_Z$ for on-shell decays}
\item{}$H^+H^- \rightarrow HHW^+W^- \rightarrow HH \bar{\nu}\nu l^+ l^-$ with two leptons which can be {well} separated.
\end{itemize}
{S}tudies of possible search strategies for the IDM were discussed in e.g. \cite{Cao:2007rm,Dolle:2009ft,Miao:2010rg,Gustafsson:2012aj}. 

{Additionally,} we also show the {\sl total width} of the unstable dark scalars in Fig.~\ref{fig:tw}. For all points, $\Gamma/M\,\sim\,0.05$, with  $\Gamma/M\,\sim\,0.01$ being a more typical value. We therefore expect that use of the Narrow Width Approximation is well justified in {the} parameter space {explored in our study}.\footnote{Full matrix elements including interference can of course be easily used in the Monte Carlo framework described above.}

\begin{figure}
\includegraphics[angle=-90,width=0.45\textwidth]{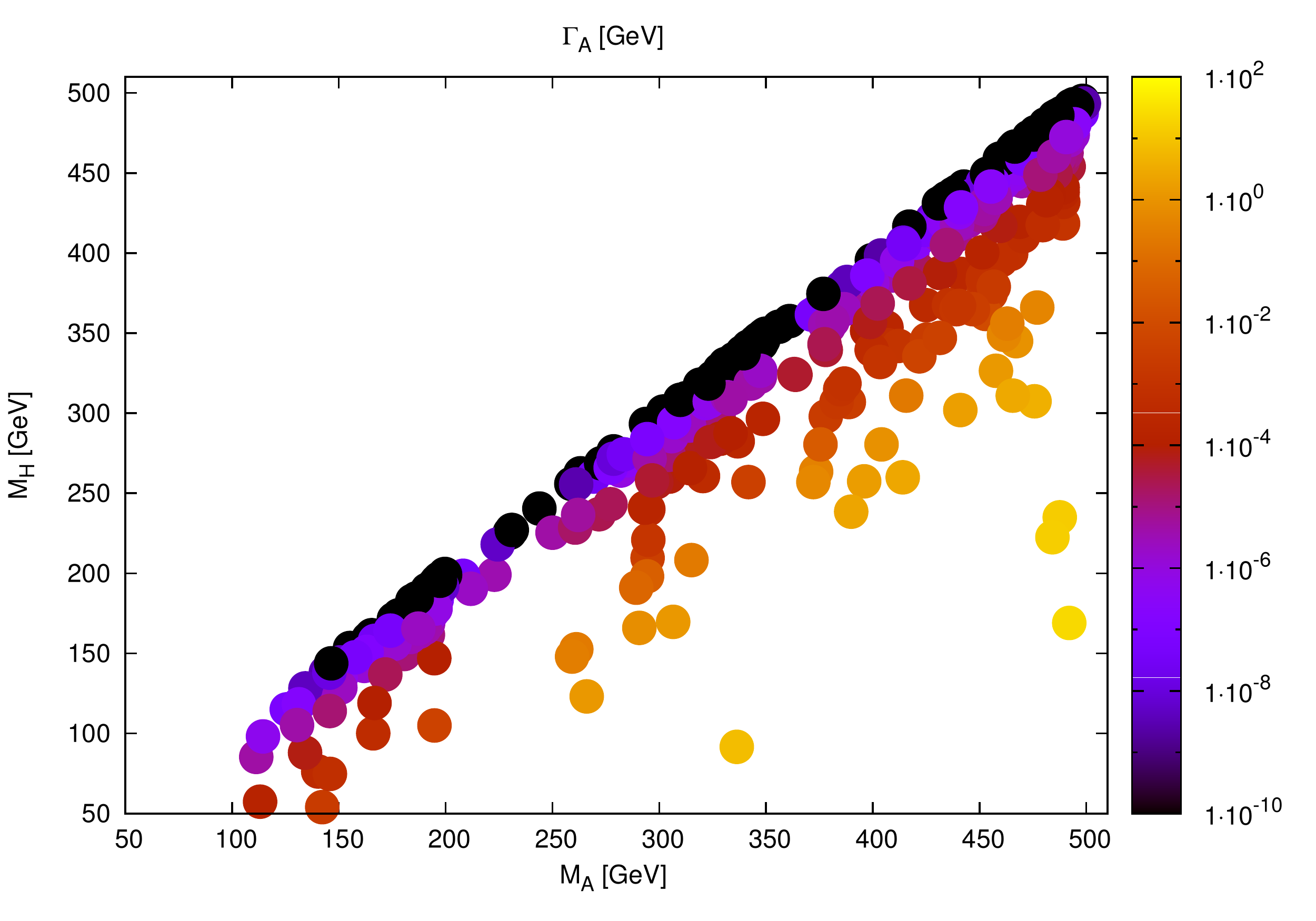}
\includegraphics[angle=-90,width=0.45\textwidth]{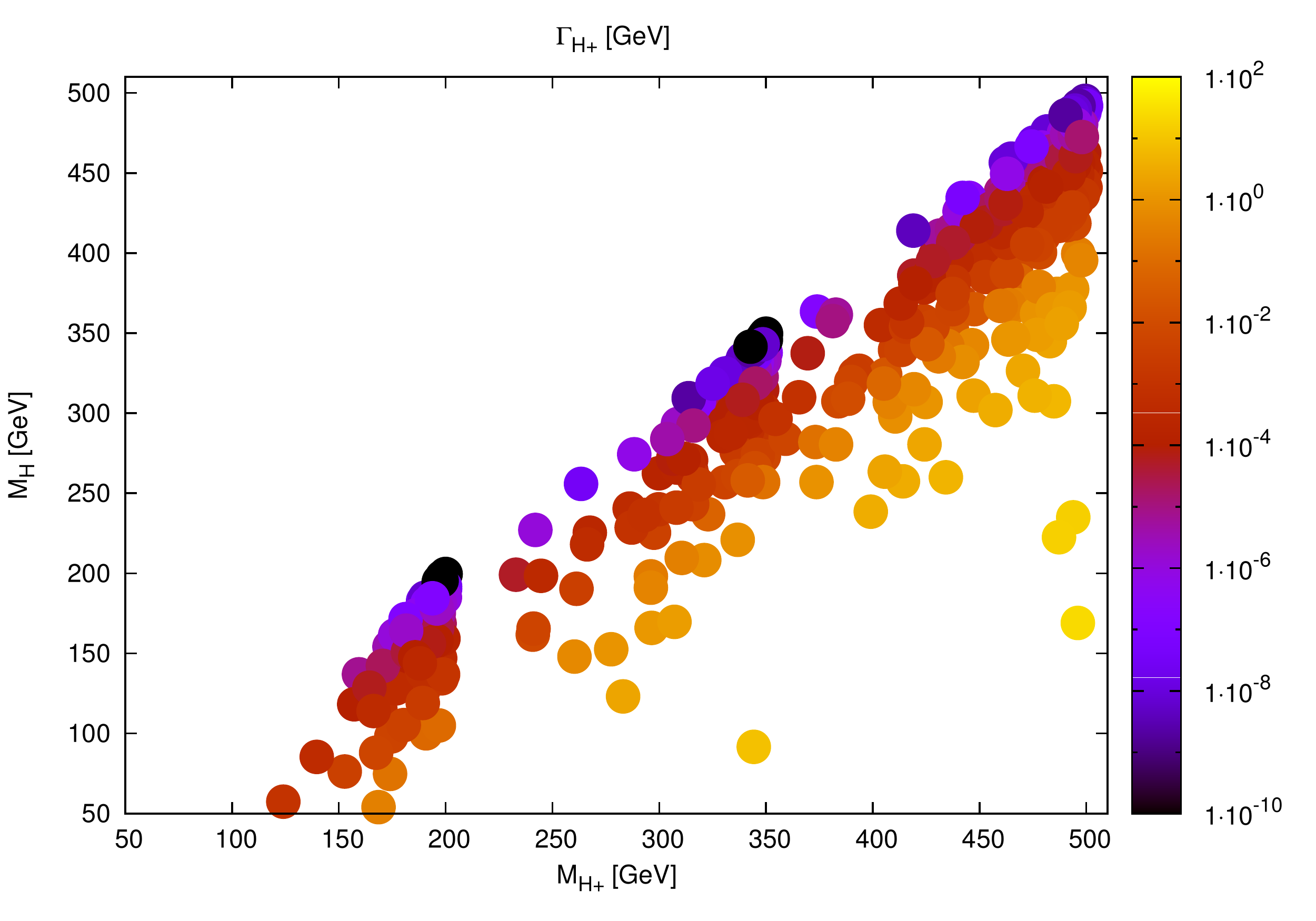}
\caption{\label{fig:tw} Total widths of unstable dark particles, {in \GeV}: A {\sl (left)} and H$^{\pm}$ {\sl (right)} in plane of their and dark matter masses. {For most points, $\Gamma/M\,\lesssim\,0.01$, with the maximal ratio $\Gamma/M\,\sim\,0.05$.}  }
\end{figure}

\section{Multi-component dark matter scenario in IDM}\label{sec:darkmc}

Before concluding, we want to briefly comment on changes in our analysis when we consider multi-component dark matter. In this case, the very strict limits from LUX in the $(M_H,\,\lam_{345})$ are significantly altered, and the allowed parameter space is greatly enhanced. This concerns especially upper values for the $\lam_{345}$ coupling. Below we plot several of our previous figures, now using the {\sl maximal} allowed value for direct detection cross-sections via rescaling {(cf. Eq. (\ref{eq:rescale}))}.
\begin{figure}[!tb]
\centering
\begin{minipage}{0.45\textwidth}
\includegraphics[width=\textwidth]{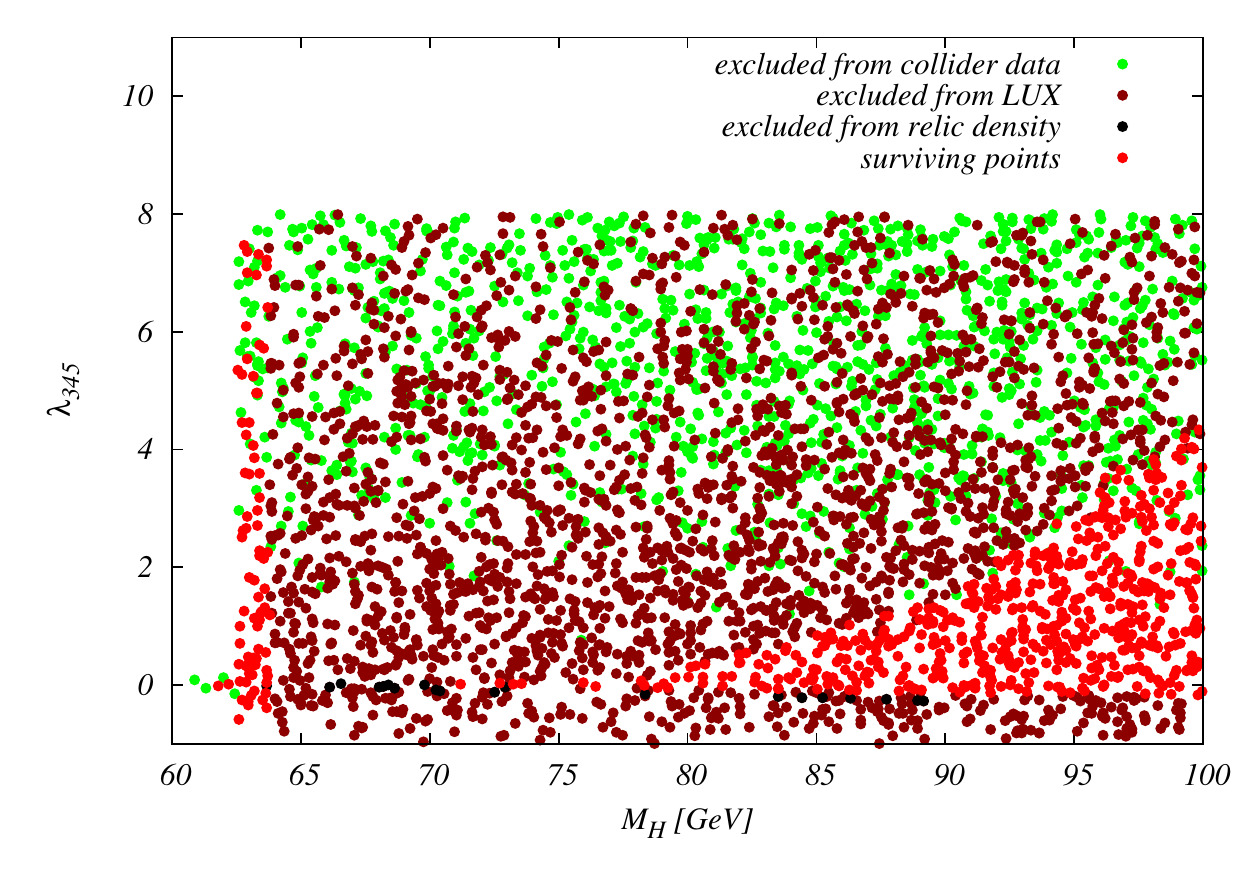}
\end{minipage}
\caption{\label{fig:l345bel1002} {Step 1} points for Dark Matter masses between 60 \GeV~ and 100 \GeV~ after the application of collider, {i.e. \HB~/\HS,} and dark matter limits. {Points below 60 \GeV~ were sampled more frequently.} }
\end{figure}%
\begin{figure}
\centering
\begin{minipage}{0.45\textwidth}
\includegraphics[width=\textwidth]{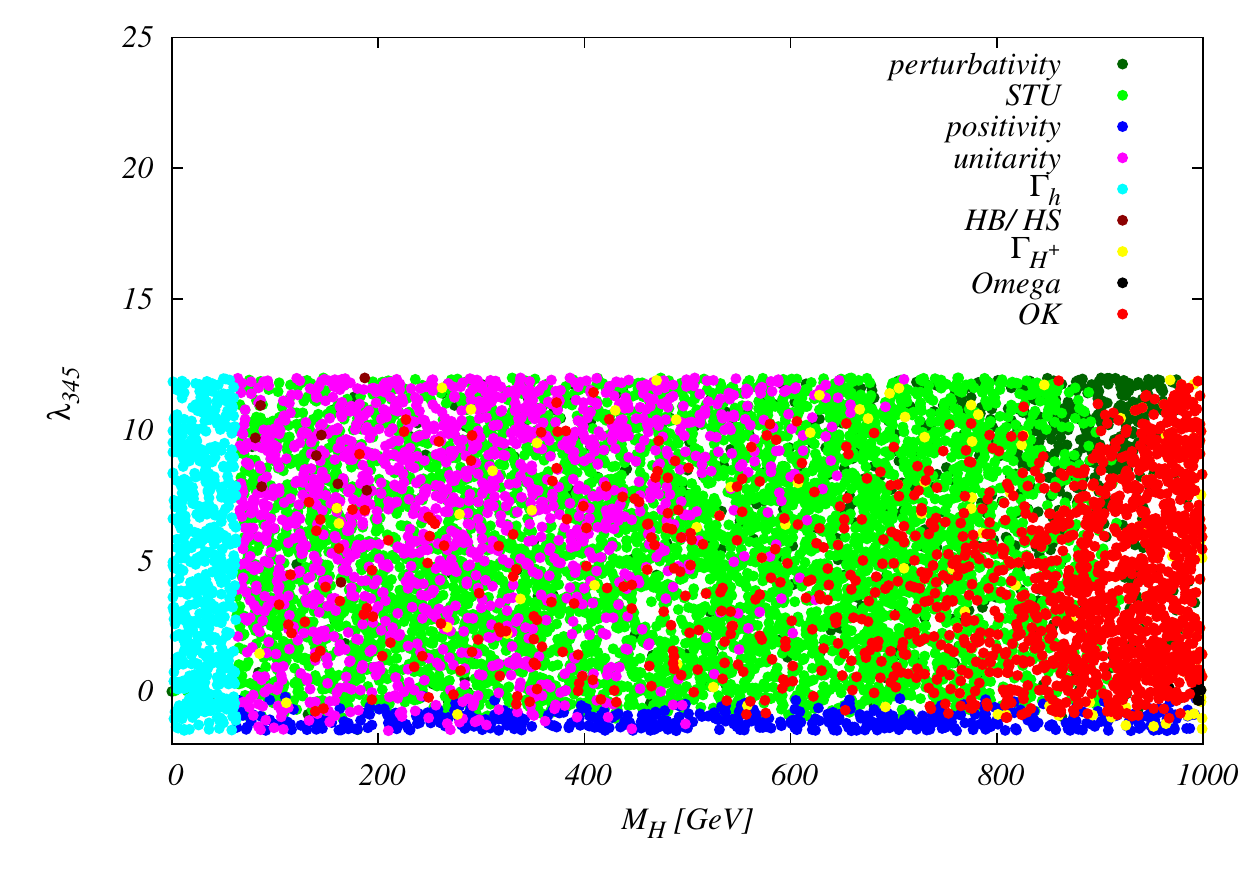}
\end{minipage}
\begin{minipage}{0.45\textwidth}
\includegraphics[width=\textwidth]{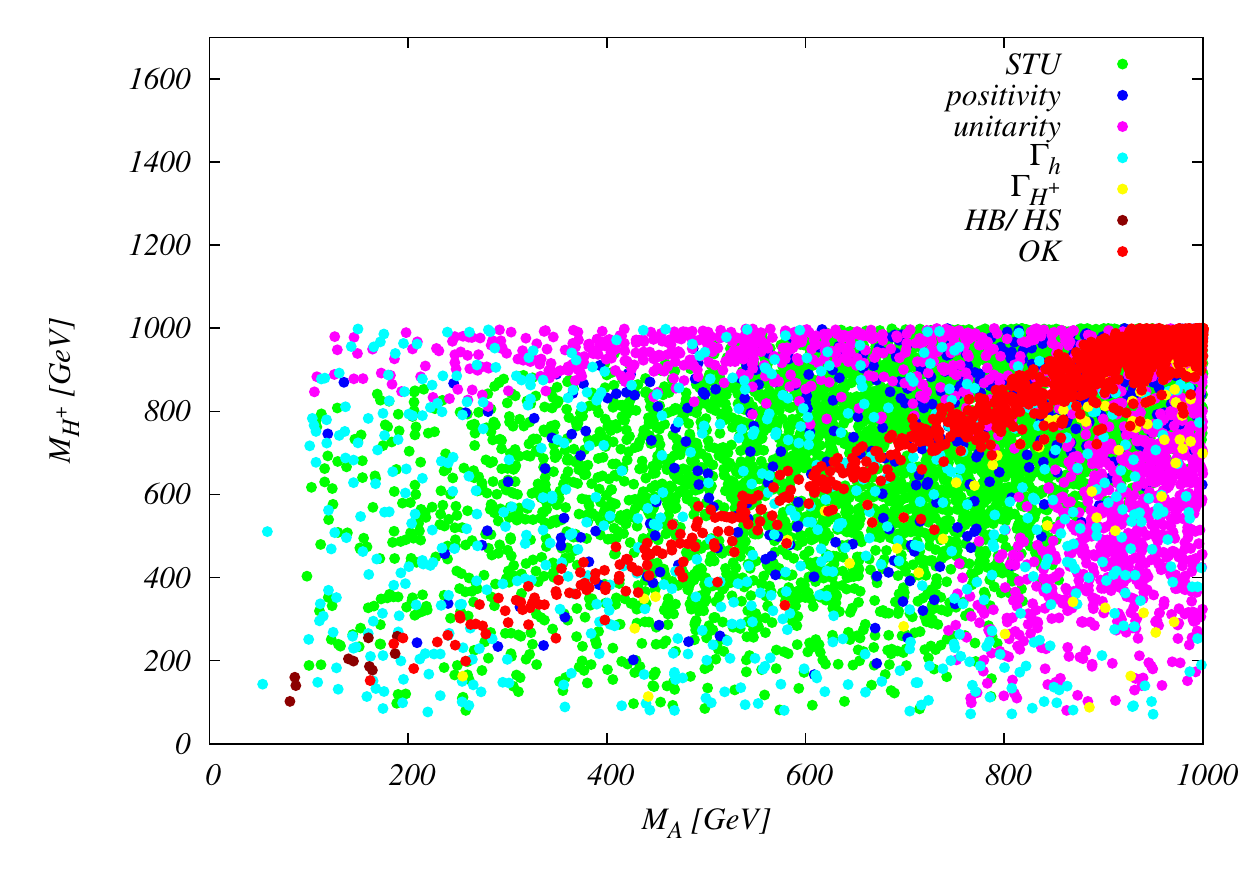}
\end{minipage}
\caption{\label{fig:l345plots1_resc} {\sl (Left):} As Fig. \ref{fig:l345plots1} {\sl (right)}; {\sl (Right):} as Fig. \ref{deltas} {\sl (right)}, with rescaled LUX limits.}
\end{figure}%
We can summarize our findings as follows
\begin{itemize}
\item{}for $M_H\,\lesssim\,60\,\GeV$, the picture basically does not change. Here, the strongest constraints on the $\lb M_H, \,\lam_{345} \rb$ plane stem from the Higgs signal strength measurements, as well as relic density, the impact of LUX limits is minimal. {However, for masses $M_H\,\in\,[50;60] \GeV$ the largest values for $|\lam_{345}|$ are slightly enhanced, from 0.02 to 0.03 as maximal values.} Note also that this is the region where in addition $\frac{\Omega^\text{LUX}}{\Omega}\,\approx\,1$.
\item{}{for masses $M_H\,\in\,[60;100] \GeV$, the picture dramatically changes. Now, $\lam_{345}$ can become as large as $\sim\,8$ for masses in the co-annihilation regime, where the dark matter relic density is relatively low, cf. fig \ref{fig:l345bel1002}. Furthermore, for masses $M_H\,\gtrsim\,80\,\GeV$ again the parameter space opens up significantly. } 
\item{}for $M_H \,\geq\,100\,\GeV$, the parameter space in the $\lb M_{H},\,\lam_{345} \rb$ plane  {also} largely opens up, allowing $\lam_{345}$ to reach values $\sim\,4\,\pi$, cf. Fig. \ref{fig:l345plots1_resc}. The upper limit on this value are now basically set by $S,T,U$, perturbativity, and unitarity constraints. {In particular, constraints from direct detection do no longer play a prominent role in the determination of allowed regions in the $\lb M_{H},\,\lam_{345} \rb$ parameter plane.}
\item{}Similarly, the relation $M_{H^\pm}\,\geq\,M_A$ no longer holds; however, we still observe the same degeneracies between the dark scalar masses {as before}.
\item{}{Regarding total widths of the dark scalars, the general behaviour {also does not significantly change, and total widths are similar to the case without rescaling} (see Fig. \ref{fig:tw}). The ratio $\Gamma/M$ can now reach up to $10\,\%$ for the second neutral scalar A. As before, the charged scalar dominantly decays into $W^+\,H$. For the decay of $A$, the additional channel $H^\pm\,W^\mp$ also opens up, cf. Fig. \ref{fig:xsec3}.}
\end{itemize}
A primary effect of opening up the additional parameter space for $\lam_{345}$ is the fact that now the production channel $A\,A$ also renders cross-section values which are of similar order of magnitude as the other channels, cf. Fig. \ref{fig:xsec3}. Therefore, a discovery in this channel could additionally give information about the nature of the dark matter scenario, which again highlights the strong connection between collider and astroparticle physics in the IDM. All other production cross-sections are mainly gauge boson induced, and the production cross-sections remain at a similar order of magnitude.

Furthermore, if we now investigate the {\sl dominant} decay channels responsible for relic density, the relaxed constraints now also allow regions where channels other than the diboson annihilation dominate, and nearly all points from Fig. \ref{fig:dm_chan} {\sl (left)} are now allowed by the LUX limit.
\begin{figure}
\begin{minipage}{0.45\textwidth}
\includegraphics[width=\textwidth]{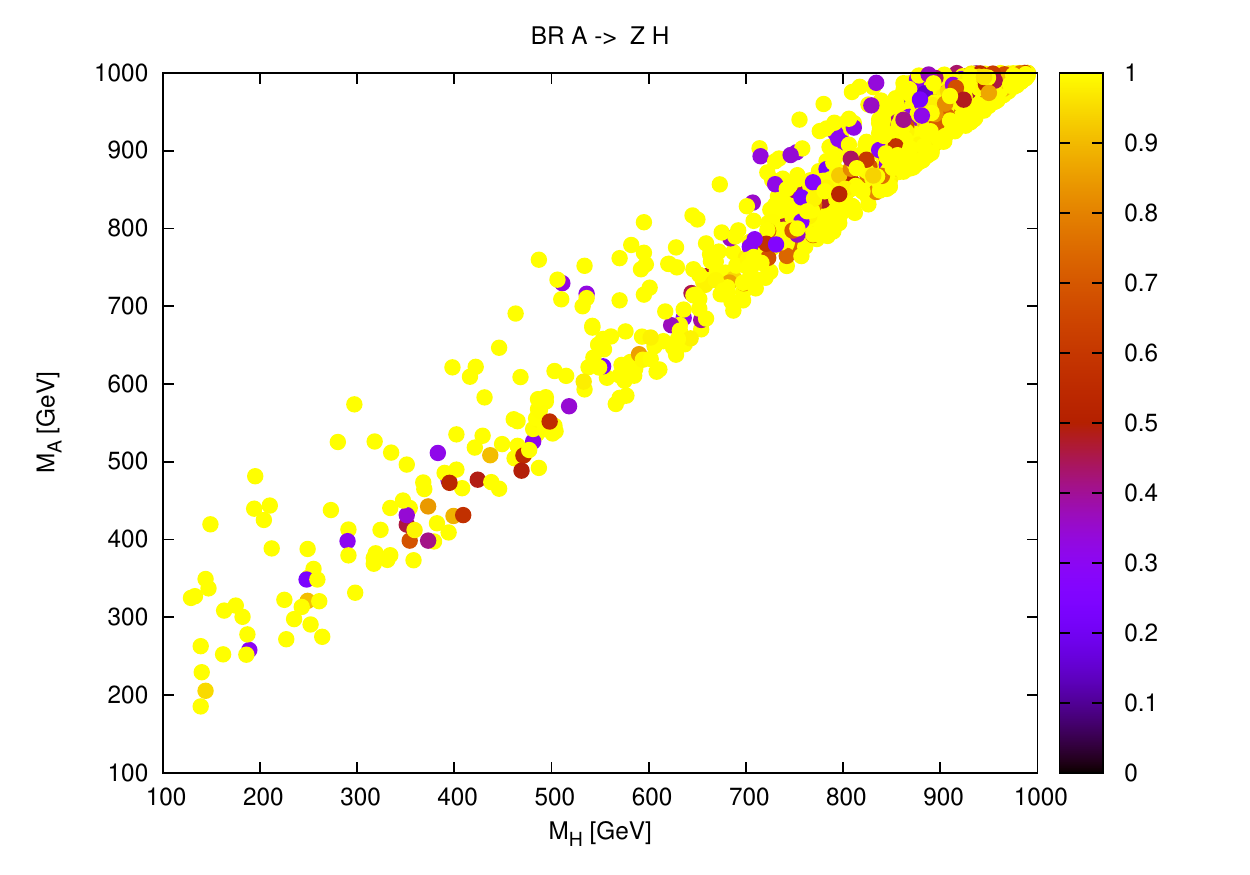}
\end{minipage}
\begin{minipage}{0.45\textwidth}
\includegraphics[width=\textwidth]{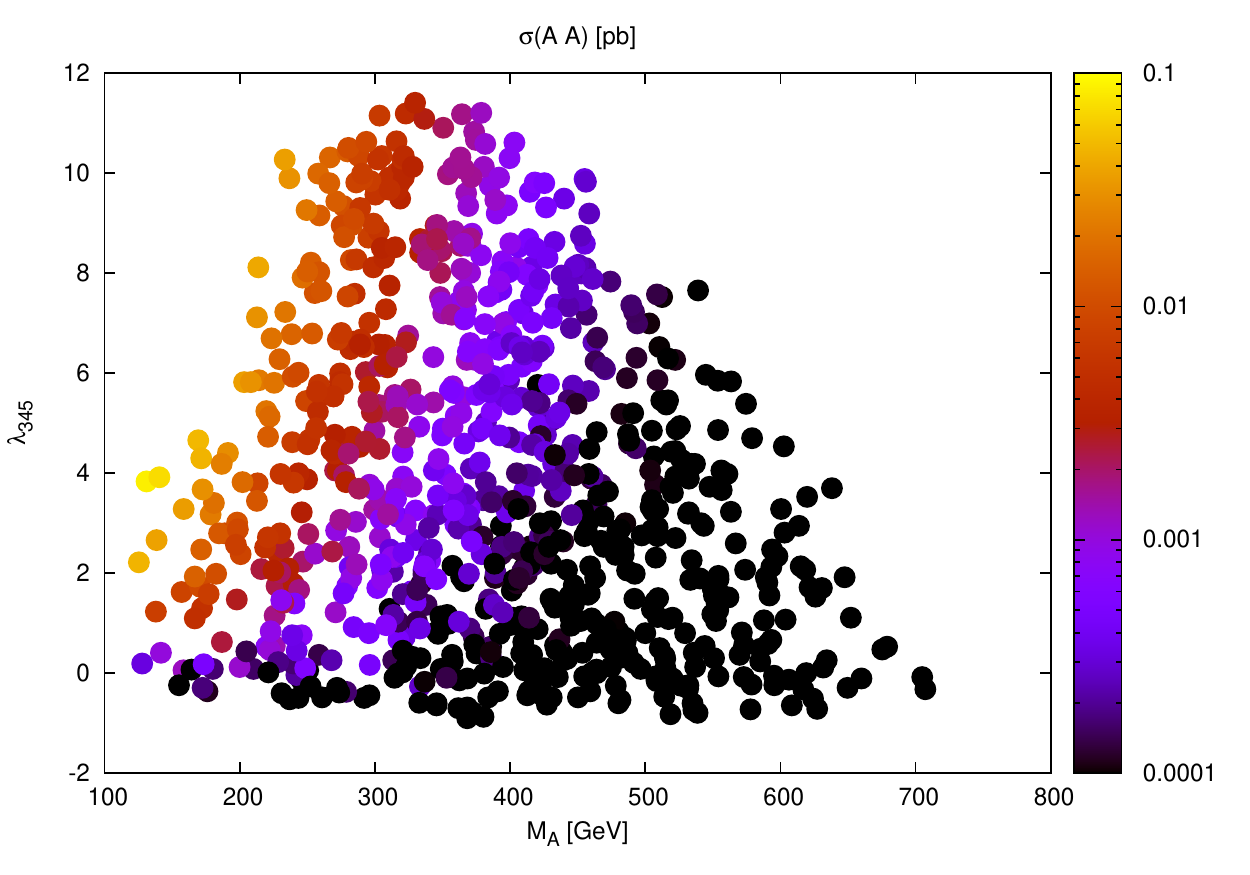}
\end{minipage}
\caption{\label{fig:xsec3}{{\sl (Left):} Branching ratio of $A$ to $Z\,H$; the other decay mode is $A\,\rightarrow\,H^\pm\,W^\mp$} {\sl (Right): }Production cross-sections in \pb~ at a 13 \TeV~ LHC for $A\,A$ in the $M_A,\,\lam_{345}$, when using rescaled LUX bounds. }
\end{figure}
\section{Conclusions}
In this work, we have revised the current theoretical as well as experimental constraints on a two Higgs doublet model with an exact $Z_2$ symmetry (Inert Doublet Model), which extends the SM Higgs sector in a {straightforward} way and in addition comes with a dark matter candidate. We have included all experimental bounds from direct searches, the 125 \GeV~Higgs coupling strength, as well as limits from {recent} astrophysical measurements such as limits from direct {DM search}{es} {as well as dark matter relic density, the latter in the form of an upper limit}. We have closely investigated the interplay of {these} bounds for dark scalar masses which are in the discovery regime of current and future collider experiments. We found that an interplay of astroparticle and collider constraints rules out regions with $M_H\,\lesssim\,45\,\GeV$, thereby setting a minimal mass scale on this model. Furthermore, the parameter space is highly constrained by direct {DM search} experiments, which supply stringent limits especially in the low mass region.\\

Apart from clear cuts on the additional couplings, we found that the combination of all constraints leads to a mass hierarchy in the dark scalar sector. {The combination of different constrains {ensures} $M_{H^\pm}\,\geq\,M_A > M_H$. {In addition}, in the flat scan more degenerate masses of dark scalars are preferred {for $M_H\,\geq\,200\,\GeV$}. {Furthermore, as we do not require the IDM to provide the full relic density,} intermediate masses of {the dark matter candidate} ($\sim 100-600\, \GeV$) become available, {which open up appealing parameter regions for the current LHC run.} {Our results therefore correspond to the most up-to-date investigation of the models parameter space, thereby providing important information about viable parameter regions for collider phenomenology.} {Furthermore, we also briefly commented on the case of multi-component dark matter. In this case, the limits from direct detection are relaxed and we find that especially for $M_H\,\geq\,M_h/2$, the parameter space opens up. This in turn leads to cross-sections for $AA$ pair-production which are of similar magnitude as $H^+H^-$ production, and could therefore in principle also be investigated at the current LHC run. This again emphasizes the complementarity of collider searches and astrophysical measurements within the IDM. } \\

 In addition, we have provided benchmark planes and cross-sections for pair-production of the dark scalars for the dominant production modes within our model in terms of the relevant new physics parameters, which should be used as input for the experimental searches. As both production and decay modes for the new particles are mainly determined by kinematics, the above degeneracies directly induce relatively small production cross-sections as the masses increase, leading to $\sigma\,\sim\,1\,\fb$ for masses $\geq\,300\,\GeV$. Nevertheless, this model constitutes an attractive extension of the SM, and provides an excellent example of complementarity between astroparticle and collider experiments. We strongly encourage the LHC experiments to investigate the benchmarks proposed here in the current LHC run.

\begin{acknowledgments}

 {TR wants to thank {P. Bechtle, J. Erler,} A. Goudelis, B. Herrmann, O. Stal and T. Stefaniak for useful discussions and R. Frederix for MG5 support.}  {We also thank B. Swiezewska, D. Sokolowska, P. Swaczyna for discussion on the IDM as well as the Higgs Cross Section working group for encouraging us to investigate possible benchmarks.} We thank S. Kraml and collaborators from Grenoble, for fruitful discussions and  S. Najjari for cross checking the Feynman rules for IDM. {This research was supported by the DAAD grant PPP Poland Project 56269947 "Dark Matter at Colliders" } and by the Polish grant NCN OPUS 2012/05/B/ST2/03306
(2012-2016). {The work of AI is supported by the 7th Framework Programme of the European Commission through the Initial Training Network HiggsTools PITN-GA-2012-316704.} 
\end{acknowledgments}

\begin{appendix}
\section{IDM Feynman rules and other relations}\label{app:fr}
The parameters $m_{22}^2,\,\lam_3,\,\lam_4,\,\lam_5$ {can be re-expressed} in terms of our input parameters: 
\begin{eqnarray}
&&m_{22}^2\,=\,\lam_{345}\,v^2-2\,M_H^2,\\
&&\lam_3\,=\,\lam_{345}-\frac{2}{v^2}\,\lb M_H^2-M_{H^\pm}^2 \rb,\,\lam_4\,=\,\frac{M_A^2+M_H^2-2\,M_{H^\pm}^2}{v^2},\,\lam_5\,=\,\frac{M_H^2-M_A^2}{v^2}
\end{eqnarray}

For completeness, we list the relevant Feynman rules including dark scalars (omitting goldstone modes, as we are working in the unitary gauge and at tree level), {see Tables \ref{tab:scal} and \ref{tab:gb} }. Note that, {since} the second doublet does not participate in electroweak symmetry breaking and fermion mass generation, the couplings of the SM-like Higgs {$h$} to electroweak gauge bosons as well as fermions are given by their SM values, {see e.g.}  \cite{Bohm:2001yx}, {with {the} convention $g_{h W^+_\mu W^-_\nu}=ie^2 v/2 {s_W}^2g_{\mu\nu}$ }.

\begin{table}[h!]
\begin{tabular}{l|c}
{vertex}& coupling\\ \hline
$hHH$&$\lam_{345}\,v$\\
$hAA$&$\bar{\lam}_{345}\,v$\\
$hhh$&$3\,\lam_1\,v$\\
$h\,H^+\,H^-$&$\lam_3\,v$\\
&\\ \hline
$hhhh$&$3\,\lam_1$\\
$H^+\,H^+\,H^-\,H^-$&$2\,\lam_2$ \\
$HHAA$&$\lam_2$\\
$HHHH$&$3\,\lam_2$\\
$AAAA$&$3\,\lam_2$\\
$H^+H^-\,AA$&$\lam_2$\\
$H^+H^-HH$&$\lam_2$\\
$hhH^+H^-$&$\lam_3$\\
$hhHH$&$\lam_{345}$\\
$hhAA$&$\bar{\lam}_{345}$
\end{tabular}
\caption{\label{tab:scal} Non SM-like couplings in the scalar sector of the IDM; a global factor of $-i$ is omitted.}
\end{table}
\begin{table}[h!]
\begin{tabular}{l|c}
{vertex}& coupling\\ \hline
$H^-\,H^+\,\gamma$&$i\,e$\\
$H^-\,H^+\,Z$&$i\,\frac{g}{2}\,\frac{\cos\,(2\theta_W)}{\cos\theta_W}$\\
$H\,H^\pm\,W^\mp$&$\mp\,i\,\frac{g}{2}$\\
$A\,H^\mp\,W^\pm$&$-\frac{g}{2}$\\
$H\,A\,Z$&$-\frac{g}{2\cos\theta_W}$
\end{tabular}
\caption{\label{tab:gb} Non-SM couplings to gauge {bosons} in the IDM; all vertices are proportional to \htb{a} kinematic factor $(p_2-p_1)_\mu$ for {a vertex of the form} $X_1(p_1)\,X_2(p_2)\,X_3(p_3)$.}
\end{table}

\section{$R_{\gamma\gamma}$ as a standalone variable}\label{app:rgg} 
As an alternative to the generic fit using \HS, and also as a consistency check, we can regard the branching ratio of the SM-like Higgs into two photons as a single constraining variable. We then take as a reference value \cite{Heinemeyer:2013tqa}
\begin{\eqn*}
\text{BR}_{h\,\rightarrow\,\gamma\gamma}^\text{SM}\,=\,\lb 2.28\,\pm\,0.11 \rb\,\times\,10^{-3}.
\end{\eqn*} 
Furthermore, we use \cite{moriond15}
\begin{\eqn*}
\mu_{\gamma\gamma}^\text{ATLAS}\,=\,1.17\,\pm\,0.28,\,\,\,\mu_{\gamma\gamma}^\text{CMS}\,=\,1.12\,\pm\,0.24
\end{\eqn*}
which yields a naively combined value of
\begin{\eqn*}
\bar{\mu}_{\gamma\gamma}\,=\,1.15\,\pm\,0.18.
\end{\eqn*}
Again combining these, we have
\begin{\eqn*}
\text{BR}_{h\,\rightarrow\,\gamma\gamma}\,=\,\lb 2.62\,\pm\,0.43 \rb\,\times\,10^{-3}
\end{\eqn*}
which at $2\,\sigma$ yields
\begin{\eqn*}
\text{BR}_{h\,\rightarrow\,\gamma\gamma}\,\in\,\left[1.76\,;3.84 \right]\,\times\,10^{-3}.
\end{\eqn*}
Obviously all errors combinations above neglect both correlations as well as non-gaussianity of the respective distributions.

\section{{Previous scans and benchmarks}}\label{app:bms}
\subsection{{Previous scans}}

{In this subsection, we give a brief overview on similar studies which have been performed after the discovery of a 125 \GeV~ Higgs, and point to differences with respect to the constraints and results achieved in this work. We specifically discuss \cite{Swiezewska:2012eh,Arhrib:2013ela, Krawczyk:2013jta, Swiezewska:2012ej,Goudelis:2013uca}.}
\begin{itemize}
\item{} Obviously, all above studies respect the theoretical bounds on the potential, i.e. positivity, as well as constraints from perturbative unitarity. Most of them also include the constraint in Eqn.~ (\ref{eq:invac}). As we did not find points in parameter space where this is the most important constraint, we can state that theoretical bounds where treated on equal footing\footnote{In \cite{Goudelis:2013uca}, scale dependent limits where also tested at larger scales. However, if we consider the IDM as an effective model at the electroweak scale, the limits coincide.}.
\item{}Similarly, decays from electroweak gauge bosons, electroweak precision constraints in terms of oblique parameter (although with slightly different central values), a lower bound of the charged Higgs mass $\sim\,70\,\GeV$, as well as recasted LEP results were also included in all above studies.
\item{}Refs.~ \cite{Swiezewska:2012eh,Krawczyk:2013jta, Swiezewska:2012ej} particularly discuss the {\sl interplay} of constraints from the 125 \GeV~ Higgs signal strength measurement, mostly in terms of possible modifications to the decay rate to diphoton final states, with constraints from dark matter detection. In fact, these will also prove important especially in the {\sl low mass} region of our scan, and have therefore been correctly identified in the above work as the most important constraints for certain parts of parameter space.
\item{}More specifically, in ref.~ \cite{Swiezewska:2012ej,Swiezewska:2012eh} the effect of including the by that time newly measured Higgs mass $M_h$ is discussed, and predictions for the SM-like Higgs decay into diphoton or $\gamma\,Z$ final states are considered, leading to additional constraints.  The authors present their results in terms of two-dimensional planes, and determine allowed/ forbidden regions in parameter space from the constraints included. Obviously, a more detailed analysis using Higgs coupling strength measurements and LUX data was not possible at that time, so these have not been included in the above study.
\item{}In ref.~\cite{Krawczyk:2013jta}, the above studies are combined with constraints from XENON \cite{Aprile:2012nq} direct detection limits. In this work, the authors concentrate on low or middle regions for the dark matter mass, and mainly consider the case where $M_H\,\lesssim\,100\,\GeV$. Furthermore, they concentrate on identifying regions where the IDM constitutes the full dark matter content of the universe. Our study clearly differs insofar as we allow for lower relic density values, which in turn opens up the parameter space for higher dark matter masses. In addition, we provide results in terms of a full random scan, which is also not done in this work.
\item{}Ref.~ \cite{Goudelis:2013uca} is closely related to our work, where the authors in addition consider vacuum stability and perturbativity at arbitrary scales using renormalization group equations (RGEs). {As expected, requiring validity at higher scales constrains the allowed parameter space considerably. However, we find that after all constraints have been taken into account, parameters which are important for collider phenomenology are in similar ranges\footnote{{The only difference is that we allow for slightly larger negative values of $\lam_{345}$ for dark masses $\gtrsim\,600\,\GeV$, and a larger range of $\lam_2$. As discussed below, these parameters are of no relevance concerning LHC phenomenology.}}}. {T}he direct detection limits applied here follow from XENON \cite{Aprile:2012nq}. {Furthermore, out of the benchmark points presented here only one survives the updated constraints presented in our study.}
\item{}Ref.~ \cite{Arhrib:2013ela} seems to be closely related to our study. In this work, the authors basically include all above constraints apart from perturbativity of the Higgs self couplings, the total decay width of the 125 \GeV~ Higgs and the charged scalar; the inclusion of direct search bounds was persecuted on a slightly different footing.  Furthermore, limits from signal strength measurements were not done on the level of a global fit as implemented in \HS. However, here again the authors require {\sl exact} agreement with the dark matter relic density, while we include the latter requiring an upper bound only which immediately opens up larger parts of parameter space. The main focus of the authors {is} a presentation of the successive combination of constraints on profile likelihood fits, which they discuss in detail in their work. In our study, this is performed in a slightly different way, allowing the possibility to present results in terms of several individual two dimensional parameter planes directly showing the impact of separate constraints (cf. section \ref{sec:results}).
Whenever applicable, the regions they find consistent with all their bounds are indeed similar to the results obtained here, {which aligns with our results.} Furthermore, the authors do not provide predictions for production cross sections at the current LHC run, which we consider imminent.
\end{itemize}
\subsection{Previous benchmarks in the literature and exclusions}

In this section, we present benchmarks which have been discussed in the literature for the IDM in the context of LHC phenomenology, with Higgs masses roughly $\sim\,125\,\GeV$ (allowing for a range of $\pm\,5\,\GeV$ or similar), and the reasons why these are by now excluded, following the step of constraints as discussed in the main body of the manuscript. We do not claim this to be an exhaustive list of all benchmarks ever presented. We furthermore restrict ourselves to scalar masses $\,\leq\,1\,\TeV$, in order to maintain cross sections which might be probed at current colliders.
\begin{itemize}
\item{}Benchmarks "LH2, LH3, LH4, LH5" from \cite{Dolle:2009ft}\\
LH2: excluded by S,T,U, and total width of h\\
all others: excluded from ATLAS search \cite{Aad:2014ioa}
\item{}additional benchmarks "LH6, LH8" from \cite{Miao:2010rg}\\
LH6: excluded by total width of h\\
LH8: excluded from ATLAS search \cite{Aad:2014ioa} 
\item{}IDM benchmarks suggested in \cite{Abercrombie:2015wmb}\\
BM 1,3 : from STU\\
all others: from direct detection
\item{}benchmarks from \cite{Arhrib:2013ela}\\
Table II: first benchmark excluded from combination of LEP recast and decay width of electroweak gauge bosons; second benchmark excluded from dark matter relic density\footnote{Using micromegas 4.2.3, with exactly the same input values}, benchmark 3 from signal strength (too large discrepancy in $M_h$), benchmark 4 from perturbativity\\
Table III: first two benchmarks from perturbativity, last from STU
\item{}benchmarks from \cite{Cao:2007rm}\\
all ruled out from combination of LEP recast and electroweak gauge boson decay widths
\item benchmarks from \cite{Goudelis:2013uca}\\
BM I: excluded from direct detection\\
BM III: excluded from STU\\
BM II allowed
\item benchmark IV from \cite{Gustafsson:2007pc}\\
excluded by combination of LEP recast and electroweak gauge boson decays
\item benchmarks from table VII in \cite{Gustafsson:2012aj}\\
Benchmarks 1,6: excluded from direct detection\\
Benchmark 2: excluded from STU\\
all others allowed
\end{itemize}
For completeness, we provide production cross sections and dominant decay modes for scenarios which have been discussed above in table \ref{tab:gf}.
\begin{table}
\begin{tabular}{c|c|c|c}
&$\sigma_{AH} [\pb]$&$\sigma_{H^+H^-} [\pb]$& BR $\lb H^+\,\rightarrow\,W^+\,H \rb$\\ \hline
2&0.1275 (1)&0.03949(4)& 1.0\\
3&0.9208 (5)&0.1063 (2)& 0.68\\
4&0.2673 (3)&0.04908 (1) & 1.0\\ \hline
II&$2.1\,\times\,10^{-4}$  & $8.6\,\times\,10^{-6}$&0.81\\ \hline
1&$1.7\,\times\,10^{-4}$&$1.8\,\times\,10^{-4}$&0.91\\
2&$7.2\,\times\,10^{-6}$  &$7.7\,\times\,10^{-6}$&1.0\\
\end{tabular}
\caption{\label{tab:gf} production cross sections for valid benchmarks from \cite{Gustafsson:2012aj} (first three entries), \cite{Goudelis:2013uca} (middle entry, with roles of A and H interchanged), \cite{Garcia-Cely:2013zga} (last two entries) with an adjusted mass $M_h\,=\,125.1\,\GeV$}
\end{table}

\end{appendix}

\end{document}